\documentclass[fleqn,usenatbib]{mnras}

\usepackage{newtxtext,newtxmath}

\usepackage[T1]{fontenc}

\DeclareRobustCommand{\VAN}[3]{#2}
\let\VANthebibliography\thebibliography
\def\thebibliography{\DeclareRobustCommand{\VAN}[3]{##3}\VANthebibliography}

\usepackage{graphicx}	
\usepackage{amsmath}	
\usepackage{bm}        
\usepackage{physics}    
\usepackage{booktabs}
\usepackage{longtable} 
\usepackage{multirow}  
\usepackage{subcaption} 

\title[Crust-superfluid coupling from timing noise]{
    Measuring the crust-superfluid coupling time-scale for 105 UTMOST pulsars with a Kalman filter 
}

\author[Dong et al.]{Wenhao Dong,$^{1,2}$\thanks{E-mail: wddong@student.unimelb.edu.au}
Andrew Melatos,$^{1,2}$
Nicholas J. O'Neill,$^{1,2}$
Patrick M. Meyers$^{3,4}$
and Daniel K. Boek$^{1,2}$
\\
$^{1}$School of Physics, University of Melbourne, Parkville, VIC 3010, Australia. \\
$^{2}$ARC Centre of Excellence for Gravitational Wave Discovery (OzGrav), University of Melbourne, Parkville, VIC 3010, Australia. \\
$^{3}$Theoretical Astrophysics Group, California Institute of Technology, Pasadena, CA 91125, USA. \\
$^{4}$Institute for Particle Physics and Astrophysics, ETH Zurich, Wolfgang-Pauli-Strasse 27, 8093 Zurich, Switzerland.
}

\date{Accepted XXX. Received YYY; in original form ZZZ}

\pubyear{\the\year{}}

\begin{document}
\label{firstpage}
\pagerange{\pageref{firstpage}--\pageref{lastpage}}
\maketitle
\begin{abstract}
Crust-superfluid coupling plays an important role in neutron star rotation, particularly with respect to timing noise and glitches.
Here, we present new timing-noise-based estimates of the crust-superfluid coupling time-scale \(\tau\) for 105 radio pulsars in the UTMOST dataset, by Kalman filtering the pulse times of arrival.
The 105 objects are selected because they favor a two-component, crust-superfluid model over a one-component model with log Bayes factor \(\ln \mathfrak{B}_{\rm BF} \geq 5\).
The median estimate of \(\tau\) ranges from \(10^{4.6\pm0.4}\)\,s for PSR J2241$-$5236 to \(10^{7.7^{+0.7}_{-0.4}}\)\,s for PSR J1644$-$4559 among 28 out of 105 objects with sharply peaked \(\tau\) posteriors.
A hierarchical Bayesian analysis is performed on 101 out of 105 objects that are canonical (i.e.\ neither recycled nor magnetars) and reside in the populous core of the \(\Omega_{\rm c}\)-\(\dot{\Omega}_{\rm c}\) plane. 
It returns the population-level scaling \(\tau \propto \Omega_{\rm c}^{0.19^{+0.50}_{-0.52}} |\dot{\Omega}_{\rm c}|^{0.18^{+0.18}_{-0.19}}\), where \(\Omega_{\rm c}\) and \(\dot{\Omega}_{\rm c}\) are the angular velocity and spin-down rate of the crust respectively.
The variances of the stochastic crust and superfluid torques are also estimated hierarchically, with \(Q_{\rm c} \propto \Omega_{\rm c}^{1.23^{+0.80}_{-0.75}} |\dot{\Omega}_{\rm c}|^{0.49^{+0.27}_{-0.32}}\) and \(Q_{\rm s} \propto \Omega_{\rm c}^{0.71^{+0.76}_{-0.78}} |\dot{\Omega}_{\rm c}|^{1.27^{+0.30}_{-0.28}}\) respectively.
Implications for the physical origin of crust-superfluid coupling, e.g.\ through mutual friction, are discussed briefly.
\end{abstract}




\section{Introduction} \label{sec:introduction}

Multi-component models of neutron star interiors were first proposed to explain the rotational glitches\footnote{
    See e.g.\ \citet{HaskellMelatos2015,AntonelliEtAl2022,AntonopoulouEtAl2022,ZhouEtAl2022} for recent reviews of neutron star glitches.
} observed in the Vela pulsar \citep{BaymEtAl1969}, and have been generalized subsequently to describe modifications to secular spin down \citep{GoglichidzeEtAl2015,GoglichidzeBarsukov2019}, glitch rise times \citep{GraberEtAl2018,AshtonEtAl2019a,CeloraEtAl2020}, and multi-component glitch recoveries \citep{AlparEtAl1993,AlparEtAl1996,GraberEtAl2018,GugercinogluAlpar2020,PizzocheroEtAl2020,SourieChamel2020}.
In a two-component model, for example, the neutron star is assumed to comprise a solid crust and a superfluid core, which couple frictionally.
The angular velocity lag between the two components relaxes exponentially over a characteristic time-scale \(\tau\), termed the crust-superfluid coupling time-scale.
Crust-superfluid coupling is mediated by various mechanisms, e.g.\ viscosity in the Ekman layer at the crust-superfluid boundary \citep{Easson1979,AbneyEpstein1996,VanEysdenMelatos2010}, magnetic stresses \citep{Easson1979,Melatos2012,GlampedakisLasky2015,BransgroveEtAl2018}, and mutual friction arising from scattering between superfluid vortices and electrons \citep{AlparSauls1988,Mendell1991,AnderssonEtAl2006,GavassinoEtAl2020}.

Previous measurements of the crust-superfluid coupling time-scale typically involve measurements of glitch recovery time-scales \citep{McCullochEtAl1983,DodsonEtAl2001,WongEtAl2001} or glitch rates interpreted in terms of recoupling, vortex creep, or a state-dependent Poisson process \citep{AlparEtAl1984,AlparEtAl1993,AlparEtAl1996,GugercinogluAlpar2017,MelatosMillhouse2023}.
Alternatively, they involve measurements of timing noise statistics, which contain signatures of mean reversion on a time-scale related to \(\tau\), e.g.\ the autocorrelation time-scales in PSR J1136+1551 (B1133+16) and PSR J1935+1616 (B1933+16) \citep{PriceEtAl2012}, and the Kalman filter analysis of PSR J1359$-$6038 \citep{ONeillEtAl2024}.
However, until now, timing noise has not been exploited widely to measure the coupling time-scale, because mean reversion is obscured by other, unrelated noise processes, including interstellar medium fluctuations and pulse shape changes \citep{LiuEtAl2011,GoncharovEtAl2021}.

In this paper, we apply a Kalman filter \citep{MeyersEtAl2021,MeyersEtAl2021a,ONeillEtAl2024} to detect mean reversion in pulsar timing noise data and hence measure \(\tau\).
To this end, we analyze 286 pulsars from the UTMOST timing campaign conducted with the upgraded Molongolo Observatory Synthesis Telescope \citep{BailesEtAl2017}.
The 286 objects are selected by excluding pulsars that glitched during the UTMOST observing time-span, as recorded in the Australia Telescope National Facility (ATNF) pulsar catalog \citep{ManchesterEtAl2005}.
The UTMOST dataset delivers good timing precision (about \(5\,\mu\)s), and a spread of timing baselines ($1.0$--$4.8$\,yr) and observing cadences (daily to monthly) \citep{JankowskiEtAl2019,LowerEtAl2020}.
We combine the Kalman filter with a nested sampler to estimate the parameters of the two-component model including the crust-superfluid coupling time-scale \(\tau\) and the normalized crust and superfluid noise variances, denoted by \(Q_{\rm c}\) and \(Q_{\rm s}\) respectively.
We then test how \(\tau, Q_{\rm c}\), and \(Q_{\rm s}\) scale with the angular velocity \(\Omega_{\rm c}\) and spin-down rate \(\dot{\Omega}_{\rm c}\) of the crust across the sample.

The paper is organized as follows.
In Section~\ref{sec:two_component_model}, we review briefly the two-component model and associated Kalman filter used in this paper and previous work \citep{MeyersEtAl2021,MeyersEtAl2021a,ONeillEtAl2024}. 
In Section~\ref{sec:hierarchical_model-outline}, we describe and motivate two complementary Bayesian approaches for inferring \(\tau, Q_{\rm c}\), and \(Q_{\rm s}\) from the UTMOST data at the per-pulsar and population (i.e.\ hierarchical) levels.
The first approach, summarized in Section~\ref{subsec:hm-indiv_pop_simul}, is a general, hierarchical model which infers the properties of individual pulsars and their population simultaneously.
The second approach, summarized in Section~\ref{subsec:hm-indiv_pop_separate}, infers pulsar- and population-level properties separately and is more tractable computationally.
In Section~\ref{sec:model_selection_psr_level}, we perform Bayesian model selection between one- and two-component models and present posterior distributions for \(\tau, Q_{\rm c}\), and \(Q_{\rm s}\) for the 105 objects, whose timing data favor the two-component model.
These are the first timing-noise-based estimates of \(\tau\) (i.e.\ unrelated to glitch recoveries) in 102 UTMOST pulsars, including three millisecond pulsars.
In Section~\ref{sec:hierarchical_regression}, we estimate how \(\tau, Q_{\rm c}\), and \(Q_{\rm s}\) scale with \(\Omega_{\rm c}\) and \(\dot{\Omega}_{\rm c}\) within the framework of a population-level, hierarchical Bayesian model.
Comparisons with glitch-based inferences of \(\tau\), where available, are discussed in Section~\ref{sec:glitch_relaxation_comparison}.
Astrophysical implications are canvassed briefly in Section~\ref{sec:discussion_conclusion}.

\section{Two-component model} \label{sec:two_component_model}
In this paper, we analyze the coupling between the crust and superfluid of a neutron star in terms of the classic two-component model introduced by \citet{BaymEtAl1969}.
We review briefly the variables in the model and their equations of motion in Section~\ref{subsec:equations_of_motion}.
We then explain how the equations of motion map one-to-one onto the mathematical structure of a Kalman filter, which can be combined with a nested sampler to infer the model parameters given a sequence of pulse times of arrival (TOAs) generated by a pulsar timing experiment, e.g.\ UTMOST data.

\subsection{Equations of motion} \label{subsec:equations_of_motion}
The two-component model comprises a solid crust (labelled by the subscript `c') and a superfluid core (labelled by the subscript `s'), which rotate uniformly with angular velocities \(\Omega_{\rm c}\) and \(\Omega_{\rm s} > \Omega_{\rm c}\) respectively.
The equations of motion for the crust and superfluid take the form
\begin{align}
    I_{\rm c} \dot{\Omega}_{\rm c}
    &= -\frac{I_{\rm c}}{\tau_{\rm c}} (\Omega_{\rm c} - \Omega_{\rm s}) + N_{\rm c} + \xi_{\rm c}(t) 
    \label{eq:EOM_c}
    \\
    I_{\rm s} \dot{\Omega}_{\rm s}
    &= -\frac{I_{\rm s}}{\tau_{\rm s}} (\Omega_{\rm s} - \Omega_{\rm c}) + N_{\rm s} + \xi_{\rm s}(t) ,
    \label{eq:EOM_s}
\end{align}
where \(I_{\rm c}\) and \(I_{\rm s}\) denote effective moments of inertia, \(\tau_{\rm c}\) and \(\tau_{\rm s}\) are coupling time-scales related to friction coefficients, and \(N_{\rm c}\) and \(N_{\rm s}\) are deterministic external (e.g.\ electromagnetic) torques.
The stochastic torques \(\xi_{\rm c}(t)\) and \(\xi_{\rm s}(t)\) are modelled as white noise processes with zero means, \(\expval{\xi_{\rm c,s}(t)} = 0\), and variances $\sigma_{\rm c}^2$ and $\sigma_{\rm s}^2$ respectively, satisfying
\begin{align}
    \expval{\xi_{\rm c}(t) \xi_{\rm c}(t')} &= \sigma_{\rm c}^2 \delta(t-t') 
    \label{eq:xi_c_autocorr} \\
    \expval{\xi_{\rm s}(t) \xi_{\rm s}(t')} &= \sigma_{\rm s}^2 \delta(t-t') .
    \label{eq:xi_s_autocorr}
\end{align}
The angular brackets \(\langle \dots \rangle\) denote the ensemble average over many random realizations of \(\xi_{\rm c}(t)\) and \(\xi_{\rm s}(t)\).
No cross-correlation between \(\xi_{\rm c}(t)\) and \(\xi_{\rm s}(t)\) is assumed as a first pass, i.e. \(\langle\xi_{\rm c}(t) \xi_{\rm s}(t') \rangle = 0\), but it is easy to generalize to \(\langle \xi_{\rm c}(t) \xi_{\rm s}(t')\rangle \neq 0\) in the future, if data warrant. We define \(Q_{\rm s,c} = \sigma_{\rm s,c}^2 / I_{\rm s,c}^2\) for notation convenience in what follows.

Impulsive changes in \(\Omega_{\rm c}\) and \(\Omega_{\rm s}\), e.g.\ during a glitch, decay exponentially on the time-scale
\begin{align}
    \tau = \left(\frac{1}{\tau_{\rm c}} + \frac{1}{\tau_{\rm s}}\right)^{-1} .
    \label{eq:tau}
\end{align}
The composite parameter \(\tau\) is identifiable formally, whereas \(\tau_{\rm c}\) and \(\tau_{\rm s}\) individually are not identifiable.
That is, \(\tau\) can be inferred from a sequence of pulse TOAs in the formal sense stipulated in the electrical engineering literature \citep{BellmanAstrom1970}, e.g.\ with a Kalman filter, whereas \(\tau_{\rm c}\) and \(\tau_{\rm s}\) cannot be disentangled and inferred separately; see Section~2.3 in \citet{MeyersEtAl2021a} and Section~3.2 in \citet{ONeillEtAl2024}.
Throughout the rest of this paper, we focus on \(\tau\) instead of \(\tau_{\rm c}\) and \(\tau_{\rm s}\), except briefly in Section~\ref{subsec:healing_parameter}, where we relate \(\tau_{\rm s} / \tau_{\rm c}\) to glitch-based estimates of \(I_{\rm s} / I_{\rm c}\).

In this paper, we apply the equations of motion \eqref{eq:EOM_c}--\eqref{eq:xi_s_autocorr} to analyze radio pulsar timing data in the form of TOAs generated by standard timing software like \textsc{tempo2} \citep{HobbsEtAl2006,EdwardsEtAl2006}.
TOAs are related to the rotational phase of the crust, \(\phi_{\rm c}\), via a linear transformation.
Therefore, we supplement \eqref{eq:EOM_c}--\eqref{eq:xi_s_autocorr} with an additional equation of motion,
\begin{align}
    \frac{\mathrm{d} \phi_{\rm c}}{\mathrm{d} t} = \Omega_{\rm c} .
    \label{eq:phi_c_EOM}
\end{align}
For the sake of completeness, we also solve the analogous equation for the rotational phase of the superfluid, \(\phi_{\rm s}(t)\), which cannot be observed in pulsar timing experiments, because the radio beam is tied to the stellar crust rather than the core.
By tracking \(\phi_{\rm c}(t)\) instead of \(\Omega_{\rm c}(t)\), we generalize the analysis of PSR J1359$-$6038 by \citet{ONeillEtAl2024}, who constructed \(\Omega_{\rm c}(t)\) from a sequence of TOAs by finite differencing instead of Kalman filtering the TOAs directly.

The measurement errors \(\Delta t\) are mapped to the errors in \(\phi_{\rm c}(t)\) via \(\Delta \phi_{\rm c} = \Omega_{\rm c}(t=0) \Delta t\).
The approximation \(\Omega_{\rm c}(t) \approx \Omega_{\rm c}(t=0)\) in this context introduces discrepancies of order \(\dot{\Omega}_{\rm c} t \Delta t\), which are negligible for \(T_{\rm obs} \ll \Omega_{\rm c} / |\dot{\Omega}_{\rm c}|\), where \(T_{\rm obs}\) is the total observing time-span.

\subsection{Kalman filter} \label{subsec:kalman_filter}
The Kalman filter is a recursive estimator, which applies to any dynamical system, whose evolution is described by stochastic differential equations like \eqref{eq:EOM_c}--\eqref{eq:xi_s_autocorr} and \eqref{eq:phi_c_EOM}.
It discretizes the evolution, so that the state vector \(\bm{X}_{n} = [\phi_{\rm c}(t_n), \phi_{\rm s}(t_n), \Omega_{\rm c}(t_n), \Omega_{\rm s}(t_n)]^T\) is sampled at \(N_{\rm TOA}\) discrete time steps \(t_n\), with \(1 \leq n \leq N_{\rm TOA}\).
The discretized forms of \eqref{eq:EOM_c}--\eqref{eq:xi_s_autocorr} are modified from Appendix~C in \citet{MeyersEtAl2021a} and \citet{ONeillEtAl2024} to include the phase evolution via \eqref{eq:phi_c_EOM}.
The equations of motion are supplemented by a measurement equation
\begin{align}
    \bm{Y}_{n} = \mathbfss{H} \bm{X}_{n} + \bm{v}_{n} ,
    \label{eq:measurement_equation}
\end{align}
which relates a vector of measured data \(\bm{Y}_{n}\) at time \(t_n\) to \(\bm{X}_{n}\).
The vectors \(\bm{Y}_{n}\) and \(\bm{X}_{n}\) do not contain the same components, nor do they share the same dimension in general.
Here, for example, the pulse TOAs are the only observables, and \(\bm{Y}_{n} = \phi_{\rm c}(t_n)\) is a scalar, with
\begin{align}
    \mathbfss{H} = \begin{bmatrix} 1 & 0 & 0 & 0 \end{bmatrix} .
    \label{eq:measurement_matrix}
\end{align}
In \eqref{eq:measurement_equation}, \(\bm{v}_{n}\) denotes the measurement noise, which is assumed to be additive and Gaussian in this paper.
The Kalman filter returns an estimated time series \(\hat{\bm{X}}_{n}\), which minimizes the mean square error at each \(t_n\) \citep{Kalman1960}.
The Kalman filter is the optimal linear filter, provided that the dynamics are linear [e.g.\ equations~\eqref{eq:EOM_c}--\eqref{eq:EOM_s} and \eqref{eq:phi_c_EOM}], and the process and measurement noises [e.g.\ \(\xi_{\rm c,s}\) and \(\bm{v}_{n}\)] have finite first and second moments, regardless of whether their statistics are Gaussian \citep{AndersonMoore2005,Simon2006,UhlmannJulier2022}.\footnote{
    The Kalman filter is also optimal among linear filters, when the noises have finite second order moments, but the dynamics are nonlinear.
    Nonlinear generalizations of the linear Kalman filter include the extended Kalman filter and unscented Kalman filter \citep{JulierUhlmann1997,WanVanDerMerwe2000,AndersonMoore2005,Simon2006}.
    We do not analyze nonlinear dynamics in this paper, because the standard two-component model \eqref{eq:EOM_c}--\eqref{eq:phi_c_EOM} is linear.
}

Recursion relations for the predict-correct steps in the Kalman filter are given in Appendix~C of \citet{MeyersEtAl2021a} and \citet{ONeillEtAl2024}.
The filter runs through the data for \(1 \leq n \leq N_{\rm TOA}\) with fixed trial values of the static parameters \(\tau_{\rm s} / \tau_{\rm c}, \tau, Q_{\rm c}, Q_{\rm s}, \langle\dot{\Omega}_{\rm c}\rangle, \langle\Omega_{\rm c} - \Omega_{\rm s}\rangle\) and \(\Omega_{\rm c,0} = \Omega_{\rm c}(t=0)\) and computes a likelihood.
A separate sampling routine then seeks to maximize the likelihood by stepping to a revised set of trial values nearby and executing the filter for \(1 \leq n \leq N_{\rm TOA}\) again, using gradient methods or their equivalent to perform the maximization.
In this paper, we perform the parameter search using a nested sampler \citep{Skilling2006,AshtonEtAl2019a}, which maximizes the Kalman log-likelihood
\begin{align}
    &\ln \mathcal{L}(\{\bm{Y}_{n}\}_{1 \leq n \leq N_{\rm TOA}} | \bm{\theta}) 
    \nonumber \\
    &= -\frac{1}{2} \sum_{n=1}^{N_{\rm TOA}} \left[ \mathrm{dim}_{\bm{Y}} \ln(2\pi) + \ln \det(\mathbfss{S}_{n}) + \bm{\epsilon}_{n}^{\rm T} \mathbfss{S}_{n}^{-1} \bm{\epsilon}_{n} \right] ,
    \label{eq:kalman_likelihood}
\end{align}
where \(\{\bm{Y}_{n}\}_{1 \leq n \leq N_{\rm TOA}}\) denotes the sequence of measurements up to \(t_{N_{\rm TOA}}\), dim$_{\bm{Y}}$ refers to the dimension of \(\bm{Y}_{n}\) (here unity), \(\bm{\theta} = (\tau_{\rm s} / \tau_{\rm c}, \tau, Q_{\rm c}, Q_{\rm s}, \langle\dot{\Omega}_{\rm c}\rangle, \langle\Omega_{\rm c} - \Omega_{\rm s}\rangle, \Omega_{\rm c,0})\) is the static parameter vector, and we write \(\bm{\epsilon}_{n} = \bm{Y}_{n} - \mathbfss{H}\bm{X}_{n | n-1}\) and \(\mathbfss{S}_{n} = \mathrm{var}(\bm{\epsilon}_{n})\). 
Here, \(\bm{X}_{n | n-1}\) denotes the predicted state at time \(t_n\) given the estimated state \(\hat{\bm{X}}_{n-1}\).
Full details about how to implement the Kalman filter in this context are set out in Appendix~C in \citet{ONeillEtAl2024}.

\subsection{Physical interpretation of the model variables}
The physical meanings of the variables \(\Omega_{\rm c, 0}\), \(\langle \dot{\Omega}_{\rm c} \rangle = \langle \dot{\Omega}_{\rm s} \rangle\), and \(\langle \Omega_{\rm c} - \Omega_{\rm s} \rangle\) are straightforward.
The variables correspond to the initial angular velocity of the crust, long-term spin-down rate of the crust or superfluid, and the steady-state lag between the crust and superfluid, respectively.
The composite timescale \(\tau\) is the $e$-folding timescale over which \(\Omega_{\rm c} - \Omega_{\rm s}\) responds after impulsive changes in \(\Omega_{\rm c}\) and \(\Omega_{\rm s}\), and describes how quickly the crust couples to the superfluid.
We discuss its relation to the glitch recovery timescale \(\tau_{\rm g}\) in Section~\ref{subsec:glitch_tau_comparison} and Appendix~\ref{app:healing_in_2C}.
The ratio \(\tau_{\rm s} / \tau_{\rm c}\) can be interpreted as the ratio of the moments of inertia of the superfluid and crust, \(I_{\rm c} / I_{\rm s}\), if the internal torques \(I_{\rm c} (\Omega_{\rm c} - \Omega_{\rm s}) / \tau_{\rm c}\) and \(I_{\rm s} (\Omega_{\rm c} - \Omega_{\rm s}) / \tau_{\rm s}\) form an action-reaction pair.

The normalized noise variances \(Q_{\rm c}\) and \(Q_{\rm s}\) have units of rad\(^2\)s\(^{-3}\).
Dimensionally, they may be approximated as \(Q_{\rm c,s} \sim \langle \delta W_{\rm c,s}^2 \rangle \tau_{\rm corr} / I_{\rm c,s}^2\), where \(\delta W_{\rm c}\) and \(\delta W_{\rm s}\) are the amplitudes of stochastic torque fluctuations acting on the crust and superfluid, respectively, and \(\tau_{\rm corr}\) is the correlation time-scale of the fluctuations.
The physical origin of \(\delta W_{\rm c}\) and \(\delta W_{\rm s}\) depends on the astrophysical context.
For example, in isolated objects, sporadic avalanches of pinned superfluid vortices transfer angular momentum to and from the crust in an erratic yet persistent manner, resulting in \(\delta W_{\rm s}\) and \(\delta W_{\rm c} \sim -I_{\rm s} \delta W_{\rm s} / I_{\rm c}\) \citep{LambEtAl1978,DrummondMelatos2018}.
Quantum mechanical simulations show that the detailed form of the angular momentum transfer is governed by the pinning strength and the angle between the rotation and magnetic axes \citep{WarszawskiMelatos2011,DrummondMelatos2018,LonnbornEtAl2019}.
In addition, \(\delta W_{\rm c}\) may also arise from variations in the electromagnetic braking torque, which is associated with pair production in the magnetosphere \citep{Cheng1987,StairsEtAl2019}.
Hydrodynamic turbulence in the superfluid \citep{MelatosLink2014} and gravitational waves from stellar oscillation modes \citep[e.g.\ excited during a glitch, see][]{SideryEtAl2010} also contribute to \(\delta W_{\rm s}\).
These processes are idealized as white noises in this paper, assuming \(\tau_{\rm corr} \ll \tau\) [e.g.\ \(\tau_{\rm corr} \sim 10\,\rm{ms}\) for outer-gap pair production \citep{Cheng1987}].

\section{Hierarchical Bayesian model} \label{sec:hierarchical_model-outline}
At the time of writing, \(\tau\) has been inferred from timing noise data in just a handful of pulsars, e.g.\ PSR J1136+1551, PSR J1935+1616, and PSR J1359$-$6038 \citep{PriceEtAl2012,ONeillEtAl2024}.
Consequently, little is known about what objects are described well by the two-component model and what objects are not, and little is known about the distribution of \(\tau\) across the pulsar population.
More is known about the \(\tau\) distribution inferred from glitch recoveries \citep{McCullochEtAl1983,AlparEtAl1993,AlparEtAl1996,DodsonEtAl2001,WongEtAl2001,Gugercinoglu2017}, but there is no guarantee that the same relaxation physics operates during glitches and in timing noise. 
Any hypothetical connection should be tested observationally without being presumed.

The twin goals of this paper are to measure \(\tau\) on a per-object basis, for as many UTMOST objects as possible, as well as infer the population-level distribution of \(\tau\).
Both goals can be achieved simultaneously through a hierarchical Bayesian analysis \citep{GelmanEtAl2013}.
It is prudent to split the analysis into two parts, as this is the first timing noise measurement of \(\tau\) in many objects.
Firstly, we analyze the selected sample of 286 UTMOST pulsars individually and independently, within a non-hierarchical Bayesian framework, to check whether each object is described better by the one- or two-component model (see Section~\ref{subsec:selection-workflow}), and to infer per-object posteriors on \(\tau\), \(Q_{\rm c}\), and \(Q_{\rm s}\), when the two-component model is preferred.
Secondly, we combine the per-pulsar posteriors through a hierarchical reweighting scheme to infer population-level posteriors for \(\tau\), \(Q_{\rm c}\), and \(Q_{\rm s}\).
The reweighting scheme is a standard and widely used approximation to an exact hierarchical scheme, in situations where the number of objects is moderate (\(\sim 10^2\) here), and computational resources are limited.

In Section~\ref{subsec:hm-indiv_pop_simul}, we introduce an exact hierarchical Bayesian model, with which one can jointly infer pulsar- and population-level posteriors for \(\tau\), \(Q_{\rm c}\), and \(Q_{\rm s}\).
We then show in Section~\ref{subsec:hm-indiv_pop_separate} how to approximate the exact model through a reweighting scheme \citep{ThraneTalbot2019,MooreGerosa2021,EssickEtAl2022,GoncharovSardana2025}, to generate pulsar- and population-level posteriors separately while simultaneously selecting between the one- and two-component models.
The approach in Section~\ref{subsec:hm-indiv_pop_separate} is applied to UTMOST timing data in Section~\ref{sec:model_selection_psr_level} to perform pulsar-level model selection and parameter estimation, partly as a consistency check, and partly to orient the reader.
The pulsar-level results are then combined through reweighting to infer hierarchical population-level posteriors in Section~\ref{sec:hierarchical_regression}.

\subsection{Joint inference at the pulsar and population levels} \label{subsec:hm-indiv_pop_simul}
Let us assume that the \(N_{\rm psr}\) UTMOST pulsars are drawn from a single population-level distribution, as far as the parameters \(\tau\), \(Q_{\rm c}\), and \(Q_{\rm s}\) are concerned.
That is, they all share the same internal frictional response to stochastic crust and superfluid torques, as in Section~\ref{subsec:equations_of_motion}.
Let \(\bm{\theta}^{(i)} = [\tau^{(i)}, Q_{\rm c}^{(i)}, Q_{\rm s}^{(i)}]\) denote the parameters of the \(i\)-th pulsar, let \(\bm{d}^{(i)} = [Y_1^{(i)}, \ldots, Y_{N_{\rm TOA}^{(i)}}^{(i)}]\) denote the timing data gathered for the \(i\)-th pulsar (viz.\ \(N_{\rm TOA}^{(i)}\) TOAs \(t_1, \ldots, t_{N_{\rm TOA}^{(i)}}\)), and let \(\bm{\Lambda}\) denote the hyperparameters of the population-level distribution.
The full joint posterior for such a hierarchical model can be written as
\begin{align}
    p[\{\bm{\theta}^{(i)}\}, \bm{\Lambda} \,|\, \{\bm{d}^{(i)}\}]
    &\propto \mathcal{L}[\{\bm{d}^{(i)}\} \,|\, \{\bm{\theta}^{(i)}\}, \bm{\Lambda}] \, p[\{\bm{\theta}^{(i)}\}, \bm{\Lambda}],
    \label{eq:full_hyperposterior-decomposition-1}
    \\
    &= \pi(\bm{\Lambda}) \prod_{i=1}^{N_{\rm psr}} \mathcal{L}[\bm{d}^{(i)} \,|\, \bm{\theta}^{(i)}] p[\bm{\theta}^{(i)} \,|\, \bm{\Lambda}] .
    \label{eq:full_hyperposterior-decomposition-2}
\end{align}
In \eqref{eq:full_hyperposterior-decomposition-2}, \(\pi(\bm{\Lambda})\) is the prior distribution of the hyperparameters, \(p[\bm{\theta}^{(i)} \,|\, \bm{\Lambda}]\) is the population-level distribution of the per-pulsar parameters \(\bm{\theta}^{(i)}\) given \(\bm{\Lambda}\), \(\mathcal{L}[\bm{d}^{(i)} \,|\, \bm{\theta}^{(i)}]\) is the likelihood of the data \(\bm{d}^{(i)}\), and curly braces denote sets built from the union of the \(N_{\rm psr}\) objects labelled by \(1 \leq i \leq N_{\rm psr}\).
One passes from \eqref{eq:full_hyperposterior-decomposition-1} to \eqref{eq:full_hyperposterior-decomposition-2} by assuming exchangeability \citep{GelmanEtAl2013}, i.e.\ the joint distribution of \(\{\bm{\theta}^{(i)}\}\) is invariant upon permuting the index \(i\), which is consistent with the factorisations
\begin{align}
    \mathcal{L}[\{\bm{d}^{(i)}\} \,|\, \{\bm{\theta}^{(i)}\}, \bm{\Lambda}] 
    &= \mathcal{L}[\{\bm{d}^{(i)}\} \,|\, \{\bm{\theta}^{(i)}\}] 
    \label{eq:joint_likelihood_hyperparamDrop}
    \\
    &= \prod_{i=1}^{N_{\rm psr}} \mathcal{L}[\bm{d}^{(i)} \,|\, \bm{\theta}^{(i)}],
    \label{eq:joint_likelihood_factorisation}
\end{align}
and
\begin{align}
    p[\{\bm{\theta}^{(i)}\}, \bm{\Lambda}] 
    &= \pi(\bm{\Lambda}) \, \prod_{i=1}^{N_{\rm psr}} p[\bm{\theta}^{(i)} \,|\, \bm{\Lambda}] .
\end{align}
That is, exchangeability amounts to assuming a priori (before analyzing the data) that all \(N_{\rm psr}\) objects satisfy the same physical model (see Section~\ref{subsec:equations_of_motion}) and are observed through identical timing experiments.
These assumptions can be relaxed partly or completely in future work, if the data or improved prior knowledge demand, or when a gravitational wave background is considered \citep{AgazieEtAl2023a,AntoniadisEtAl2023,DiMarcoEtAl2023,ReardonEtAl2023,XuEtAl2023,Haasteren2024}.
The simplification in \eqref{eq:joint_likelihood_hyperparamDrop} holds, because we assume that the hyperparameters \(\bm{\Lambda}\) affect \(\bm{d}^{(i)}\) only through \(\bm{\theta}^{(i)}\).

The joint posterior distribution \(p[\{\bm{\theta}^{(i)}\}, \bm{\Lambda} \,|\, \{\bm{d}^{(i)}\}]\) is (\(N_{\rm psr} N_{\theta} + N_{\Lambda}\))-dimensional, where \(N_{\theta}\) is the number of parameters per pulsar and \(N_{\Lambda}\) is the number of hyperparameters.
It can be calculated by sampling the (\(N_{\rm psr} N_{\theta} + N_{\Lambda}\))-dimensional region of prior support.
In this paper, we use the \textsc{dynesty} nested sampler \citep{Speagle2020} to perform the task.
With \(p[\{\bm{\theta}^{(i)}\}, \bm{\Lambda} \,|\, \{\bm{d}^{(i)}\}]\) in hand, we calculate posterior distributions of the hyperparameters and per-pulsar parameters respectively by marginalizing over the other parameters in both cases, viz.\
\begin{align}
    p[\bm{\Lambda} \,|\, \{\bm{d}^{(i)}\}] = \int \mathrm{d} \{\bm{\theta}^{(i)}\} \, p[\{\bm{\theta}^{(i)}\}, \bm{\Lambda} \,|\, \{\bm{d}^{(i)}\}]
    \label{eq:marginalised_hyperposterior_def}
\end{align}
and
\begin{align}
    p[\bm{\theta}^{(i)} \,|\, \{\bm{d}^{(i)}\}] = \int \mathrm{d}\bm{\Lambda} \, \int \mathrm{d}\{\bm{\theta}^{(j)}\}_{j\neq i} \, p[\{\bm{\theta}^{(j)}\}, \bm{\Lambda} \,|\, \{\bm{d}^{(i)}\}] .
    \label{eq:marginalised_individual_posterior_def}
\end{align}
In this paper, we have \(N_{\rm psr} \sim 10^2\), \(N_{\theta} \sim 10\), and \(N_{\Lambda} \sim 10\); see Section~\ref{sec:model_selection_psr_level} and Section~\ref{sec:hierarchical_regression} for how \(\bm{\theta}^{(i)}\) and \(\bm{\Lambda}\) are defined for the models we study.
To render the computational burden manageable, we separate the pulsar- and population-level sampling with the aid of a standard reweighting approximation, as described in Section~\ref{subsec:hm-indiv_pop_separate}.

\subsection{Reweighting to separate the pulsar and population levels} \label{subsec:hm-indiv_pop_separate}
To evaluate efficiently the marginalized posteriors \eqref{eq:marginalised_hyperposterior_def} and \eqref{eq:marginalised_individual_posterior_def}, while the exchangeability condition holds, we approximate them using weighted samples from the individual per-pulsar posteriors \(p[\bm{\theta}^{(i)} \,|\, \bm{d}^{(i)}]\) \citep{ThraneTalbot2019,MooreGerosa2021,EssickEtAl2022,GoncharovSardana2025}.
Notationally, the per-pulsar posteriors drop the curly braces around \(\bm{d}^{(i)}\), and the estimate of \(\bm{\theta}^{(i)}\) depends only on the data from the \(i\)-th pulsar.
Reweighting proceeds as follows.
\begin{enumerate}
    \item[Step 1.] Sample the individual per-pulsar posteriors \(p[\bm{\theta}^{(i)} \,|\, \bm{d}^{(i)}]\) with uninformative or weakly informative prior distributions \(\pi[\bm{\theta}^{(i)}]\) for \(1 \leq i \leq N_{\rm psr}\), e.g.\ using a nested sampler. No hierarchical structure is assumed at this stage.
    \item[Step 2.] Sample the hyperparameter posterior, \(p[\bm{\Lambda} \,|\, \{\bm{d}^{(i)}\}]\), by performing importance reweighting as part of the Monte-Carlo marginalization of the likelihood \(\mathcal{L}(\{\bm{d}^{(i)}\} \,|\, \bm{\Lambda})\), through the formula \citep{ThraneTalbot2019}
    \begin{align}
        p[\bm{\Lambda} \,|\, \{\bm{d}^{(i)}\}] 
        &\propto \pi(\bm{\Lambda}) 
        \prod_{i=1}^{N_{\rm psr}} \frac{1}{n_{\mathrm{s}, i}} 
        \sum_{j=1}^{n_{\mathrm{s}, i}} \frac{p[\bm{\theta}^{(i)}_{j} \,|\, \bm{\Lambda}]}{\pi[\bm{\theta}^{(i)}_{j}]} .
        \label{eq:hyperparam_posterior_reweighting}
    \end{align}
    In \eqref{eq:hyperparam_posterior_reweighting}, we draw \(n_{\mathrm{s}, i}\) random samples \(\theta_{j}^{(i)}\) from \(p[\bm{\theta}^{(i)} \,|\, \bm{d}^{(i)}]\), calculated in step 1;
    the subscript \(1 \leq j \leq n_{\mathrm{s}, i}\) denotes the \(j\)-th random sample.
    The prior distribution \(\pi[\bm{\theta}^{(i)}]\) from step 1 does not encode population information and is not the same as the population-level prior \(p[\bm{\theta}^{(i)} \,|\, \bm{\Lambda}]\).
    \item[Step 3.] Draw \(n_{\mathrm{s}, \Lambda}\) samples from \(p[\bm{\Lambda} \,|\, \{\bm{d}^{(i)}\}]\) computed in step 2.
    Then reweight \(p[\bm{\theta}^{(i)} \,|\, \bm{d}^{(i)}]\) for each \(1 \leq i \leq N_{\rm psr}\) according to \citep{MooreGerosa2021}
    \begin{align}
        p[\bm{\theta}^{(i)} \,|\, \{\bm{d}^{(i)}\}]
        &\propto p[\bm{\theta}^{(i)} \,|\, \bm{d}^{(i)}] w[\bm{\theta}^{(i)}] ,
        \label{eq:indiv_posterior_reweighting}
    \end{align}
    with
    \begin{align}
        w[\bm{\theta}^{(i)}]
        &= \frac{n_{\mathrm{s}, i}}{n_{\rm s, \Lambda} \pi[\bm{\theta}^{(i)}]} 
        \sum_{k=1}^{n_{\rm s, \Lambda}} 
        \frac{
            p[\bm{\theta}^{(i)} \,|\, \bm{\Lambda}_{k}]
        }{
            \sum_{j=1}^{n_{\mathrm{s}, i}} p[\bm{\theta}^{(i)}_{j} \,|\, \bm{\Lambda}_{k}] \,/\, \pi[\bm{\theta}^{(i)}_{j}]
        }.
        \label{eq:indiv_posterior_reweighting_weight}
    \end{align}
    In \eqref{eq:indiv_posterior_reweighting_weight}, \(\bm{\Lambda}_{k}\) is the \(k\)-th random sample (out of \(n_{\rm s, \Lambda}\) random samples in total) drawn from \(p[\bm{\Lambda} \,|\, \{\bm{d}^{(i)}\}]\), and \(p[\bm{\theta}^{(i)}_{j} \,|\, \bm{\Lambda}_{k}]\) is the probability for the \(j\)-th sample of \(\bm{\theta}^{(i)}\) given \(\bm{\Lambda}_{k}\).
    The weights \(w[\bm{\theta}^{(i)}]\) capture population information contained in \(\{\bm{d}^{(j)}\}_{j \neq i}\).
    The justification of \eqref{eq:indiv_posterior_reweighting_weight}, including why it is free from double-counting, is given in Appendix~\ref{app:derivation_popinform_posterior_reweighting}.
\end{enumerate}

The reweighting approximation in this section confers benefits beyond computational efficiency.
By evaluating Bayes' theorem on a per-pulsar basis first, before overlaying population-level constraints, we can identify which of the 286 UTMOST pulsars are described better by the two-component model than the one-component model.
This intermediate step is especially valuable in the context of this paper, where we analyze many UTMOST pulsars for the first time in terms of the two-component model.
Consistency checks along the way are a prudent measure, especially as the crust-superfluid coupling physics and its population-level properties are uncertain at the time of writing \citep{HaskellMelatos2015,AntonelliEtAl2022,AntonopoulouEtAl2022,ZhouEtAl2022}.

\subsection{Two types of posterior distributions}
\label{subsec:two_types_of_posteriors}
It is important to emphasize that, in this paper, we calculate two types of posterior distributions for the per-pulsar parameters \(\bm{\theta}^{(i)}\) of the \(i\)-th pulsar.
In Section~\ref{sec:model_selection_psr_level}, we calculate the population-uninformed posterior \(p[\bm{\theta}^{(i)} \,|\, \bm{d}^{(i)}]\) by applying the data from the \(i\)-th pulsar only to infer \(\bm{\theta}^{(i)}\) without admitting the population constraints implied by the data from the other pulsars, \(\{\bm{\theta}^{(j)}\}_{j \neq i}\).
In Section~\ref{sec:hierarchical_regression}, we calculate the population-informed posterior \(p[\bm{\theta}^{(i)} \,|\, \{\bm{d}^{(i)}\}]\) by applying the data from all \(N_{\rm psr}\) pulsars simultaneously to infer \(\bm{\theta}^{(i)}\).
In practice, the latter calculation is performed approximately, by reweighting \(p[\bm{\theta}^{(i)} \,|\, \bm{d}^{(i)}]\) according to the scheme in Section~\ref{subsec:hm-indiv_pop_separate}.
Comparing the two posteriors is illuminating when diagnosing outliers in the population and discerning whether per-pulsar data or population constraints dominate the inference output at the population level.

\section{Crust-superfluid coupling at the pulsar level} \label{sec:model_selection_psr_level}
In this section, Bayesian model selection is performed separately on every nonglitching UTMOST pulsar, as foreshadowed in Section~\ref{sec:hierarchical_model-outline}, comparing the two-component model against a one-component, white-timing-noise (WTN) model.
For objects whose data favor the two-component model, we also infer per-pulsar posteriors for the two-component model parameters, without overlaying population-level constraints. (The population-level overlay is performed in Section~\ref{sec:hierarchical_regression}.)
We present the model selection workflow in Section~\ref{subsec:selection-workflow}, followed by summaries of the model selection and parameter estimation results in Sections~\ref{subsec:model_preference-psr-level} and \ref{subsec:tau-psr-level}, respectively.
Bayes factors and point (median) parameter estimates of \(\tau\), \(Q_{\rm c}\), and \(Q_{\rm s}\) with uncertainties given by the 68\% credible intervals are tabulated in Appendix~\ref{app:psr_level-model_comparison-parameter}.
The mapping from \(\tau\), \(Q_{\rm c}\), and \(Q_{\rm s}\) to traditional measures of timing noise strength is discussed in Section~\ref{subsec:mapping_to_sigma_TN}.
We highlight some noteworthy objects in Section~\ref{subsec:noteworthy_objects}, including three recycled pulsars favored by the two-component model, a magnetar (PSR J1622$-$4950), an object whose \(\tau\) posterior is bimodal (PSR J1141$-$6545), and PSR J1136+1551 and PSR J1935+1616, which are the subject of previous timing noise autocorrelation studies \citep{PriceEtAl2012}.

\subsection{Model selection workflow} \label{subsec:selection-workflow}
We use the \textsc{dynesty} sampler \citep{Speagle2020} with \textsc{Bilby} as the front end \citep{AshtonEtAl2019} to run the following two models on the phase data:
\begin{itemize}
    \item The one-component WTN model (M$_1$), which assumes that the measurement noise is white, and that the deterministic evolution \(\phi_{\rm c}(t) = \phi_{\rm c,0} + \Omega_{\rm c,0} t + \langle \dot{\Omega}_{\rm c} \rangle t^2 / 2\) is governed by \(\langle \dot{\Omega}_{\rm c} \rangle\) and \(\Omega_{\rm c,0}\).
    In this model, we fit four parameters: \(\langle \dot{\Omega}_{\rm c} \rangle, \Omega_{\rm c,0}\), EFAC and EQUAD.
    \item The two-component model (M$_2$) described in Section~\ref{subsec:equations_of_motion}.
    In this model, we fit nine parameters: \(\tau_{\rm s} / \tau_{\rm c}, \tau, Q_{\rm c}\), \(Q_{\rm s}\), \(\langle\Omega_{\rm{c}} - \Omega_{\rm{s}}\rangle\), \(\langle\dot{\Omega}_{\rm c}\rangle\), \(\Omega_{\rm c,0}\), EFAC and EQUAD.
\end{itemize}
EFAC and EQUAD are standard parameters in pulsar timing which correct for underestimating TOA uncertainties. EFAC is a multiplicative constant factor, and EQUAD adds in quadrature to the uncertainties.
We write the corrected phase error as \((\Delta \phi_{\rm c}')^2 = \textrm{EFAC} \times (\Delta \phi_{\rm c})^2 + \textrm{EQUAD}\).

Table~\ref{tab:psrlevel_priors} records the prior distributions for the parameters in M$_1$ and M$_2$.
The prior ranges are wider than those assumed by \citet{ONeillEtAl2024}, to be conservative, recognizing that the theoretical uncertainties concerning \(\tau, Q_{\rm c}\), and \(Q_{\rm s}\) are considerable \citep{HaskellMelatos2015,AntonelliEtAl2022,AntonopoulouEtAl2022,ZhouEtAl2022}.

To select between M$_1$ and M$_2$ for every pulsar, we calculate the Bayes factor 
\begin{align}
    \mathfrak{B}_{\rm BF} 
    &= \frac{\int \mathrm{d} \bm{\theta}_2 \, \mathcal{L}(\bm{d} | \bm{\theta}_2, \mathrm{M}_2) \pi(\bm{\theta}_2 | \mathrm{M}_2)}{\int \mathrm{d} \bm{\theta}_1 \, \mathcal{L}(\bm{d} | \bm{\theta}_1, \mathrm{M}_1) \pi(\bm{\theta}_1 | \mathrm{M}_1)} .
    \label{eq:bayes_factor}
\end{align}
We adopt \(\ln \mathfrak{B}_{\rm BF} \geq 5\) as the arbitrary threshold for preferring strongly the two-component model. The same criterion is adopted elsewhere in the literature \citep{LowerEtAl2020}.

\begin{table*}
    \centering
    \caption{Prior distributions of per-pulsar parameters in the one-component (WTN) and two-component models, including the functional form (column~2) and the minimum (column~3) and maximum (column~4), where there is prior support.
    The quantities \(\hat{\Omega}_{\rm c,0}, \langle\hat{\dot{\Omega}}_{\rm c}\rangle\), \(\Delta_{\Omega_{\rm c,0}}\), and \(\Delta_{\langle\dot{\Omega}_{\rm c}\rangle}\) are the estimates and uncertainties returned by fitting \(\phi_{\rm c}(t)\) with a quadratic Taylor series in $t$ using \texttt{numpy.polyfit}.}
    \label{tab:psrlevel_priors}
    \begin{tabular}{c|l|c|c|c}
        \hline
        Parameter & Distribution & Min & Max & Units \\
        \hline
        \(\tau_{\rm s} / \tau_{\rm c}\) & LogUniform & \(10^{-4}\) & \(10^4\) & --- \\
        \(\tau\) & LogUniform & \(10^2\) & \(10^9\) & s \\
        \(Q_{\rm c}\) & LogUniform & \(10^{-30}\) & \(10^{-14}\) & rad$^2$ s\(^{-3}\) \\
        \(Q_{\rm s}\) & LogUniform & \(10^{-30}\) & \(10^{-14}\) & rad$^2$ s\(^{-3}\) \\
        \(\Omega_{\rm c,0}\) & Uniform & \(\hat{\Omega}_{\rm c,0} - 2 \times 10^{3} \Delta_{\Omega_{\rm c,0}}\) & \(\hat{\Omega}_{\rm c,0} + 2 \times 10^{3} \Delta_{\Omega_{\rm c,0}}\) & rad s\(^{-1}\)\\
        \(\langle\dot{\Omega}_{\rm c}\rangle\) & Uniform & \(\langle\hat{\dot{\Omega}}_{\rm c}\rangle - 2 \times 10^{3} \Delta_{\langle\dot{\Omega}_{\rm c}\rangle}\) & \(\langle\hat{\dot{\Omega}}_{\rm c}\rangle + 2 \times 10^{3} \Delta_{\langle\dot{\Omega}_{\rm c}\rangle}\) & rad s\(^{-2}\) \\
        \(\langle\Omega_{\rm{c}} - \Omega_{\rm{s}}\rangle\) & Uniform & \(-10^{-2}\) & \(10^{-2}\) & rad s\(^{-1}\) \\
        EFAC & Uniform & 0 & 4 & --- \\
        EQUAD & LogUniform & \(10^{-40}\) & \(10^2\) & rad$^2$ \\
        \(\phi_{\rm c, 0}\), \(\phi_{\rm s, 0}\) & Delta function at \(\phi_{\rm c,s}(0)\) & --- & --- & rad \\
        \hline
    \end{tabular}
\end{table*}

\subsection{Favored model per pulsar}
\label{subsec:model_preference-psr-level}
We find that 105 out of 286 UTMOST pulsars favor the two-component model with \(\ln \mathfrak{B}_{\rm BF} \geq 5\).
We present favored models and \(\ln \mathfrak{B}_{\rm BF}\) values for all 286 pulsars in Tables~\ref{tab:model_comparison_canonical}--\ref{tab:model_comparison_magnetar} in Appendix~\ref{app:psr_level-model_comparison-parameter}.
The pulsars are classified as canonical, i.e.\ located in the populous core of the \(\Omega_{\rm c}\)-\(\dot{\Omega}_{\rm c}\) plane (Table~\ref{tab:model_comparison_canonical}); recycled, i.e.\ satisfying the condition proposed by \citet{LeeEtAl2012} (Table~\ref{tab:model_comparison_msp}); and magnetar-like, with surface magnetic field \(\geq 4.4 \times 10^{13}\,\)G (Table~\ref{tab:model_comparison_magnetar}).
They are tabulated separately, to allow for the theoretical possibility that the physics of crust-superfluid coupling is different in pulsars with different evolutionary histories.
For example, it is conceivable that magnetic stresses play a larger role in magnetars than in canonical pulsars, as discussed in Section~\ref{subsec:noteworthy_objects}.

As a consistency check, we compare the model selection results in this paper with those reported from a previous UTMOST analysis by \citet{LowerEtAl2020}, summarized in the last two columns of Tables~\ref{tab:model_comparison_canonical}--\ref{tab:model_comparison_magnetar}.
We confirm that there is broad agreement.
Specifically, 93 out of the 105 objects with \(\ln \mathfrak{B}_{\rm BF} \geq 5\) in this paper also returned \(\ln \mathfrak{B}_{\rm BF} \geq 5\) when analyzed by \citet{LowerEtAl2020}.
The 12 exceptions are marked with asterisks in Tables~\ref{tab:model_comparison_canonical} and \ref{tab:model_comparison_msp} in Appendix~\ref{app:psr_level-model_comparison-parameter}.
Conversely, five nonglitching objects which returned \(\ln \mathfrak{B}_{\rm BF} \geq 5\) when analyzed by \citet{LowerEtAl2020} yield \(\ln \mathfrak{B}_{\rm BF} < 5\) in this paper.
The five marginal cases satisfy \(2 < \ln \mathfrak{B}_{\rm BF} < 5\) and are marked with daggers in Table~\ref{tab:model_comparison_canonical} in Appendix~\ref{app:psr_level-model_comparison-parameter}.

It is interesting to ask whether the 181 out of 286 UTMOST pulsars with \(\ln \mathfrak{B}_{\rm BF} < 5\) should be interpreted as containing no (or almost no) independently rotating superfluid, as the timing data for these objects favor the one-component WTN model.
The absence of a superfluid component is an intriguing possibility which is not ruled out physically, in view of the many theoretical uncertainties about the thermodynamic phases of bulk nuclear matter which exist at the time of writing \citep{YakovlevEtAl1999,LattimerPrakash2007,BaymEtAl2018}.
However, it is not the only possible interpretation of the inference results in Section~\ref{sec:model_selection_psr_level}.
One alternative is that the star does contain a superfluid component, but the superfluid couples magnetically to the crust (and hence corotates with it) on the fast Alfv\'en time-scale \(\sim 20\,\text{s}\) \citep{Mendell1998,Melatos2012,GlampedakisLasky2015}.
Fast coupling is allowed physically, even if the superfluid is composed of uncharged neutrons, because of strong density-density and current-current interactions between superfluid vortices and superconductor flux tubes \citep{SrinivasanEtAl1990,RudermanEtAl1998}, especially when the rotation and magnetic axes are inclined, and the vortices and flux tubes are tangled \citep{DrummondMelatos2017,DrummondMelatos2018,ThongEtAl2023}.
Strong coupling (faster than the observing cadence \(\sim \text{days}\)) without magnetic mediation can also occur, if the drag via mutual friction due to vortex-electron scattering is sufficiently strong (e.g.\ in the limit of perfect vortex pinning) \citep{AnderssonEtAl2006,HaskellAntonopoulou2013,GraberEtAl2018}.
The opposite situation is also conceivable, wherein a superfluid component exists but couples weakly to the crust on a time-scale much longer than \(T_{\rm obs}\).
Such weak coupling would yield TOAs consistent with the single-component model.
Finally, there are statistical factors to take into account.
On the one hand, the white measurement noise may dominate the red physical noise in some objects, explaining the preference for the WTN model.
On the other hand, the WTN model is not favored strongly in every object with \(\ln \mathfrak{B}_{\rm BF} < 5\); it is favored strongly only in those objects with \(\ln \mathfrak{B}_{\rm BF} < -5\), of which there are none.
Moreover, the \(\ln \mathfrak{B}_{\rm BF}\) criterion is arbitrary; naturally, a different choice by the analyst leads to a different classification of the objects in the sample.
All 181 out of 286 UTMOST pulsars that satisfy the original criterion \(\ln \mathfrak{B}_{\rm BF} < 5\) do so marginally; there are 37 objects with \(0 < \ln \mathfrak{B}_{\rm BF} < 5\) and 144 objects with \(-5 < \ln \mathfrak{B}_{\rm BF} < 0\).
For all the reasons above, we remain cautious about drawing strong conclusions about the composition of the neutron star interior from the inference results in Section~\ref{sec:model_selection_psr_level}, until more data become available.

\subsection{Crust-superfluid coupling time-scale} \label{subsec:tau-psr-level}
Tables~\ref{tab:model_comparison_canonical}--\ref{tab:model_comparison_magnetar} in Appendix~\ref{app:psr_level-model_comparison-parameter} summarize the Bayesian estimates for the two-component parameters \(\tau, Q_{\rm c}\), and \(Q_{\rm s}\).
For the sake of completeness, estimates are quoted for all 286 analyzed pulsars, but naturally they are more meaningful for the 105 objects with \(\ln \mathfrak{B}_{\rm BF} \geq 5\), that favor the two-component model.
The estimates are quantified in terms of the median and the 68\% credible interval centered on the median.
The numerical values are supplemented by an approximate indicator of whether the posterior distribution of \(\tau\) is ``peaky'', as a guide to the reader.
The \(\tau\) posterior is termed peaky arbitrarily, if the half-maximum points of its kernel density estimate are at least 0.5 dex away from the boundaries of the prior support.\footnote{
    It is possible to study whether the \(Q_{\rm c}\) and \(Q_{\rm s}\) posteriors are peaky as well, of course.
    We do not do so systematically in this paper, as the focus is on \(\tau\) throughout.
    As a rule of thumb, though, half the objects with peaky \(\tau\) posteriors also exhibit peaky \(Q_{\rm c}\) and \(Q_{\rm s}\) posteriors; see Fig.~\ref{fig:venn_peaky_tau_Qc_Qs}.
    \label{footnote:peaky_QcQs_nodisplay_disclaimer}
}

\begin{figure}
    \centering
    \includegraphics[width=\columnwidth]{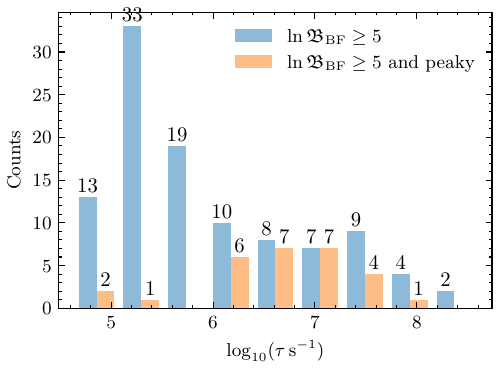}
    \caption{Crust-superfluid coupling time-scale per pulsar: histogram of the medians of the \(\log_{10} (\tau \, \rm{s}^{-1})\) posteriors for the 105 UTMOST objects with \(\ln \mathfrak{B}_{\rm BF} \geq 5\) (blue bars) and the 28 objects with \(\ln \mathfrak{B}_{\rm BF} \geq 5\) and peaky \(\tau\) posteriors (orange bars).
    The counts in each bin are recorded above each bar.
    }
    \label{fig:tau_histogram_pop_uninform}
\end{figure}

Tables~\ref{tab:model_comparison_canonical}--\ref{tab:model_comparison_magnetar} indicate that \(\tau\) covers a broad range across the UTMOST population.
Among the 105 objects with \(\ln{\mathfrak{B}_{\rm BF}} \geq 5\), we find that the per-pulsar \(\tau\) median ranges from \(\log_{10}(\tau \, \rm{s}^{-1}) = 4.6\) for PSR J2241$-$5236 (recycled) to \(\log_{10}(\tau \, \rm{s}^{-1}) = 8.6\) for PSR J1741$-$3927 (canonical).
However, only 28 of the 105 objects with \(\ln \mathfrak{B}_{\rm BF} \geq 5\) exhibit posteriors that peak sharply, in the approximate sense defined in the previous paragraph.
Among the latter 28 objects, we find that \(\tau\) ranges from \(\log_{10}(\tau \, \rm{s}^{-1}) = 4.6\) for PSR J2241$-$5236 (recycled) to \(\log_{10}(\tau \, \rm{s}^{-1}) = 7.7\) for PSR J1644$-$4559 (canonical).
In Fig.~\ref{fig:tau_histogram_pop_uninform}, we plot a histogram of the \(\tau\) posterior medians for the 105 objects with \(\ln \mathfrak{B}_{\rm BF} \geq 5\) (blue bars) and the 28 objects with \(\ln \mathfrak{B}_{\rm BF} \geq 5\) and sharply peaked \(\tau\) posteriors (orange bars).
We observe that the orange bars in Fig.~\ref{fig:tau_histogram_pop_uninform} are distributed unimodally with median \(\log_{10}(\tau \, \rm{s}^{-1}) = 6.7\), except for two objects in the leftmost bar.
The two exceptions, PSR J0437$-$4715 and PSR J2241$-$5236, have \(\log_{10}(\tau \, \rm{s}^{-1}) \approx 5\) and are recycled millisecond pulsars.
We return to recycled pulsars in Section~\ref{subsubsec:psr-millisec}.
The blue histogram is bimodal, with peaks at \(\log_{10}(\tau \, \rm{s}^{-1}) \approx 5.2\) and \(7.4\).
Many objects in the blue histogram exhibit approximately flat posteriors, so the median is near the mid-point of the posterior support, leading to the first peak.
The second peak is near the peak of the orange histogram but slightly to its right, because 13 out of the 105 \(\tau\) posteriors rail against the upper boundary of the prior support.

\begin{figure*}
    \centering
    \includegraphics[width=\textwidth]{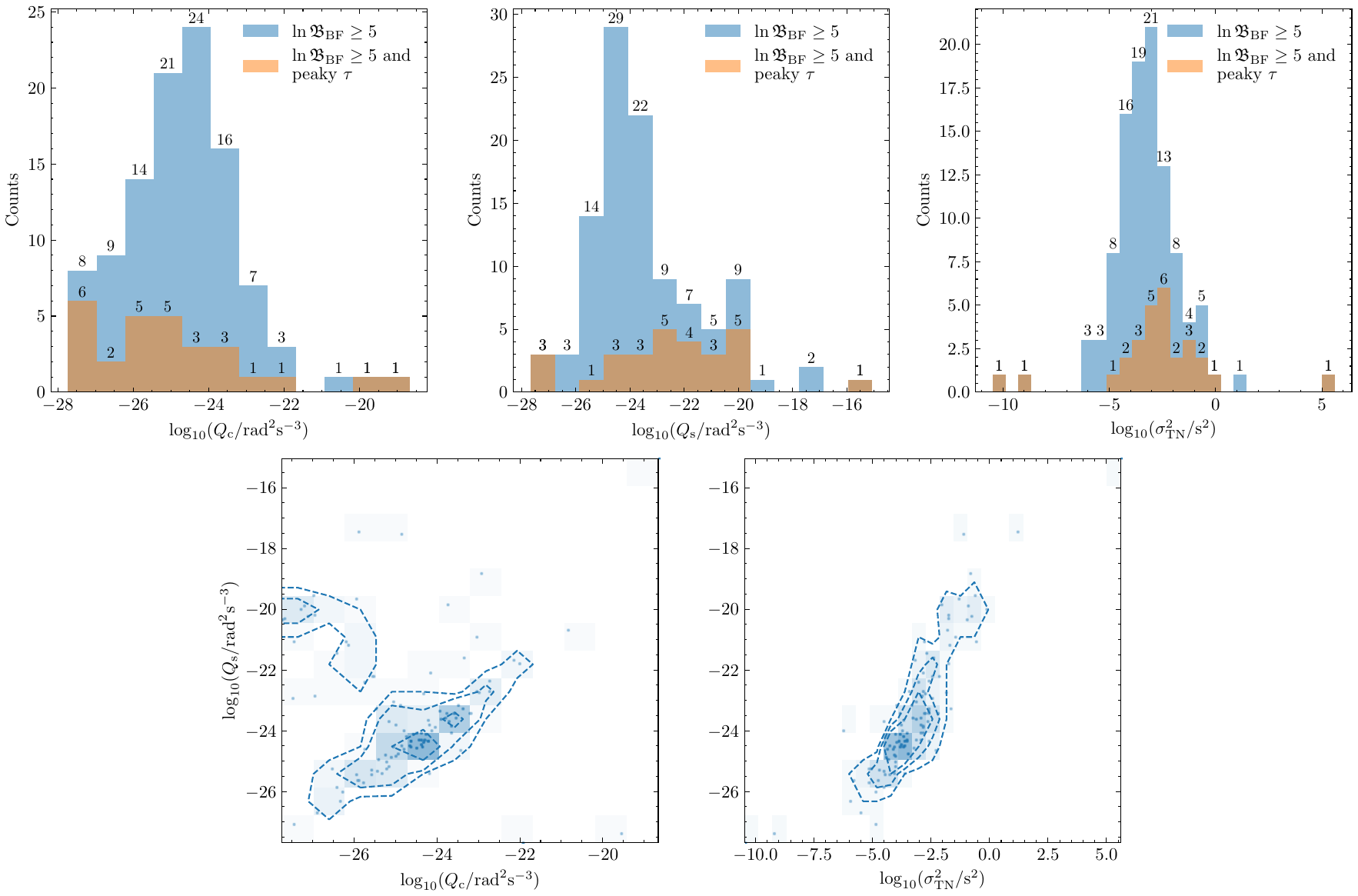}
    \caption{Crust and superfluid noise variances per pulsar: histograms of the posterior medians of \(\log_{10} Q_{\rm c}\) (top left panel; units rad\(^2\)s\(^{-3}\)), \(\log_{10} Q_{\rm s}\) (top middle panel; units rad\(^2\)s\(^{-3}\)), and \(\log_{10} \sigma_{\rm TN}^2\) (top right panel; units s\(^2\)) for the 105 UTMOST pulsars with \(\ln \mathfrak{B}_{\rm BF} \geq 5\) (blue bars) and the 28 objects with \(\ln \mathfrak{B}_{\rm BF} \geq 5\) and sharply peaked \(\tau\) posteriors (orange bars).
    The counts in each bin are recorded above each bar.
    Note that the orange histograms count objects whose posteriors are peaky in \(\tau\), not peaky in \(Q_{\rm c}\) or \(Q_{\rm s}\), to facilitate a like-for-like comparison with Fig.~\ref{fig:tau_histogram_pop_uninform}.
    The bottom panels show two-dimensional histograms of the posterior medians of \(\log_{10} Q_{\rm s}\) versus \(\log_{10} Q_{\rm c}\) (bottom left panel) and \(\log_{10} Q_{\rm s}\) versus \(\log_{10} \sigma_{\rm TN}^2\) (bottom right panel) for the 105 pulsars with \(\ln \mathfrak{B}_{\rm BF} \geq 5\).
    The contours enclose the \(1\)-, \(1.5\)-, and \(2\)-sigma confidence regions.
    Point estimates (blue dots) are overlaid for the 105 pulsars, to help visualize the probability mass density.
    }
    \label{fig:QcQsTN_corner_pop_uninform}
\end{figure*}

\subsection{Crust and superfluid noise variances} \label{subsec:mapping_to_sigma_TN}
Column six in Tables~\ref{tab:model_comparison_canonical}--\ref{tab:model_comparison_magnetar} displays the per-pulsar estimates of the normalized crust noise variance \(Q_{\rm c}\).
We find that the \(Q_{\rm c}\) posterior median spans 10 dex across the 105 objects with \(\ln \mathfrak{B}_{\rm BF} \geq 5\), from \(\log_{10} (Q_{\rm c} / \rm{rad}^2 \rm{s}^{-3}) = -27.7\) for PSR J1644$-$4559 (canonical) to \(\log_{10} (Q_{\rm c} / \rm{rad}^2 \rm{s}^{-3}) = -18.7\) for PSR J1622$-$4950 (magnetar).
Within the subset of 101 canonical pulsars, the \(Q_{\rm c}\) posterior median ranges from \(\log_{10} (Q_{\rm c} / \rm{rad}^2 \rm{s}^{-3}) = -27.7\) for PSR J1644$-$4559 to \(\log_{10} (Q_{\rm c} / \rm{rad}^2 \rm{s}^{-3}) = -20.9\) for PSR J1105$-$6107.
A histogram of the \(Q_{\rm c}\) posterior medians for the 105 objects with \(\ln \mathfrak{B}_{\rm BF} \geq 5\) (blue columns), and the subset of 28 objects with sharply peaked \(\tau\) posteriors (orange columns), is plotted in the left panel of Fig.~\ref{fig:QcQsTN_corner_pop_uninform}.\footnote{
    The orange histogram in the left panel of Fig.~\ref{fig:QcQsTN_corner_pop_uninform} counts objects whose posteriors are peaky in \(\tau\), not peaky in \(Q_{\rm c}\), even though the variable on the horizontal axis is \(Q_{\rm c}\).
    This is done to facilitate a like-for-like comparison with the same 28 objects in the orange histogram in Fig.~\ref{fig:tau_histogram_pop_uninform}, because the focus of the paper is crust-superfluid coupling; see footnote~\ref{footnote:peaky_QcQs_nodisplay_disclaimer}.
}
The blue histogram is unimodal and skewed left.
The orange histogram is also skewed left and exhibits two peaks at \(\log_{10} (Q_{\rm c} / \rm{rad}^2 \rm{s}^{-3}) \approx -27.5\) and \(-25.5\), noting that the bin counts are small.
The top two objects in the high-\(Q_{\rm c}\) tail are PSR J1622$-$4950 (magnetar) and PSR J2241$-$5236 (recycled).

Column seven in Tables~\ref{tab:model_comparison_canonical}--\ref{tab:model_comparison_magnetar} displays the per-pulsar estimates of the normalized superfluid noise variance \(Q_{\rm s}\).
As for \(Q_{\rm c}\), we find that \(Q_{\rm s}\) covers a wider range. Specifically, we find that the \(Q_{\rm s}\) posterior median spans 13 dex across the 105 objects with \(\ln \mathfrak{B}_{\rm BF} \geq 5\), from \(\log_{10} (Q_{\rm s} / \rm{rad}^2 \rm{s}^{-3}) = -27.7\) for PSR J0437$-$4715 (recycled) to \(\log_{10} (Q_{\rm s} / \rm{rad}^2 \rm{s}^{-3}) = -15.1\) for PSR J1622$-$4950 (magnetar).
Within the subset of 101 canonical pulsars, the \(Q_{\rm s}\) posterior median ranges from \(\log_{10} (Q_{\rm s} / \rm{rad}^2 \rm{s}^{-3}) = -27.1\) for PSR J1849$-$0636 to \(\log_{10} (Q_{\rm s} / \rm{rad}^2 \rm{s}^{-3}) = -17.5\) for PSR J1048$-$5832.
A histogram of the \(Q_{\rm s}\) posterior medians is plotted in the middle panel of Fig.~\ref{fig:QcQsTN_corner_pop_uninform}.
The blue histogram is weakly bimodal, with a primary peak at \(\log_{10} (Q_{\rm s} / \rm{rad}^2 \rm{s}^{-3}) \approx -25\) and a secondary peak at \(\log_{10} (Q_{\rm s} / \rm{rad}^2 \rm{s}^{-3}) \approx -20\), noting that the counts per bin are small.
The orange histogram also peaks at \(\log_{10} (Q_{\rm s} / \rm{rad}^2 \rm{s}^{-3}) \approx -23\) and skews right, with a weak, secondary peak at \(\log_{10} (Q_{\rm s} / \rm{rad}^2 \rm{s}^{-3}) \approx -20\).
The peaks and troughs in the orange histogram are hard to interpret, as the counts per bin are small.
The four outliers in the low-\(Q_{\rm s}\) and high-\(Q_{\rm s}\) tails of the orange histogram are PSR J0437$-$4715 (recycled), PSR J1849$-$0636 (canonical), and PSR J2241$-$5236 (recycled) for \(\log_{10} (Q_{\rm s} / \rm{rad}^2 \rm{s}^{-3}) \lesssim -26\), and PSR J1622$-$4950 (magnetar) for \(\log_{10} (Q_{\rm s} / \rm{rad}^2 \rm{s}^{-3}) \gtrsim -18\).
The left bottom panel in Fig.~\ref{fig:QcQsTN_corner_pop_uninform} displays the two-dimensional histogram of medians of \(Q_{\rm c}\) and \(Q_{\rm s}\), with point estimates of the medians overlaid. 
We see that objects around the secondary peak of \(Q_{\rm s}\) cluster with objects in the left tail of \(Q_{\rm c}\), while most objects fall along the \(Q_{\rm s} \approx Q_{\rm c}\) diagonal.
However, the cluster is enclosed by contours \(\gtrsim 1.5\)-sigma and is less significant statistically than the main trend along the diagonal. 
The sample size (105) and the counts per bin (\(\lesssim 7\)) in the histogram are relatively small.

\begin{figure}
    \centering
    \includegraphics[width=0.9\columnwidth]{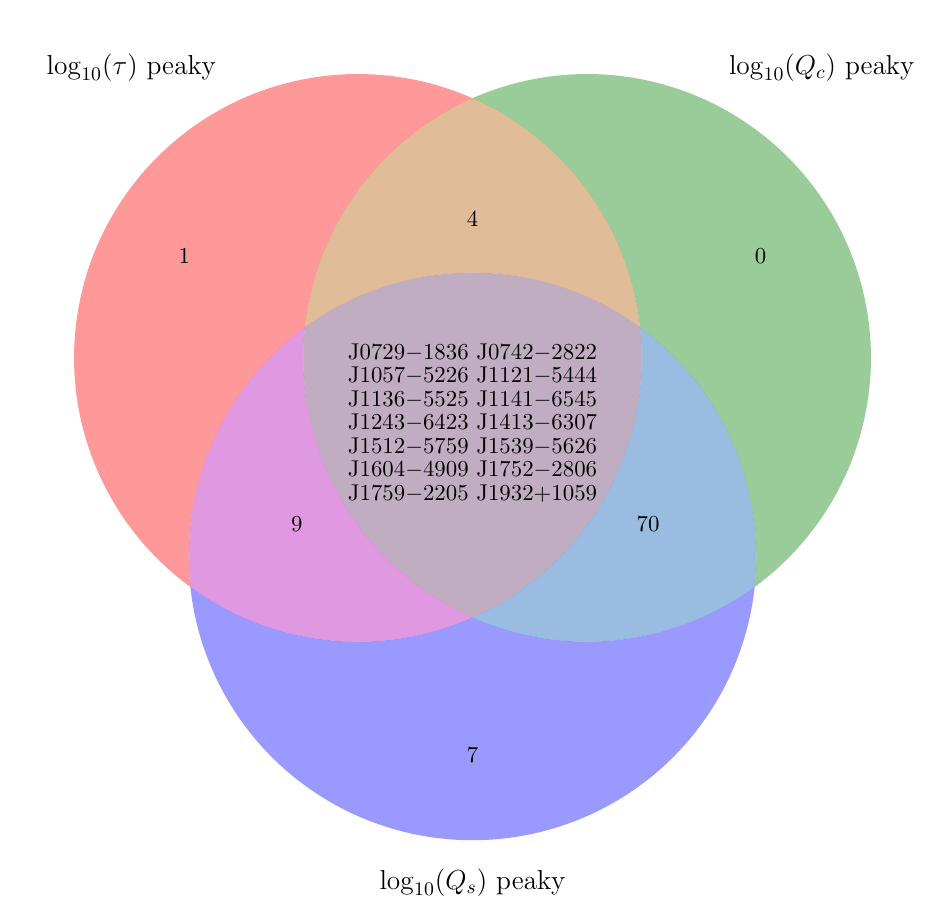}
    \caption{Venn diagram of 105 UTMOST pulsars with \(\ln \mathfrak{B}_{\rm BF} \geq 5\) that have peaky posteriors in \(\tau\) (top left circle), \(Q_{\rm c}\) (top right circle), and \(Q_{\rm s}\) (bottom circle).
    Regions are labelled with their cardinalities, except for the central region, which is labelled with the names of the 14 objects it contains.
    }
    \label{fig:venn_peaky_tau_Qc_Qs}
\end{figure}

What is the overlap between objects with peaky \(\tau\) posteriors and those with peaky \(Q_{\rm c}\) and/or \(Q_{\rm s}\) posteriors?
To illustrate this, we plot a Venn diagram of the subsets of 105 UTMOST pulsars with \(\ln \mathfrak{B}_{\rm BF} \geq 5\) that have peaky posteriors in \(\tau\), \(Q_{\rm c}\), and \(Q_{\rm s}\) in Fig.~\ref{fig:venn_peaky_tau_Qc_Qs}.
Peakiness is defined the same way for the three variables, according to the rule of thumb in the first paragraph of Section~\ref{subsec:tau-psr-level}.
The Venn diagram indicates the cardinality of each subset.
The central overlap region is labelled with the names of the 14 objects it contains.
We see that half of the 28 objects with peaky \(\tau\) posteriors also have peaky posteriors in both \(Q_{\rm c}\) and \(Q_{\rm s}\), including four objects that glitched before the UTMOST observations (see Section~\ref{sec:glitch_relaxation_comparison}).
Only one object, PSR J1849$-$0636, has a peaky posterior in \(\tau\) but not in either \(Q_{\rm c}\) or \(Q_{\rm s}\).
Among the 77 objects that do not have peaky \(\tau\) posteriors, 70 have peaky posteriors in both \(Q_{\rm c}\) and \(Q_{\rm s}\). 
Corner plots for five representative canonical pulsars, from different subsets in Fig.~\ref{fig:venn_peaky_tau_Qc_Qs}, are displayed and discussed in Appendix~\ref{app:cornerplot_selections}.

We relate the noise parameters \(Q_{\rm c}\) and \(Q_{\rm s}\) in the two-component model to \(\sigma_{\rm TN}\), the root-mean-square of the timing residuals \citep{CordesHelfand1980,LowerEtAl2020}, which is a traditional measure of timing noise strength, via \citep{AntonelliEtAl2023}
\begin{align}
    \sigma_{\rm TN}^2 
    \approx 2 \int_{1/T_{\rm obs}}^{f_{\rm s}/2} \mathrm{d}\left(\frac{\omega}{2\pi}\right) \, 
    \frac{P_{\!\delta\Omega_{\rm c}}(\omega)}{\omega^2 \Omega_{\rm c,0}^2} .
    \label{eq:sigma_TN_conversion}
\end{align}
In \eqref{eq:sigma_TN_conversion}, \(f_{\rm s} \sim N_{\rm TOA} / T_{\rm obs}\) denotes the sampling frequency, and \(P_{\!\delta\Omega_{\rm c}}(\omega) \propto \langle{|\delta \Omega_{\rm c}(\omega)|^2}\rangle\) is the power spectral density of \(\delta \Omega_{\rm c}(t) = \Omega_{\rm c}(t) - \langle\Omega_{\rm c}(t)\rangle\) \citep{MeyersEtAl2021a,AntonelliEtAl2023,ONeillEtAl2024}, with
\begin{align}
    P_{\!\delta\Omega_{\rm c}}(\omega)
    &= \frac{
        Q_{\rm c} (1 + \omega^{-2} \tau_{\rm eff}^{-2})
    }{
        \omega^2 (1 + \omega^{-2} \tau^{-2})
    } ,
    \label{eq:PSD_Omega_c}
\end{align}
and
\begin{align}
    \tau_{\rm eff}^{-2} = \tau_{\rm c}^{-2} \left(\frac{\tau_{\rm c}^2}{\tau_{\rm s}^2} + \frac{Q_{\rm s}}{Q_{\rm c}}\right) .
\end{align}
Substituting \eqref{eq:PSD_Omega_c} into \eqref{eq:sigma_TN_conversion}, we obtain
\begin{align}
    \sigma_{\rm TN}^2 
    = &\frac{Q_{\rm c} \tau^2}{3 \pi \Omega_{\rm c,0}^2 \tau_{\rm eff}^2} 
    \Bigg\{
        \left(1 - \frac{8}{N_{\rm TOA}^3}\right) T_{\rm obs}^3 + 3(\tau^2 - \tau_{\rm eff}^2)
        \nonumber \\
        &\times \left[
            \tau \left(\arctan{\frac{\tau N_{\rm TOA}}{2T_{\rm obs}}} - \arctan{\frac{\tau}{T_{\rm obs}}}\right) - \left(1 - \frac{2}{N_{\rm TOA}}\right) T_{\rm obs}
        \right]
    \Bigg\} .
    \label{eq:sigma_TN_squared}
\end{align}

Column eight in Tables~\ref{tab:model_comparison_canonical}--\ref{tab:model_comparison_magnetar} displays the per-pulsar estimates of \(\sigma_{\rm TN}^2\) calculated from \eqref{eq:sigma_TN_squared}.
We find that the \(\sigma_{\rm TN}^2\) posterior median spans 16 dex across the 105 objects with \(\ln \mathfrak{B}_{\rm BF} \geq 5\), from \(\log_{10} (\sigma_{\rm TN}^2 \, \rm{s}^{-2}) = -10.5\) for PSR J0437$-$4715 (recycled) to \(\log_{10} (\sigma_{\rm TN}^2 \, \rm{s}^{-2}) = 5.6\) for PSR J1622$-$4950 (magnetar).
Within the subset of 101 canonical pulsars, the median of \(\sigma_{\rm TN}^2\) ranges from \(\log_{10} (\sigma_{\rm TN}^2 \, \rm{s}^{-2}) = -5.9\) for PSR J1210$-$5559 to \(\log_{10} (\sigma_{\rm TN}^2 \, \rm{s}^{-2}) = 1.2\) for PSR J1048$-$5832.
The associated \(\sigma_{\rm TN}^2\) histogram for the 105 UTMOST pulsars with \(\ln \mathfrak{B}_{\rm BF} \geq 5\) (blue columns) and the 28 objects with \(\ln \mathfrak{B}_{\rm BF} \geq 5\) and sharply peaked \(\tau\) posteriors (orange columns), is displayed in the right panel of Fig.~\ref{fig:QcQsTN_corner_pop_uninform}.
The blue histogram is weakly bimodal.
The primary peak at \(\log_{10} (\sigma_{\rm TN}^2 \, \rm{s}^{-2}) \approx -3\) and the secondary cluster at \(\log_{10} (\sigma_{\rm TN}^2 \, \rm{s}^{-2}) \approx -1\) match the primary and secondary peaks in \(Q_{\rm s}\) respectively.
The primary peak in the orange histogram is near the primary peak in the blue histogram but slightly to its right, because at least five objects with left-railing \(Q_{\rm c}\) posteriors (e.g.\ five out of the seven objects with peaky posteriors only in \(Q_{\rm s}\) in Fig.\ \ref{fig:venn_peaky_tau_Qc_Qs}) are excluded from the orange histogram.
The secondary peak in the orange histogram is hard to interpret, as the counts in the peak and its neighboring bins are small.
The outliers in the orange histogram are PSR J0437$-$4715 (recycled) and PSR J2241$-$5236 (recycled) for \(\log_{10} (\sigma_{\rm TN}^2 \, \rm{s}^{-2}) \lesssim -6\), and PSR J1622$-$4950 (magnetar) for \(\log_{10} (\sigma_{\rm TN}^2 \, \rm{s}^{-2}) > 0\).

\subsection{Noteworthy objects} \label{subsec:noteworthy_objects}
In this section, we draw the reader's attention briefly to a subset of noteworthy objects in the following categories: recycled pulsars (Section~\ref{subsubsec:psr-millisec}), magnetars (Section~\ref{subsubsec:psr-magnetars}), pulsars with a bimodal \(\tau\) posterior (Section~\ref{subsubsec:psr-bimodal_tau}), and pulsars with previously published, timing-noise-based \(\tau\) measurements (Section~\ref{subsubsec:psr-published_tau}).
Corner plots for the noteworthy objects in this section are displayed in Fig.~\ref{fig:cornerplot_msps_pop_uninformed}--\ref{fig:cornerplot_autocorr-based_J1136+1551-J1935+1616_pop_uninformed} in Appendix~\ref{app:cornerplot_selections}.

\subsubsection{Recycled pulsars}
\label{subsubsec:psr-millisec}
The noise properties of recycled millisecond pulsars are a crucial factor in setting the sensitivity of nanohertz gravitational wave searches with pulsar timing arrays \citep{HellingsDowns1983,GoncharovEtAl2021,GoncharovEtAl2021a,AgazieEtAl2023a,AntoniadisEtAl2023,DiMarcoEtAl2023,ReardonEtAl2023,XuEtAl2023,DiMarcoEtAl2025}.
Typically, it is argued that recycled millisecond pulsars exhibit less timing noise than nonrecycled pulsars in the bulk of the population (i.e.\ in the populous core of the period-period-derivative plane), as measured by \(\sigma_{\rm TN}\) for example \citep{HobbsEtAl2010,ShannonCordes2010}.
The hierarchical Bayesian analysis in this paper offers a complementary perspective on this topic.
Specifically, in this section, we test whether the noise properties of recycled millisecond pulsars and nonrecycled pulsars differ systematically, when they are interpreted through the lens of the two-component model.

Table~\ref{tab:model_comparison_msp} shows that most recycled pulsars favor the WTN model over the two-component model.
There are 18 recycled pulsars in the UTMOST sample, according to the definition in Section~\ref{subsec:model_preference-psr-level} and proposed by \citet{LeeEtAl2012}.
Three out of the 18 satisfy \(\ln \mathfrak{B}_{\rm BF} \geq 5\), namely PSR J0437$-$4715, PSR J2145$-$0750, and PSR J2241$-$5236.\footnote{
    Two other millisecond pulsars, PSR J1909$-$3744 and PSR J2051$-$0827, have \(3 < \ln \mathfrak{B}_{\rm BF} < 5\).
}
That is, pulsars that favor the two-component model are less common among recycled than canonical pulsars.
Two of the three return unimodal, peaky \(\tau\) posteriors, with \(\log_{10} (\tau \, \rm{s}^{-1}) = 4.9^{+0.5}_{-1.0}\) and \(4.6^{+0.5}_{-0.6}\) for PSR J0437$-$4715 and PSR J2241$-$5236 respectively.
The crust noise dominates in these two objects, with \(\log_{10} (Q_{\rm c} / \rm{rad}^2 \rm{s}^{-3}) = -21.9^{+2.0}_{-0.9}\) and \(-19.6^{+1.3}_{-1.0}\) respectively, compared to \(\log_{10} (Q_{\rm s} / \rm{rad}^2 \rm{s}^{-3}) = -27.7^{+1.6}_{-1.5}\) and \(-27.4^{+1.8}_{-1.8}\) respectively.
In contrast, 14 out of the 15 recycled pulsars that favor the WTN model in Table~\ref{tab:model_comparison_msp} do not return peaky \(\tau\) posteriors and satisfy \(\log_{10} (Q_{\rm c} / \rm{rad}^2 \rm{s}^{-3}) \sim \log_{10} (Q_{\rm s} / \rm{rad}^2 \rm{s}^{-3}) \lesssim -23\) and \(\log_{10} (\sigma_{\rm TN}^2 \, \rm{s}^{-2}) \lesssim -6\).

Fig.~\ref{fig:cornerplot_msps_pop_uninformed} displays corner plots for PSR J0437$-$4715 and PSR J2241$-$5236.
We see that \(\tau\) and \(Q_{\rm c}\) are anti-correlated, which is approximately consistent with the regime \(\tau_{\rm eff}^{-1} \sim 10^{-7} \mathrm{s}^{-1} \lesssim \omega \lesssim 10^{-5} \mathrm{s}^{-1} \lesssim \tau^{-1}\) in Table~A1 in \citet{MeyersEtAl2021a}.
Moreover, \(P_{\!\delta\Omega_{\rm c}}(\omega) \approx Q_{\rm c} \tau^2\) is flat (i.e.\ independent of \(\omega\)) in this regime.
This leads to \(\beta = 2\) for the power spectral density of the TOA residuals, \(P_{\!\delta t}(\omega) \propto \omega^{-\beta}\), as we have \(P_{\!\delta t}(\omega) \propto \omega^{-2} P_{\!\delta\Omega_{\rm c}}(\omega)\).

The two-component model~\eqref{eq:EOM_c}--\eqref{eq:phi_c_EOM} does not describe chromatic timing noise arising from propagation effects in the pulsar magnetosphere \citep{LyneEtAl2010,ShannonEtAl2016} or the interstellar medium \citep{KeithEtAl2013,ColesEtAl2015}, e.g.\ dispersion measure (DM) fluctuations.
It is known independently that chromatic timing noise is important in pulsar timing array experiments \citep{GoncharovEtAl2021,DiMarcoEtAl2025}.
Wide-band, multi-wavelength observations are required to correct accurately for DM fluctuations, yet the bandwidth of the UTMOST experiment is relatively narrow (\(\sim 30\,\textrm{MHz}\)) \citep{BailesEtAl2017,JankowskiEtAl2019,LowerEtAl2020}.
In the absence of DM corrections, the results for recycled pulsars in Section~\ref{subsubsec:psr-millisec} should be regarded as approximate.
The reader is referred to Appendix~\ref{app:DM-variations} for a quantitative discussion about the impact of unmodelled DM variations.

\subsubsection{Magnetars} \label{subsubsec:psr-magnetars}
Magnetars are magnetically active neutron stars, whose internal magnetic fields evolve continuously through a combination of relaxation processes, such as Ohmic dissipation and Hall drift \citep{PonsGeppert2007,ViganoEtAl2012,GourgouliatosCumming2014,MereghettiEtAl2015}.
Hence, it is possible theoretically that their timing noise properties are governed by a different magnetically-driven mechanism, e.g.\ crustal failure and Hall wave generation \citep{BransgroveEtAl2025} or magnetar outbursts \citep{LivingstoneEtAl2011,DibKaspi2014}, which does not operate in ordinary pulsars \citep{TurollaEtAl2015,KaspiBeloborodov2017}.
Indeed, there is even an argument that the strong internal field in a magnetar hinders the development of a crust-superfluid angular velocity lag, to the point where the two-component model becomes a poor approximation.
It is therefore interesting to examine the model selection and parameter estimation results in Tables~\ref{tab:model_comparison_canonical}--\ref{tab:model_comparison_magnetar} to look for evidence of a systematic difference between magnetars and other objects.

In this paper, we categorize magnetars as pulsars with a surface magnetic field strength \(B_{\rm surf} \geq 4.4 \times 10^{13} \, \rm{G}\) \citep{ThompsonDuncan1995}.
Among the 286 analyzed objects, one is a magnetar, namely PSR J1622$-$4950.
Table~\ref{tab:model_comparison_magnetar} lists its properties.
The object favors strongly the two-component model with \(\ln \mathfrak{B}_{\rm BF} = 424.7\).
Its parameter estimates are \(\log_{10} (Q_{\rm c} / \rm{rad}^2 \rm{s}^{-3}) = -18.7^{+0.3}_{-5.7}\), \(\log_{10} (Q_{\rm s} / \rm{rad}^2 \rm{s}^{-3}) = -15.1^{+0.7}_{-0.8}\) (which rails against the upper edge of the prior support), and \(\log_{10} (\tau \, \rm{s}^{-1}) = 7.6^{+0.4}_{-0.4}\).\footnote{
    The \(\tau\) posterior is peaky, because \(Q_{\rm s}\) rails against the upper boundary of the prior support and cuts off the \(\tau\) posterior for \(\log_{10} (\tau \, \rm{s}^{-1}) \gtrsim 8.5\), noting the positive correlation between \(\tau\) and \(Q_{\rm s}\) (see Fig.~\ref{fig:cornerplot_magnetar_J1622-4950-pop_uninformed}).
}
The estimate of \(\sigma_{\rm TN}\) is the largest among the 286 objects in Tables~\ref{tab:model_comparison_canonical}--\ref{tab:model_comparison_magnetar}, with \(\log_{10} (\sigma_{\rm TN}^2 \, \rm{s}^{-2}) = 5.6^{+0.2}_{-0.4}\), and the estimate of EFAC is the smallest, with \(\textrm{EFAC} = 0.077^{+0.070}_{-0.029}\).
Note that the variable denoted by EFAC in this paper is the square of the EFAC variable defined elsewhere \citep{LentatiEtAl2016}.
PSR J1622$-$4950 occupies the regime \(\tau^{-1} \ll 10^{-6.8} \rm{s}^{-1} \lesssim \omega \lesssim 10^{-5.2} \rm{s}^{-1} \lesssim \tau_{\rm eff}^{-1}\), consistent with \(\beta = 6\) according to Table~A1 in \citet{MeyersEtAl2021a}.
The latter finding is consistent with \(\beta = 7.3^{+3.4}_{-3.6}\) measured by \citet{LowerEtAl2020}.

\subsubsection{Bimodal $\tau$ posterior} \label{subsubsec:psr-bimodal_tau}
The posterior distribution of \(\tau\) is bimodal for PSR J1141$-$6545. 
The mode \(\log_{10} (\tau \, \rm{s}^{-1}) \approx 7\) is favored, when we have \(\log_{10} (Q_{\rm c} / \rm{rad}^2 \rm{s}^{-3}) \geq \log_{10} (Q_{\rm s} / \rm{rad}^2 \rm{s}^{-3}) \approx -25\) and \(\tau_{\rm s} / \tau_{\rm c} \sim 1\).
The mode \(\log_{10} (\tau \, \rm{s}^{-1}) \approx 6\) is favored, when we have \(\log_{10} (Q_{\rm c} / \rm{rad}^2 \rm{s}^{-3}) \leq \log_{10} (Q_{\rm s} / \rm{rad}^2 \rm{s}^{-3}) \approx -25\) and \(\tau_{\rm s} / \tau_{\rm c} \gtrsim 1\).
The two modes are visible in the corner plot of PSR J1141$-$6545 in Fig.~\ref{fig:cornerplot_bimodal-tau_J1141-6545-pop_uninformed}.

The bimodality corresponds to two different \(\beta\) regimes.
The mode at \(\log_{10} (\tau \, \rm{s}^{-1}) \approx 6\) implies \(1 / \tau \lesssim \omega \lesssim 1 / \tau_{\rm eff}\) and hence \(\beta = 6\), while the mode at \(\log_{10} (\tau \, \rm{s}^{-1}) \approx 7\) implies \(1 / \tau_{\rm eff} \lesssim 1 / \tau \ll \omega\) and hence \(\beta = 4\) \citep[][Table~A1]{MeyersEtAl2021a}.
The spectrum measured by \citet{LowerEtAl2020} has \(\beta = 4.7^{+3.2}_{-1.0}\), intermediate between the two modes.\footnote{
    Upon marginalizing over the two modes, i.e.\ weighting by the relative heights of the peaks in Fig.~\ref{fig:cornerplot_bimodal-tau_J1141-6545-pop_uninformed}, we obtain \(\beta \approx 5.2\). 
}

Interestingly, PSR J1141$-$6545 glitched on MJD \(54274\pm20\), before the start of the UTMOST observations. The glitch recovery timescale is measured as \(\tau_{\rm g} = 495 \pm 140\,\)d \citep{ManchesterEtAl2010}, consistent with the \(\log_{10} (\tau \, \rm{s}^{-1}) \approx 7\) mode.
However, this may be accidental: the relaxation physics may not be the same during timing noise and glitch recoveries.
We return to this object and several others that glitched before the UTMOST timing campaign in Section~\ref{sec:glitch_relaxation_comparison}.

\subsubsection{Autocorrelation-based \(\tau\) measurements} \label{subsubsec:psr-published_tau}
Until now, timing-noise-based measurements of \(\tau\) have usually been performed by autocorrelating the time series of TOA residuals.
Notably, \citet{PriceEtAl2012} analyzed Jodrell Bank data from the 12.8-m Lovell Radio Telescope for PSR J1136+1551 and PSR J1935+1616 and inferred the autocorrelation time-scales to be \(\tau_{\rm corr} = 10 \pm 1\,\)d and \(20 \pm 1\,\)d respectively, with the errors being half the width of the sampling window.
It is natural to ask whether these values are consistent with the analysis in this paper.
Unfortunately, there are only 36 and 59 TOAs in the UTMOST dataset for PSR J1136+1551 and PSR J1935+1616 respectively, so we cannot recover \(\tau\) accurately; see the corresponding corner plots in Fig.~\ref{fig:cornerplot_autocorr-based_J1136+1551-J1935+1616_pop_uninformed}.
It turns out that the posteriors of \(Q_{\rm c}\) and \(Q_{\rm s}\) are sharply peaked and return \(\log_{10} (Q_{\rm c} / \rm{rad}^2 \rm{s}^{-3}) = -25.3^{+1.3}_{-1.5}\) and \(\log_{10} (Q_{\rm s} / \rm{rad}^2 \rm{s}^{-3}) = -25.4^{+3.2}_{-2.8}\) for PSR J1136+1551, and \(\log_{10} (Q_{\rm c} / \rm{rad}^2 \rm{s}^{-3}) = -25.9^{+0.5}_{-1.8}\) and \(\log_{10} (Q_{\rm s} / \rm{rad}^2 \rm{s}^{-3}) = -25.7^{+3.5}_{-2.3}\) for PSR J1935+1616.
It may be a fortuitous accident, that the posteriors of \(Q_{\rm c}\) and \(Q_{\rm s}\) are peaky, but we include the results regardless for the sake of completeness.

\section[Crust-superfluid coupling at the population level]{Crust-superfluid coupling at the population level: trends in the populous core of the \(\Omega_{\rm \lowercase{c}}\)-\(\dot{\Omega}_{\rm \lowercase{c}}\) plane} \label{sec:hierarchical_regression}

The inference analysis in Section~\ref{sec:model_selection_psr_level} treats every UTMOST pulsar individually.
In this section, we treat the UTMOST pulsars collectively by overlaying population-level constraints.
In Section~\ref{subsec:hypermodel_scaling}, we incorporate scalings of \(\tau, Q_{\rm c}\), and \(Q_{\rm s}\) with \(\Omega_{\rm c}\) and \(\dot{\Omega}_{\rm c}\) into the hierarchical Bayesian framework formulated in Section~\ref{sec:hierarchical_model-outline}.
We present the inferred values of the hyperparameters defining the scalings in Sections~\ref{subsec:hyper_tau} (for \(\tau\)) and \ref{subsec:hyper_QcQs} (for \(Q_{\rm c}\) and \(Q_{\rm s}\)), as well as the population-informed posterior distributions.
In this section, unlike in Section~\ref{sec:model_selection_psr_level}, we restrict attention to the 101 objects with \(\ln \mathfrak{B}_{\rm BF} > 5\) that are canonical, i.e.\ that lie in the populous core of the \(\Omega_{\rm c}\)-\(\dot{\Omega}_{\rm c}\) plane and are neither recycled pulsars nor magnetars.
We do so because it is plausible that a single \(\Omega_{\rm c}\)-\(\dot{\Omega}_{\rm c}\) scaling applies to canonical objects in the populous core but does not extend to objects with different evolutionary status on the periphery, such as recycled pulsars and magnetars.

\begin{table}
    \centering
    \caption{Prior distributions for the hyperparameters \(C_{2, \theta}\), \(a_{\theta}\), \(b_{\theta}\), and \(\sigma_{\theta}\), with \(\theta \in \{\tau, Q_{\rm c}, Q_{\rm s}\}\).
    The quantities \(\min_{\theta}\) and \(\max_{\theta}\) governing \(C_{2, \theta}\) are the minimum and maximum values of the \(\theta\) priors in Table~\ref{tab:psrlevel_priors}.
    }
    \label{tab:hypermodel_priors}
    \begin{tabular}{c|c|c|c}
        \hline
        Hyperparameter & Distribution & Min & Max \\
        \hline
        \(C_{2, \theta}\) & Uniform & \(\log_{10}(\min_{\theta}) - 30\) & \(\log_{10}(\max_{\theta}) + 30\) \\
        \(a_{\theta}, b_{\theta}\) & Uniform & \(-10\) & 10 \\
        \(\sigma_{\theta}\) & Uniform & 0 & 10 \\
        \hline
    \end{tabular}
\end{table}

\subsection[Gaussian linear regression for the population-level prior]
{Gaussian linear regression for the population-level prior \(p[\bm{\theta}^{(i)} \,|\, \bm{\Lambda}]\)} \label{subsec:hypermodel_scaling}

We wish to investigate how the pulsar parameters \(\tau, Q_{\rm c}\), and \(Q_{\rm s}\) scale with the spin-down variables \(\Omega_{\rm c} \approx \Omega_{\rm c,0}\) and \(\dot{\Omega}_{\rm c} \approx \langle \dot{\Omega}_{\rm c} \rangle\).
To accomplish this consistently in a hierarchical Bayesian framework, we select a specific analytic form (here, Gaussian) for the population-level prior \(p[\bm{\theta}^{(i)} \,|\, \bm{\Lambda}]\) and combine it with a linear regression to describe its first moment as a function of \(\Omega_{\rm c}\) and \(\dot{\Omega}_{\rm c}\)\footnote{
    The reader is referred to Chapter~14 in \citet{GelmanEtAl2013} for a thorough introduction to regression in hierarchical Bayesian models.
}.
Specifically, we assume that \(p(\bm{\theta} \,|\, \bm{\Lambda}) \propto \mathcal{N}(\log_{10} \bm{\theta} \,|\, \bm{\mu}, \bm{\Sigma})\) for \(\bm{\theta} = \{\tau, Q_{\rm c}, Q_{\rm s}\}\) is a trivariate normal distribution, with mean vector \(\bm{\mu} = (\mu_{\tau}, \mu_{Q_{\rm c}}, \mu_{Q_{\rm s}})\) and diagonal covariance matrix \(\bm{\Sigma} = \text{diag}(\sigma_{\tau}^2, \sigma_{Q_{\rm c}}^2, \sigma_{Q_{\rm s}}^2)\).
The \(\theta\)-component of \(\bm{\mu}\) is assumed to be a log-linear function of \(\Omega_{\rm c}\) and \(\dot{\Omega}_{\rm c}\), viz.\
\begin{align}
    \mu_{\theta} = C_{2, \theta} + a_{\theta} \log_{10} (\Omega_{\rm c} / 1 \, \mathrm{rad\,s}^{-1}) + b_{\theta} \log_{10} (|\dot{\Omega}_{\rm c}| / 1 \, \mathrm{rad\,s}^{-2}) ,
    \label{eq:linear_regression_model}
\end{align}
where \(C_{2, \theta}\), \(a_{\theta}\), and \(b_{\theta}\) are hyperparameters.
In formal Bayesian terminology, \(\Omega_{\rm c}\) and \(\dot{\Omega}_{\rm c}\) are explanatory variables, and \(\tau, Q_{\rm c}, Q_{\rm s}\) are response variables \citep{GelmanEtAl2013}.
The covariance matrix \(\bm{\Sigma}\) is assumed to be diagonal as a first pass but can be generalized easily if required.
For simplicity, we also take \(\bm{\Sigma}\) to be fixed, i.e.\ independent of \(\Omega_{\rm c}\) and \(\dot{\Omega}_{\rm c}\).
Models like \eqref{eq:linear_regression_model} are called ordinary linear regression models \citep{GelmanEtAl2013}.
Other parametrizations are possible, such as \(\mu_{\theta} = \mu_{\theta}(B_{\rm surf}, T_{\rm age})\), where \(B_{\rm surf} \propto \Omega_{\rm c}^{-3/2} |\dot{\Omega}_{\rm c}|^{1/2}\) is the magnetic field strength at the stellar surface, and \(T_{\rm age} = \Omega_{\rm c} / (2 |\dot{\Omega}_{\rm c}|)\) is the characteristic age.
One can map \(\mu_{\theta}(\Omega_{\rm c}, \dot{\Omega}_{\rm c})\) to \(\mu_{\theta}(B_{\rm surf}, T_{\rm age})\) by applying the chain rule to the likelihood function.

The prior distributions (hyperpriors) for the hyperparameters \{\(C_{2,\theta}, a_{\theta}, b_{\theta}, \sigma_{\theta}\)\} with \(\theta \in \{\tau, Q_{\rm c}, Q_{\rm s}\}\) are listed in Table~\ref{tab:hypermodel_priors}. 
The hyperprior for \(\sigma_{\theta}\) is chosen to be uniform over a wide range, viz.\ Uniform\((0, 10)\).\footnote{
    A wide uniform prior for \(\sigma_{\theta}\) may overestimate \(\sigma_{\theta}\), when the number of objects \(n\) within the population is small.
    In this paper, \(n=101\) is not small in the latter sense.
    Alternative weakly informative priors exist if desired, such as the inverse-\(\chi^2\) distribution for \(\sigma_{\theta}^2\) or the half-Cauchy distribution for \(\sigma_{\theta}\) \citep{Gelman2006}.
}

\begin{figure*}
    \centering
    \includegraphics[width=\textwidth]{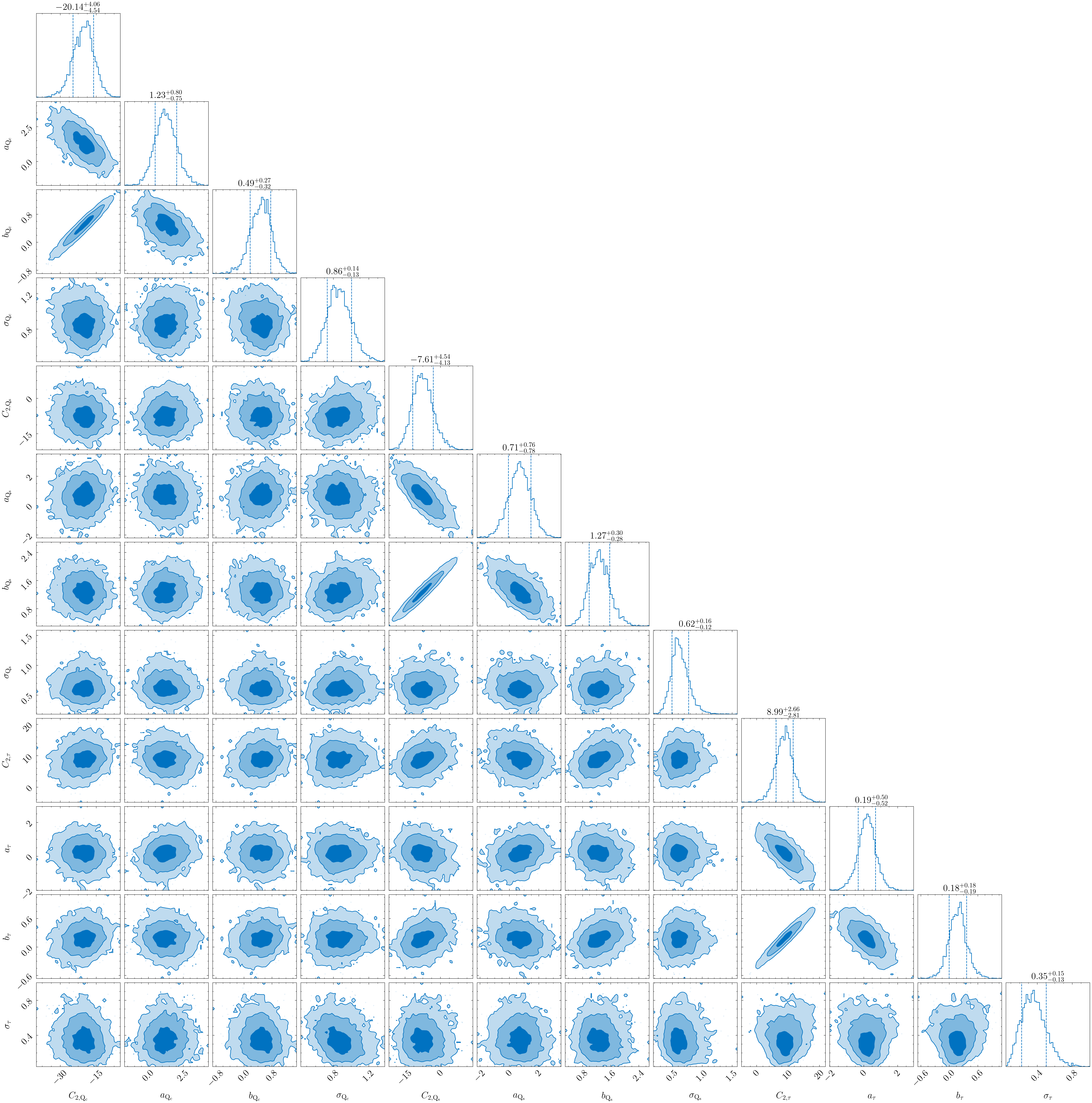}
    \caption{
        Corner plot of the posterior distribution of the hyperparameters \(C_{2, \theta}\), \(a_{\theta}\), \(b_{\theta}\), and \(\sigma_{\theta}\) for the hierarchical regression model described by \eqref{eq:linear_regression_model} for \(\theta \in \{\tau, Q_{\rm c}, Q_{\rm s}\}\) after analyzing 101 canonical pulsars with \(\ln \mathfrak{B}_{\rm BF} \geq 5\) in the populous core of the \(\Omega_{\rm c}\)-\(\dot{\Omega}_{\rm c}\) plane, i.e.\ excluding recycled pulsars and magnetars.
        The contours indicate the 1-, 2-, and 3-sigma credible regions for a two-dimensional slice of the posterior after marginalizing over the other 10 parameters.
        The vertical dashed lines in the one-dimensional histograms, after marginalizing over the other 11 parameters, bracket the 68\% confidence intervals.
        The heading above each rightmost panel quotes the median and 68\% credible interval.
        Every panel confirms that the posterior distribution converges numerically and is unimodal.}
    \label{fig:cornerplot_QcQstau_scaling}
\end{figure*}

\subsection{Crust-superfluid coupling time-scale} \label{subsec:hyper_tau}
Fig.~\ref{fig:cornerplot_QcQstau_scaling} displays a traditional corner plot for the hyperparameters \(C_{2,\theta}, a_{\theta}, b_{\theta}\), and \(\sigma_{\theta}\) with \(\theta \in \{\tau, Q_{\rm c}, Q_{\rm s}\}\) for the 101 canonical pulsars with \(\ln \mathfrak{B}_{\rm BF} \geq 5\) in the populous core of the \(\Omega_{\rm c}\)-\(\dot{\Omega}_{\rm c}\) plane.
The posterior distribution converges and is unimodal.
There is no evidence of railing against the prior bounds.
We find that \(C_{2, \theta}\) is correlated negatively with \(a_{\theta}\), correlated positively with \(b_{\theta}\), and uncorrelated with \(\sigma_{\theta}\) for all \(\theta \in \{\tau, Q_{\rm c}, Q_{\rm s}\}\).
Apart from a weak positive correlation between \(b_{\tau}\) and \(b_{Q_{\rm s}}\), a weak negative correlation between \(\sigma_{\tau}\) and \(\sigma_{Q_{\rm c}}\), and negative correlations between \(a_{\theta}\) and \(b_{\theta}\) for all \(\theta\), no significant correlation is found between \(a_{\theta}\), \(b_{\theta}\), and \(\sigma_{\theta}\) for all \(\theta\).
The median estimates and 68\% credible intervals of \(C_{2,\theta}, a_{\theta}, b_{\theta}\), and \(\sigma_{\theta}\) are summarized in Table~\ref{tab:hypermodel_comparison_Qc-Qs-tau_101psr}.

As a first step, it is interesting to ask if the data favor the hierarchical model in Section~\ref{subsec:hypermodel_scaling} over the \(N_{\rm psr}=101\) independent per-pulsar models.
We address this question in detail in terms of the relevant Bayes factor and comparison plots in Appendix~\ref{app:relating_per_pulsar_population_inference}.
In short, we find \(\ln \mathfrak{B}_{\rm BF, \Lambda} \approx 135\), strongly favoring the hierarchical population model over the independent per-pulsar models.
The population-informed estimates of \(\tau, Q_{\rm c},\) and \(Q_{\rm s}\) are mostly consistent with their population-uninformed counterparts for the 25 canonical objects, whose population-uninformed \(\tau\) posteriors are sharply peaked.

\begin{figure*}
    \centering
    \includegraphics[width=\textwidth]{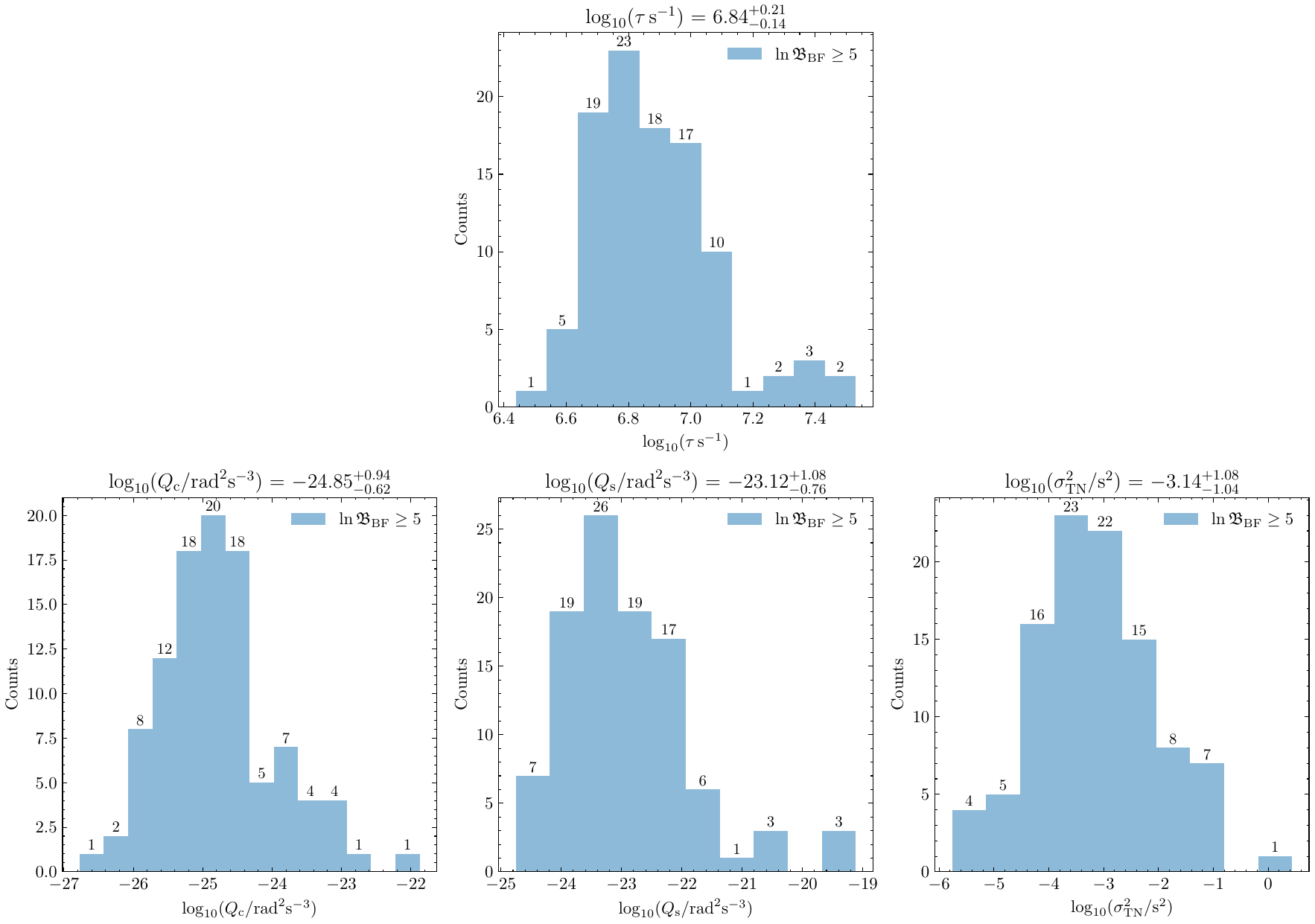}
    \caption{
        Histograms of the population-informed posterior medians of \(\log_{10} \tau\) (top panel; units s), \(\log_{10} Q_{\rm c}\) (bottom left panel; units \(\rm{rad}^2 \rm{s}^{-3}\)), \(\log_{10} Q_{\rm s}\) (bottom middle panel; units \(\rm{rad}^2 \rm{s}^{-3}\)), and \(\log_{10} \sigma_{\rm TN}^2\) (bottom right panel; units \(\rm{s}^{2}\)) for 101 canonical pulsars with \(\ln \mathfrak{B}_{\rm BF} \geq 5\) in the populous core of the \(\Omega_{\rm c}\)-\(\dot{\Omega}_{\rm c}\) plane, assuming the hierarchical regression model \eqref{eq:linear_regression_model}.
        The counts in each bin are recorded above each bar.
        The heading above each panel quotes the median and 68\% credible interval.}
    \label{fig:histogram_median_of_pop_informed_posteriors}
\end{figure*}

\begin{table}
    \centering
    \caption{Inferred values of the hyperparameters \(C_{2, \theta}\), \(a_{\theta}\), \(b_{\theta}\) and \(\sigma_{\theta}\) for \(\theta \in \{\tau, Q_{\rm c}, Q_{\rm s}\}\). 
    Estimates are quantified by the median and 68\% credible interval. 
    The hypermodel is applied to 101 canonical pulsars in the populous core of the \(\Omega_{\rm c}\)-\(\dot{\Omega}_{\rm c}\) plane with $\ln \mathfrak{{B}} \geq 5$.}
    
\label{tab:hypermodel_comparison_Qc-Qs-tau_101psr}
\begin{tabular}{llll}
\toprule
 & $\tau$ & $Q_{\rm c}$ & $Q_{\rm s}$ \\
\midrule
$C_{2, \theta}$ & $8.99^{+2.66}_{-2.81}$ & $-20.14^{+4.06}_{-4.54}$ & $-7.61^{+4.54}_{-4.13}$ \\
$a_{\theta}$ & $0.19^{+0.50}_{-0.52}$ & $1.23^{+0.80}_{-0.75}$ & $0.71^{+0.76}_{-0.78}$ \\
$b_{\theta}$ & $0.18^{+0.18}_{-0.19}$ & $0.49^{+0.27}_{-0.32}$ & $1.27^{+0.30}_{-0.28}$ \\
$\sigma_{\theta}$ & $0.35^{+0.15}_{-0.13}$ & $0.86^{+0.14}_{-0.13}$ & $0.62^{+0.16}_{-0.12}$ \\
\bottomrule
\end{tabular}

\end{table}

In this section, we focus on the crust-superfluid coupling time-scale.
Table~\ref{tab:hypermodel_comparison_Qc-Qs-tau_101psr} indicates that \(\tau\) is approximately independent of \(\Omega_{\rm c}\) at the population level, with \(10^{\mu_{\tau}} \propto \Omega_{\rm c}^{0.19^{+0.50}_{-0.52}}\).
Table~\ref{tab:hypermodel_comparison_Qc-Qs-tau_101psr} also indicates that \(\tau\) depends weakly on \(\dot{\Omega}_{\rm c}\), with \(10^{\mu_{\tau}} \propto |\dot{\Omega}_{\rm c}|^{0.18^{+0.18}_{-0.19}}\).
Possible physical interpretations are discussed in Section~\ref{sec:discussion_conclusion}.
The population-level information also modifies the \(\tau\) posteriors at the per-pulsar level presented in Section~\ref{sec:model_selection_psr_level}.
The top panel of Fig.~\ref{fig:histogram_median_of_pop_informed_posteriors} displays a histogram of the medians of the 101 population-informed \(\tau\) posteriors for the 101 objects under analysis.
The median of the population-informed posterior medians is \(\log_{10} (\tau \, \rm{s}^{-1}) = 6.9\), which is higher than the per-pulsar results without the population-level overlay in Section~\ref{subsec:tau-psr-level} and Fig.~\ref{fig:tau_histogram_pop_uninform}.
However, both the median and mode of medians of the population-informed posterior are lower than the median and mode of the orange histogram (excluding two recycled pulsars and one magnetar) in Fig.~\ref{fig:tau_histogram_pop_uninform}.
The 68\% quantile range of medians of the population-informed posterior is \(6.7 \leq \log_{10} (\tau \, \rm{s}^{-1}) \leq 7.1\), tighter than that of the orange histogram (excluding two recycled pulsars and one magnetar) in Fig.~\ref{fig:tau_histogram_pop_uninform}, which is \(6.8 \leq \log_{10} (\tau \, \rm{s}^{-1}) \leq 7.6\).
A secondary peak occurs in the histogram of the population-informed \(\tau\) posterior medians in Fig.~\ref{fig:histogram_median_of_pop_informed_posteriors}, at \(\log_{10} (\tau \, \rm{s}^{-1}) \approx 7.4\).
Five out of the seven pulsars around this secondary mode are pulsars that have glitch histories (see Section~\ref{sec:glitch_relaxation_comparison}).

\subsection{Crust and superfluid noise variances} \label{subsec:hyper_QcQs}
We turn now to population-level trends obeyed by the normalized crust and superfluid noise variances.
Columns three and four in Table~\ref{tab:hypermodel_comparison_Qc-Qs-tau_101psr} present the estimates of \(C_{2, \theta}\), \(a_{\theta}\), \(b_{\theta}\), and \(\sigma_{\theta}\) for \(\theta = Q_{\rm c}\) and \(Q_{\rm s}\) respectively.
We find that \(Q_{\rm c}\) increases with \(\Omega_{\rm c}\) and \(\dot{\Omega}_{\rm c}\), scaling as \(10^{\mu_{Q_{\rm c}}} \propto \Omega_{\rm c}^{1.23^{+0.80}_{-0.75}} |\dot{\Omega}_{\rm c}|^{0.49^{+0.27}_{-0.32}}\).
Likewise, \(Q_{\rm s}\) increases with both \(\Omega_{\rm c}\) and \(\dot{\Omega}_{\rm c}\), scaling as \(10^{\mu_{Q_{\rm s}}} \propto \Omega_{\rm c}^{0.71^{+0.76}_{-0.78}} |\dot{\Omega}_{\rm c}|^{1.27^{+0.30}_{-0.28}}\).

The population-level information also modifies the \(Q_{\rm c}\) and \(Q_{\rm s}\) posteriors at the per-pulsar level in Section~\ref{sec:model_selection_psr_level}.
The bottom left and middle panels in Fig.~\ref{fig:histogram_median_of_pop_informed_posteriors} display histograms of the medians of the 101 population-informed posteriors of \(Q_{\rm c}\) and \(Q_{\rm s}\) respectively.
The median of the population-informed posterior medians of \(Q_{\rm c}\) is \(\log_{10} (Q_{\rm c} / \rm{rad}^2 \rm{s}^{-3}) = -24.9\), which is smaller than the per-pulsar results for \(Q_{\rm c}\) without the population-level overlay in Section~\ref{subsec:mapping_to_sigma_TN} and Fig.~\ref{fig:QcQsTN_corner_pop_uninform} by 0.2 dex.
The median of the population-informed posterior medians of \(Q_{\rm s}\) is \(\log_{10} (Q_{\rm s} / \rm{rad}^2 \rm{s}^{-3}) = -23.1\), which is larger than the per-pulsar results for \(Q_{\rm s}\) without the population-level overlay in Section~\ref{subsec:mapping_to_sigma_TN} and Fig.~\ref{fig:QcQsTN_corner_pop_uninform} by 0.7 dex.
The 68\% credible intervals of the population-informed posterior medians for \(Q_{\rm c}\) and \(Q_{\rm s}\) are narrower than the per-pulsar counterparts by a factor of \(0.6\) and \(0.5\) respectively.
Secondary peaks are observed at \(\log_{10} (Q_{\rm c} / \rm{rad}^2 \rm{s}^{-3}) \approx -24\) and \(\log_{10} (Q_{\rm s} / \rm{rad}^2 \rm{s}^{-3}) \approx -20.5\) and \(-19.5\) in Fig.~\ref{fig:histogram_median_of_pop_informed_posteriors}.
However, it is unclear whether the secondary peaks are physical (e.g.\ originating from a distinct subpopulation) or artifacts of small counts.

The scaling of \(\sigma_{\rm TN}\) with \(\Omega_{\rm c}, |\dot{\Omega}_{\rm c}|\) and \(T_{\rm obs}\) has been examined in the literature, e.g.\ red noise strength in TOA residuals satisfies \(\chi_{\rm RN} \propto \Omega_{\rm c}^{-0.84^{+0.47}_{-0.49}} |\dot{\Omega}_{\rm c}|^{0.97^{+0.16}_{-0.19}} T_{\rm obs}^{1.0\pm1.2}\) in \citet{LowerEtAl2020}. 
For the two-component model, we map \(Q_{\rm c}\) and \(Q_{\rm s}\) to \(\sigma_{\rm TN}\) using \eqref{eq:sigma_TN_conversion}.
With the population-informed \(\tau\) satisfying \(\log_{10} (\tau\,\rm{s}^{-1}) \lesssim 7.5 \lesssim T_{\rm obs}\), we can write approximately \(\sigma_{\rm TN}\propto \Omega_{\rm c}^{-1} [x_{\rm s}^2 Q_{\rm s} + (1 - x_{\rm s})^2 Q_{\rm c}]^{1/2} T_{\rm obs}^{3/2}\), where we define \(x_{\rm s} = (1 + \tau_{\rm c} / \tau_{\rm s})^{-1}\).
The hierarchical analysis implies that \(\tau_{\rm s} / \tau_{\rm c}\), and hence \(x_{\rm s}\), are approximately independent of \(\Omega_{\rm c}\) and \(\dot{\Omega}_{\rm c}\) (not plotted for brevity).
It also implies \(Q_{\rm s} \sim 10^2 Q_{\rm c}\) for most objects (see Fig.~\ref{fig:histogram_median_of_pop_informed_posteriors}). 
Together these results imply \(\sigma_{\rm TN} \propto \Omega_{\rm c}^{-0.65^{+0.38}_{-0.39}} |\dot{\Omega}_{\rm c}|^{0.64^{+0.15}_{-0.14}} T_{\rm obs}^{3/2}\) for \(\tau_{\rm s} / \tau_{\rm c} \gg 10^{-1}\), and \(\sigma_{\rm TN} \propto \Omega_{\rm c}^{-0.39^{+0.40}_{-0.38}} |\dot{\Omega}_{\rm c}|^{0.25^{+0.14}_{-0.16}} T_{\rm obs}^{3/2}\) for \(\tau_{\rm s} / \tau_{\rm c} \ll 10^{-1}\).
The per-pulsar analysis in Section~\ref{sec:model_selection_psr_level} yields \(\tau_{\rm s} / \tau_{\rm c}\) values in individual objects of the latter and former regimes, with \(10^{-4} \lesssim \tau_{\rm s} / \tau_{\rm c} \lesssim 10^{4}\). 
The \(\sigma_{\rm TN}\) scaling for \(\tau_{\rm s} / \tau_{\rm c} \gg 10^{-1}\) agrees broadly with previous measurements \(\log_{10} \sigma_{\rm TN} \propto \log_{10}(\Omega_{\rm c} |\dot{\Omega}_{\rm c}|^{-0.9 \pm 0.2})\) \citep{ParthasarathyEtAl2019}, 
\(\sigma_{\rm TN} \propto \Omega_{\rm c}^{-0.84^{+0.47}_{-0.49}} |\dot{\Omega}_{\rm c}|^{0.97^{+0.16}_{-0.19}} T_{\rm obs}^{1.0\pm1.2}\) \citep{LowerEtAl2020}, and \(\sigma_{\rm TN} \propto \Omega_{\rm c}^{-0.9\pm 0.2} |\dot{\Omega}_{\rm c}|^{1.0 \pm 0.05} T_{\rm obs}^{1.9 \pm 0.2}\) \citep{ShannonCordes2010}.
Another recent measurement with UTMOST-NS data reports \(\sigma_{\rm TN} \propto \Omega_{\rm c}^{-0.85^{+0.38}_{-0.35}} |\dot{\Omega}_{\rm c}|^{0.56\pm0.16} T_{\rm obs}^{1.2\pm0.8}\) \citep{DunnEtAl2025}, which aligns more closely with the \(\sigma_{\rm TN}\) scaling in this paper for \(\tau_{\rm s}/\tau_{\rm c} \gg 10^{-1}\).
We summarize the \(\sigma_{\rm TN}\) scalings obtained by various authors in Table~\ref{tab:regression_scaling_comparison} for the reader's convenience.
For completeness, the bottom right panel of Fig.~\ref{fig:histogram_median_of_pop_informed_posteriors} displays a histogram of the 101 population-informed \(\sigma_{\rm TN}^2\) posterior medians.

\begin{table}
    \setlength{\tabcolsep}{2.15pt}
    \centering
    \caption{Comparison between the scaling \(\sigma_{\rm TN} \propto \Omega_{\rm c}^{a_{\rm TN}} |\dot{\Omega}_{\rm c}|^{b_{\rm TN}} T_{\rm obs}^{\gamma_{\rm TN}}\) (column~2: \(\tau_{\rm s} / \tau_{\rm c} \gg 10^{-1}\) and column~3: \(\tau_{\rm s} / \tau_{\rm c} \ll 10^{-1}\)) inferred in this paper and equivalent results from \citet{ShannonCordes2010} (SC10; column~4), \citet{LowerEtAl2020} (Lo+20; column~5), and \citet{DunnEtAl2025} (Du+25; column~6). The scaling in this paper splits into two regimes, depending on \(\tau_{\rm s} / \tau_{\rm c}\). 
    We have \(\sigma_{\rm TN} \propto \Omega_{\rm c}^{-1} [x_{\rm s}^2 Q_{\rm s} + (1 - x_{\rm s})^2 Q_{\rm c}]^{1/2} T_{\rm obs}^{3/2}\) with \(x_{\rm s} = (1 + \tau_{\rm c} / \tau_{\rm s})^{-1}\) and \(\tau_{\rm s} / \tau_{\rm c}\), which is not modelled explicitly at the population level due to a lack of theoretical information, governs the relative contribution of \(Q_{\rm s}\) and \(Q_{\rm c}\) to \(\sigma_{\rm TN}\).}
    \label{tab:regression_scaling_comparison}
    \begin{tabular}{rccccc}
        \hline
         & \(\tau_{\rm s} / \tau_{\rm c} \gg 10^{-1}\) & \(\tau_{\rm s} / \tau_{\rm c} \ll 10^{-1}\) & SC10 & Lo+20 & Du+25 \\
        \hline
        \(a_{\rm TN}\) & \( -0.65^{+0.38}_{-0.39} \) & \( -0.39^{+0.40}_{-0.38} \) & \( -0.9\pm0.2 \) & \( -0.84^{+0.47}_{-0.49} \) & \( -0.85^{+0.38}_{-0.35} \) \\
        \(b_{\rm TN}\) & \( 0.64^{+0.15}_{-0.14} \) & \( 0.25^{+0.14}_{-0.16} \) & \( 1.0\pm0.05 \) & \( 0.97^{+0.16}_{-0.19} \) & \( 0.56\pm0.16 \) \\
        \(\gamma_{\rm TN}\) & \(3/2\) & \(3/2\) & \(1.9\pm0.2\) & \(1.0\pm1.2\) & \(1.2\pm0.8\) \\
        \hline
    \end{tabular}
\end{table}

\section{Glitch recoveries} \label{sec:glitch_relaxation_comparison}
Most published measurements of the crust-superfluid coupling time-scale are derived from the characteristic time-scale of quasi-exponential glitch recoveries \citep{McCullochEtAl1983,DodsonEtAl2001,WongEtAl2001}.
It is natural to ask whether the glitch recovery time-scale \(\tau_{\rm g}\) agrees or disagrees with the timing noise time-scale \(\tau\) measured in Sections~\ref{sec:model_selection_psr_level} and \ref{sec:hierarchical_regression}, in those pulsars where glitches have been observed.
The question is important theoretically, because it is unclear at the time of writing whether the same relaxation physics operates before, during, and after a glitch \citep{HaskellMelatos2015,AntonelliEtAl2022,AntonopoulouEtAl2022,ZhouEtAl2022}.
In Section~\ref{subsec:glitch_tau_comparison}, we compare \(\tau_{\rm g}\) and \(\tau\) in the five UTMOST pulsars, that glitched before the UTMOST timing campaign, for which \(\tau_{\rm g}\) is resolved.
In Section~\ref{subsec:healing_parameter}, we examine the inferred value of \(\tau_{\rm s} / \tau_{\rm c}\) and relate it to the healing parameter \(q_{\rm heal}\) measured in the five objects studied in Section~\ref{subsec:glitch_tau_comparison}, to check whether the data are in accord with the relation between \(\tau_{\rm s} / \tau_{\rm c}\) and \(q_{\rm heal}\) predicted theoretically.

\begin{table*}
    \caption{
        Comparison between glitch-based and timing-noise-based measurements of the crust-superfluid coupling: population-informed hierarchical Bayesian estimates of \(\tau\) (column~2; units s) and \(\tau_{\rm s} / \tau_{\rm c}\) (column~3; dimensionless) in this work compared against \(\tau_{\rm g}\) (column~5; units s) and \(q_{\rm heal} \approx (1 + \tau_{\rm c} / \tau_{\rm s})^{-1}\) (column~6; dimensionless) measured directly from glitch recoveries, for the five pulsars in the UTMOST dataset identified to have resolved glitch recoveries before the UTMOST timing campaign.
        The associated glitch epochs (column~4; units Modified Julian Date) are also listed, together with references (column~7) for the glitch-based measurements.
        Glitch data are taken from the ATNF Pulsar Catalog \citep{ManchesterEtAl2005} and the references in column~7.}
    \label{tab:tau_comparison_with_glitch_recovery}
    \begin{tabular}{lcccccl}
\hline
PSR J & \(\log_{10} (\tau\,\rm{s}^{-1})\) & \(\log_{10} [(1 + \tau_{\rm c}/\tau_{\rm s})^{-1}]\) & Epoch & \(\log_{10} (\tau_{\rm g}\,\rm{s}^{-1})\) & \(\log_{10} q_{\rm heal}\) & References \\
\hline \multirow{3}{*}{1048$-$5832} & \multirow{3}{*}{$7.4^{+0.26}_{-0.23}$} & \multirow{3}{*}{$-0.0036^{+0.0035}_{-0.064}$} & 49034(9) & $7.1^{+0.1}_{-0.14}$ & $-1.6^{+0.09}_{-0.11}$ & \citet{WangEtAl2000} \\
 &  &  & 50788(3) & $6.7^{+0.12}_{-0.18}$ & $-2.1^{+0.14}_{-0.2}$ & \citet{WangEtAl2000} \\
 &  &  & 56756(4) & $6.6^{+0.23}_{-0.5}$ & $-2.4^{+0.18}_{-0.3}$ & \citet{LowerEtAl2021} \\
\hline \multirow{1}{*}{1141$-$6545} & \multirow{1}{*}{$7.0^{+0.3}_{-0.3}$} & \multirow{1}{*}{$-0.63^{+0.57}_{-0.81}$} & 54277(20) & $7.6^{+0.11}_{-0.14}$ & $-2.4^{+0.07}_{-0.084}$ & \citet{ManchesterEtAl2010} \\
\hline \multirow{1}{*}{1452$-$6036} & \multirow{1}{*}{$6.9^{+0.26}_{-0.43}$} & \multirow{1}{*}{$-0.0037^{+0.0036}_{-0.15}$} & 55055.22(4) & $8.3^{+0.12}_{-0.16}$ & $-0.92^{+0.15}_{-0.23}$ & \citet{LowerEtAl2021} \\
\hline \multirow{5}{*}{1803$-$2137} & \multirow{5}{*}{$7.4^{+0.39}_{-0.38}$} & \multirow{5}{*}{$-0.0046^{+0.0044}_{-0.099}$} & 48245(11) & $7.1^{+0.0084}_{-0.0085}$ & $-1.9^{+0.0094}_{-0.0096}$ & \citet{EspinozaEtAl2011}  \\
 &  &  & 50777(4) & $6.0^{+0.067}_{-0.079}$ & $-2.0^{+0.048}_{-0.054}$ & \citet{EspinozaEtAl2011,YuEtAl2013} \\
 &  &  & 50777(4) & $6.8^{+0.075}_{-0.091}$ & $-2.5^{+0.19}_{-0.36}$ & \citet{YuEtAl2013} \\
 &  &  & 53429(1) & $7.1^{+0.035}_{-0.037}$ & $-2.2^{+0.011}_{-0.011}$ & \citet{EspinozaEtAl2011,YuEtAl2013} \\
 &  &  & 55775(2) & $6.5^{+0.051}_{-0.058}$ & $-2.1^{+0.03}_{-0.032}$ & \citet{LowerEtAl2021} \\
\hline \multirow{1}{*}{1833$-$0827} & \multirow{1}{*}{$7.4^{+0.46}_{-0.59}$} & \multirow{1}{*}{$-1.5^{+1.2}_{-1.8}$} & 48051(4) & $7.2^{+0.079}_{-0.097}$ & $-3.0^{+0.087}_{-0.11}$ & \citet{EspinozaEtAl2011}  \\
\hline
\end{tabular}

\end{table*}

\subsection[Are glitch-based tau and timing-noise-based tau equal?]{Are \(\tau_{\rm g}\) and \(\tau\) equal?} \label{subsec:glitch_tau_comparison}
Among all 286 UTMOST pulsars, a total of 46 pulsars are known to have glitched before or after the UTMOST timing campaign, according to the Australian Telescope National Facility (ATNF) Pulsar Catalog \citep{ManchesterEtAl2005}.\footnote{\url{https://www.atnf.csiro.au/research/pulsar/psrcat/glitchTbl.html}}
Half of the 46 glitching objects have \(\ln \mathfrak{B}_{\rm BF} < 5\).
Among the 101 canonical pulsars with \(\ln \mathfrak{B}_{\rm BF} \geq 5\), a total of 23 have glitched before or after the UTMOST timing campaign.
Objects with pre- or post-observation glitches are marked with a superscript `g' in Table~\ref{tab:model_comparison_canonical}.\footnote{
    None of the pulsars selected for analysis in this paper glitched during the UTMOST timing campaign.
    The selection policy is deliberate; the two-component model \eqref{eq:EOM_c}--\eqref{eq:phi_c_EOM} and the WTN model do not describe glitches without modification.
}
Only nine out of the 46 objects have \(\tau_{\rm g}\) measurements to date, specifically PSR J1048$-$5832, PSR J1141$-$6545, PSR J1452$-$6036, PSR J1803$-$2137, and PSR J1833$-$0827 (all with \(\ln \mathfrak{B}_{\rm BF} \geq 5\)), as well as PSR J1123$-$6259, PSR J1757$-$2421, PSR J1841$-$0425, and PSR J1852$-$0635 (all with \(\ln \mathfrak{B}_{\rm BF} < 5\)).

Columns two and five in Table~\ref{tab:tau_comparison_with_glitch_recovery} summarize the population-informed, timing-noise-based estimates of \(\tau\) and the glitch recovery time-scales \(\tau_{\rm g}\) measured for the five objects with \(\ln \mathfrak{B}_{\rm BF} \geq 5\).
Some objects have glitched several times, so we tabulate \(\tau_{\rm g}\) for each event (epoch in the fourth column).
The percentage differences between median \(\log_{10} (\tau\,\rm{s}^{-1})\) and median \(\log_{10} (\tau_{\rm g}\,\rm{s}^{-1})\) range from 2\% for PSR J1833$-$0827 to 23\% for PSR J1803$-$2137.
For objects with multiple glitches, such as PSR J1048$-$5832 and PSR J1803$-$2137, the glitch-to-glitch variations in \(\log_{10} \tau_{\rm g}\)
are comparable to the discrepancies between \(\log_{10} (\tau\,\rm{s}^{-1})\) and \(\log_{10} (\tau_{\rm g}\,\rm{s}^{-1})\), when the uncertainties in \(\tau\) are taken into account.

The \(\tau_{\rm g}\) measurements for the four pulsars with \(\ln \mathfrak{B}_{\rm BF} < 5\) range from \(15\pm10\)\,d to \(840\pm100\)\,d (see, the ATNF Pulsar Catalog and references therein).
PSR J1757$-$2421, specifically, has three exponential relaxation components reported, with \(\tau_{\rm g} = 15\pm10\)\,d, \(98\pm18\)\,d, and \(672\pm162\)\,d \citep{YuanEtAl2017,LowerEtAl2021}.
The population-uninformed \(\tau\) posteriors are approximately flat, with \(\log_{10} (\tau\,\text{s}^{-1}) \approx 5.5^{+2.5}_{-2.5}\); see panel~(d) in Fig.~\ref{fig:cornerplot_exemplary_canonical_uninformed} for an example.
The gap between the start of the UTMOST observation and any glitch epoch is at least three times the relevant \(\tau_{\rm g}\) value for the four pulsars.
The healing parameters for the four pulsars satisfy \(q_{\rm heal} \lesssim 10^{-3}\) (see the ATNF Pulsar Catalog).
We define \(q_{\rm heal}\) and interpret \(q_{\rm heal} \ll 1\) in Section~\ref{subsec:healing_parameter}.

No systematic differences in \(\tau\) (as distinct from \(\tau_{\rm g}\)) are found between UTMOST pulsars that glitch and those that do not, except for PSR J1141$-$6545, whose pulsar-level \(\tau\) posterior is bimodal (see Section~\ref{subsubsec:psr-bimodal_tau}).

\subsection{Healing parameter} \label{subsec:healing_parameter}

Observations of post-glitch relaxation yield the healing parameter \(q_{\rm heal} = \Delta \nu_{\rm d} / \Delta \nu_{\rm g}\), where \(\Delta \nu_{\rm d}\) is the impulsive increment in the spin frequency that decays exponentially, and \(\Delta \nu_{\rm g}\) is the total glitch size.
In the two-component model of pulsar glitches, the healing parameter satisfies \(q_{\rm heal} \approx I_{\rm s} / (I_{\rm s} + I_{\rm c})\) \citep{AlparEtAl1993,AntonelliEtAl2022}.
As the terms \(I_{\rm c} (\Omega_{\rm c} - \Omega_{\rm s}) / \tau_{\rm c}\) and \(I_{\rm s} (\Omega_{\rm c} - \Omega_{\rm s}) / \tau_{\rm s}\) in \eqref{eq:EOM_c} and \eqref{eq:EOM_s} form an action-reaction pair, we have \(\tau_{\rm s} / \tau_{\rm c} = I_{\rm s} / I_{\rm c}\).
One therefore arrives at the theoretical prediction \(q_{\rm heal} \approx x_{\rm s} = (1 + \tau_{\rm c} / \tau_{\rm s})^{-1}\).
In this section, we apply the Bayesian inference results in Sections~\ref{sec:model_selection_psr_level} and \ref{sec:hierarchical_regression} to test this theoretical relation.

Column three in Table~\ref{tab:tau_comparison_with_glitch_recovery} lists \(x_{\rm s} = (1 + \tau_{\rm c} / \tau_{\rm s})^{-1}\) inferred from the hierarchical Bayesian analysis in Section~\ref{sec:hierarchical_regression} after the population-level information is overlaid.\footnote{
    Note that the log uniform prior on \(\tau_{\rm s} / \tau_{\rm c}\) at the pulsar level in Section~\ref{sec:model_selection_psr_level} significantly weights \(x_{\rm s} = (1 + \tau_{\rm c} / \tau_{\rm s})^{-1}\) towards \(x_{\rm s} \ll 10^{-1}\) and \(x_{\rm s} \approx 1\), with a dip at intermediate \(x_{\rm s}\).
    The hierarchical Bayesian analysis in Section~\ref{sec:hierarchical_regression} does not model \(\tau_{\rm s} / \tau_{\rm c}\) explicitly at the population level; see \eqref{eq:linear_regression_model}.
}
Column six in Table~\ref{tab:tau_comparison_with_glitch_recovery} lists \(q_{\rm heal}\) drawn from the literature for 11 glitches in the pulsars PSR J1048$-$5832, PSR J1141$-$6545, PSR J1452$-$6036, PSR J1803$-$2137, and PSR J1833$-$0827.
In every object, the timing-noise-based estimate of \(x_{\rm s}\) is one to two orders of magnitude greater than the glitch-based estimate of \(q_{\rm heal}\).
That is, the theoretical relation \(q_{\rm heal} \approx x_{\rm s}\) appears to be violated.
We find that three out of five objects have \(x_{\rm s} \approx 1\) and \(q_{\rm heal} \lesssim 10^{-1}\). 
One object has \(10^{-2} \lesssim x_{\rm s} \lesssim 0.9\) and \(q_{\rm heal} \approx 4\times 10^{-3}\), and one object has \(10^{-4} \lesssim x_{\rm s} \lesssim 0.5\) and \(q_{\rm heal} \approx 1\times 10^{-3}\).
The above results suggest that angular momentum is stored and transported differently between and during glitches, and that the two-component model must be generalized to accommodate this.
Specifically, the results suggest that the superfluid and crust are the only angular momentum reservoirs accessible through the timing noise mechanism between glitches, implying \(\tau_{\rm c} / \tau_{\rm s} \lesssim 1\), whereas one or more additional angular momentum reservoirs become accessible during a glitch and its immediate aftermath, implying \(q_{\rm heal} \ll 1\) in general (and therefore \(\tau_{\rm c} / \tau_{\rm s} \gg 1\)).
The implications for the two-component model are discussed in Appendix~\ref{app:healing_in_2C}.

The ratio \(\tau_{\rm s} / \tau_{\rm c} = I_{\rm s} / I_{\rm c}\) is hard to predict theoretically from first principles.
One can have \(I_{\rm s} / I_{\rm c} \ll 1\), if the rigid crust and superfluid core are tightly coupled on a time-scale shorter than \(\tau_{\rm g}\), e.g.\ magnetically \citep{Mendell1998}, and only the superfluid pinned to the inner crust contributes to \(I_{\rm s}\) \citep{AlparEtAl1984}.
This scenario corresponds to \(x_{\rm s} \ll 1\).
Alternatively, one can have \(I_{\rm c} / I_{\rm s} \ll 1\), when \(I_{\rm c}\) is composed of the rigid crust alone, and the rest of the star makes up \(I_{\rm s}\).
This scenario corresponds to \(x_{\rm s} \approx 1\).
Furthermore, the moment of inertia of the neutron superfluid pinned to the inner crust may not be sufficient to explain accumulated glitch sizes, when entrainment is considered \citep{AnderssonEtAl2012,Chamel2012}.
Measurements from the 2016 Vela glitch yield \(x_{\rm s} \approx 0.6\) and \(x_{\rm s} \approx 0.15\) for weakly and strongly pinned superfluid respectively within a three-component model, implying that the crustal reservoir is insufficient \citep{MontoliEtAl2020}.

\subsection{Effect of pre-observation glitches}
Are objects in Table~\ref{tab:tau_comparison_with_glitch_recovery} affected by pre-observation glitches?
Because the two-component model \eqref{eq:EOM_c}--\eqref{eq:phi_c_EOM} does not describe glitches, which possibly involve additional angular momentum components for \(q_{\rm heal} \neq 1\), we cannot apply the analysis in Section~\ref{sec:model_selection_psr_level} as it stands to infer the presence or absence of a timing signature lagging a glitch that occurred before the start of the UTMOST observation.
Instead, we estimate roughly how much variance in the phase residuals is expected from long-term glitch recoveries and compare it with \(\sigma_{\rm TN}\) from \eqref{eq:sigma_TN_conversion}.
To do so, we model the post-glitch recovery as \(\delta \nu_{\rm d} (t) = \delta \nu_{\rm d}(t_{0}) \exp[-(t-t_{0})/\tau_{\rm g}]\) with \(\delta \nu_{\rm d}(t_{0}) = \Delta \nu_{\rm d} \exp[-(t_{0}-t_{\rm g})/\tau_{\rm g}]\), where \(t_{\rm g}\) and \(t_{0}\) are the glitch epoch and the start of the UTMOST timing campaign, respectively.
Then the glitch-induced timing residuals \(r(t) = (2\pi / \Omega_{\rm c,0}) \int \text{d}t'\,\delta\nu_{\rm d}(t')\) split into two parts: 
(i) a term linear in \(t\), which is absorbed by \(\Omega_{\rm c, 0}\), viz.\ \(\Omega_{\rm c, 0} \mapsto \Omega_{\rm c, 0} + 2 \pi \delta \nu_{\rm d}(t_{0})\); and 
(ii) the remainder that is misattributed effectively to \(\xi_{\rm c}(t)\) and \(\xi_{\rm s}(t)\) in \eqref{eq:EOM_c}--\eqref{eq:phi_c_EOM}, denoted by \(r_{2}(t)\).
With this split, one can write
\begin{align}
    r_{2}(t) 
    &= \frac{2\pi \delta \nu_{\rm d}(t_{0})}{\Omega_{\rm c,0}} \int_{t_{0}}^{t} \text{d}t'\, \left\{ \exp[-(t-t_{0})/\tau_{\rm g}] - 1\right\} \,.
    \label{eq:r2_glitch_recovery}
\end{align}

Let us evaluate \eqref{eq:r2_glitch_recovery} for the glitches in Table~\ref{tab:tau_comparison_with_glitch_recovery}. 
In the three objects where \(r_{2}(t_{0}+T_{\rm obs})\) is significant, we find \(r_{2}(t_{0}+T_{\rm obs}) = 0.33^{+7.0}_{-0.3}\,\text{ms}\) for PSR J1048$-$5832 (glitch epoch MJD\,56756), \(0.48^{+1.1}_{-0.42}\,\text{ms}\) for PSR J1141$-$6545 (glitch epoch MJD\,54277), and \(3.6^{+1.6}_{-1.7}\,\text{ms}\) for PSR J1452$-$6036 (glitch epoch MJD\,55055.22), where the quoted uncertainties are the 68\% credible intervals.
The other glitches in Table~\ref{tab:tau_comparison_with_glitch_recovery} have \((t_{0} - t_{\rm g}) / \tau_{\rm g} > 34\) and hence \(r_{2}(t) \lesssim 10^{-15}\,\text{s}\).
Upon comparing with the population-uninformed \(\sigma_{\rm TN}\) from \eqref{eq:sigma_TN_conversion}, we obtain \(\log_{10} [r_{2}(t_{0}+T_{\rm obs}) / \sigma_{\rm TN}] = -4.1^{+1.5}_{-4.5}\), \(-1.3^{+0.62}_{-1.0}\), and \(-0.88^{+0.27}_{-0.47}\) for PSR J1048$-$5832, PSR J1141$-$6545, and PSR J1452$-$6036 respectively.
Hence, long-term glitch recoveries contribute negligibly to the timing residuals except perhaps for PSR J1141$-$6545 and PSR J1452$-$6036, where \(r_{2}(t_{0}+T_{\rm obs})\) contributes \(\lesssim 20\%\) of \(\sigma_{\rm TN}\).
This may explain the bimodal \(\tau\) posterior for PSR J1141$-$6545 (see Section~\ref{subsubsec:psr-bimodal_tau}).
For PSR J1452$-$6036, the population-uninformed \(\tau\) posterior is approximately flat with a slight bump around \(\log_{10} (\tau\,\text{s}^{-1}) \sim 7.5\), which may arise because \(T_{\rm obs} \approx \tau_{\rm g} / 5\) is too short to resolve \(\tau\) or \(\tau_{\rm g}\).

\section{Conclusion} \label{sec:discussion_conclusion}

The crust-superfluid coupling time-scale \(\tau\) and crust and superfluid noise amplitudes, \(Q_{\rm c}\) and \(Q_{\rm s}\) respectively, are analyzed for a sample of UTMOST pulsars with a Kalman filter.
We perform Bayesian model selection between the two-component model and a one-component WTN model at the pulsar level and then overlay population-level constraints through a hierarchical Bayesian regression.
We find that 105 UTMOST objects favor the two-component model with \(\ln \mathfrak{B}_{\rm BF} \geq 5\), and 28 out of the 105 objects have unimodal and sharply peaked per-pulsar posteriors in \(\tau\).
The posterior medians span the range \(4.6 \leq \log_{10} (\tau \, \rm{s}^{-1}) \leq 8.6\) and have been measured for 26 out of 28 objects for the first time.
Additionally, the per-pulsar posterior medians of \(Q_{\rm c}\) and \(Q_{\rm s}\) span \(-27.7 \leq \log_{10} (Q_{\rm c} / \rm{rad}^2 \rm{s}^{-3}) \leq -18.7\) and \(-27.7 \leq \log_{10} (Q_{\rm s} / \rm{rad}^2 \rm{s}^{-3}) \leq -15.1\) respectively.
At the population level, the 101 canonical pulsars with \(\ln \mathfrak{B}_{\rm BF} \geq 5\) in the populous core of the \(\Omega_{\rm c}\)-\(\dot{\Omega}_{\rm c}\) plane (excluding recycled pulsars and magnetars) obey \(\tau \propto \Omega_{\rm c}^{0.19^{+0.50}_{-0.52}} |\dot{\Omega}_{\rm c}|^{0.18^{+0.18}_{-0.19}}\), \(Q_{\rm c} \propto \Omega_{\rm c}^{1.23^{+0.80}_{-0.75}} |\dot{\Omega}_{\rm c}|^{0.49^{+0.27}_{-0.32}}\), and \(Q_{\rm s} \propto \Omega_{\rm c}^{0.71^{+0.76}_{-0.78}} |\dot{\Omega}_{\rm c}|^{1.27^{+0.30}_{-0.28}}\).
The population-level information shifts the per-pulsar posterior medians by at most \(\pm2.1\) dex, \(\pm2.6\) dex, and \(\pm3.5\) dex for \(\tau\), \(Q_{\rm c}\), and \(Q_{\rm s}\) respectively.

We compare the timing-noise-based estimates of \(\tau\) and \((1 + \tau_{\rm c} / \tau_{\rm s})^{-1}\) with glitch-based measurements of \(\tau_{\rm g}\) and \(q_{\rm heal}\) respectively in five UTMOST pulsars.
We find agreement within 24\% or better between \(\log_{10} (\tau\,\rm{s}^{-1})\) and \(\log_{10} (\tau_{\rm g}\,\rm{s}^{-1})\), implying that the frictional crust-superfluid coupling mechanism operates similarly during and between glitches.
On the other hand, timing-noise-based \((1 + \tau_{\rm c} / \tau_{\rm s})^{-1}\) values are typically one to two orders of magnitude larger than \(q_{\rm heal}\), implying that the crust and superfluid are the only angular momentum reservoirs accessible between glitches, yet one or more additional reservoirs become accessible during glitches.

It is interesting to ask whether the hierarchical Bayesian results in this paper shed new light on the physical origin of the crust-superfluid coupling.
Several mechanisms have been proposed to explain the coupling theoretically, including 
Ekman pumping at the crust-superfluid boundary \citep{Easson1979,AbneyEpstein1996,VanEysdenMelatos2010}, 
magnetic stresses \citep{Easson1979,Melatos2012,GlampedakisLasky2015,BransgroveEtAl2018}, 
and mutual friction due to scattering of superfluid vortices against electrons or magnetic flux lines \citep{AlparEtAl1984a,AlparSauls1988,Mendell1991,AnderssonEtAl2006,SideryAlpar2009}.
An exhaustive discussion of this theoretical issue lies outside the scope of this empirical paper.
As one example, however, let us consider mutual friction between neutron vortices and electrons, which is quantified by a dimensionless mutual friction parameter \(\mathcal{B}_{\rm mf} = \mathcal{R} / (1 + \mathcal{R}^2)\), where \(\mathcal{R}\) is the drag-to-lift ratio without vortex pinning \citep{AnderssonEtAl2006}, viz.\
\begin{align}
    \mathcal{R}
    \approx 4 \times 10^{-4} 
    \left(\frac{\delta m_{\rm p}^{*}}{m_{\rm p}}\right)^2 
    \left(\frac{m_{\rm p}}{m_{\rm p}^{*}}\right)^{1/2} 
    \left(\frac{x_{\rm p}}{0.05}\right)^{7/6} 
    \left(\frac{\rho}{10^{14} \rm g\,cm^3}\right)^{1/6}.
    \label{eq:R_drag_to_lift_ratio}
\end{align}
In \eqref{eq:R_drag_to_lift_ratio}, \(m_{\rm p}\) and \(m_{\rm p}^{*}\) are the bare and effective proton masses due to entrainment, \(x_{\rm p} = \rho_{\rm p} / \rho\) is the proton fraction, \(\rho\) is the density, and we write \(\delta m_{\rm p}^{*} = m_{\rm p} - m_{\rm p}^{*}\).
The coupling time-scale associated with~\eqref{eq:R_drag_to_lift_ratio} is given by \citep{HaskellMelatos2015,MontoliEtAl2020}
\begin{align}
    \tau_{\rm mf} 
    &= \frac{1}{2 \Omega_{\rm c} \mathcal{B}_{\rm mf}}
    \label{eq:tau_mf}
    \\
    &= 1.25\times10^{3}
    \left(\frac{\Omega_{\rm c}}{1\,\rm rad\,s^{-1}}\right)^{-1}
    \left(\frac{B_{\rm mf}}{4\times10^{-4}}\right)^{-1} \, \rm{s}.
    \label{eq:tau_mf_scaling}
\end{align}
Consider a typical UTMOST pulsar with \(\log_{10} (\Omega_{\rm c}\,\rm rad^{-1}\,s) = 1\) and assume the 2004 Vela glitch value \(B_{\rm mf} = 4\times10^{-5}\) \citep{NewtonEtAl2015}.
Equation~\eqref{eq:tau_mf_scaling} implies \(\tau_{\rm mf} = 1.25\times10^{3}\,\rm{s}\), which is four orders of magnitude smaller than typical \(\tau\) estimates in this work.
Equation~\eqref{eq:tau_mf_scaling} also disagrees with the scaling \(\tau \propto \Omega_{\rm c}^{0.19^{+0.50}_{-0.52}}\) inferred in Section~\ref{subsec:hyper_tau}.
There are many plausible reasons for the above discrepancies.
First and foremost, it is likely that vortex pinning plays a role in the crust-superfluid coupling \citep{HaskellAntonopoulou2013,SourieChamel2020}.
Pinning modifies \(\tau_{\rm mf}\), with \(\tau_{\rm mf} \rightarrow \infty\) as \(\mathcal{R} \rightarrow \infty\) and \(\mathcal{B}_{\rm mf} \rightarrow 0\) in the limit of perfect pinning \citep{HaskellAntonopoulou2013,SourieChamel2020}.
Moreover, the form of \(\tau_{\rm mf}\) in \eqref{eq:tau_mf} changes, if a superfluid vortex tangle exists, and the linear coupling in \eqref{eq:EOM_c} and \eqref{eq:EOM_s} is replaced by the nonlinear Gorter-Mellink torque \citep{PeraltaEtAl2006,AnderssonEtAl2007}.

It is also interesting to compare the population-level trends for \(Q_{\rm c}\) and \(Q_{\rm s}\), or equivalently \(\sigma_{\rm TN}\), in this paper with existing theory and empirical studies.
Theoretically, superfluid turbulence in the core yields \(\sigma_{\rm TN} \propto \Omega_{\rm c}^{-1} |\dot{\Omega}_{\rm c}|^{1/6}\) \citep{MelatosLink2014}, which approaches but disagrees marginally with \(\sigma_{\rm TN} \propto \Omega_{\rm c}^{-0.39^{+0.40}_{-0.38}} |\dot{\Omega}_{\rm c}|^{0.25^{+0.14}_{-0.16}}\) for \(\tau_{\rm s} / \tau_{\rm c} \ll 10^{-1}\) in Section~\ref{subsec:hyper_QcQs}.
Seismic activity such as crustquakes caused by mechanical failure is expected to be a source of noise in the crust \citep{MiddleditchEtAl2006,KerinMelatos2022,BransgroveEtAl2025}, but no first-principle theory of crustquake noise statistics has been developed at the time of writing.
Recent measurements of \(\sigma_{\rm TN}\), such as \(\sigma_{\rm TN} \propto \Omega_{\rm c}^{-0.9\pm 0.2} |\dot{\Omega}_{\rm c}|^{1.0 \pm 0.05}\) \citep{ShannonCordes2010}, \(\sigma_{\rm TN} \propto \Omega_{\rm c}^{-0.84^{+0.47}_{-0.49}} |\dot{\Omega}_{\rm c}|^{0.97^{+0.16}_{-0.19}}\) \citep{LowerEtAl2020}, and \(\sigma_{\rm TN} \propto \Omega_{\rm c}^{-0.85^{+0.38}_{-0.35}} |\dot{\Omega}_{\rm c}|^{0.56\pm0.16}\) \citep{DunnEtAl2025}, are broadly consistent with \(\sigma_{\rm TN} \propto \Omega_{\rm c}^{-0.65^{+0.38}_{-0.39}} |\dot{\Omega}_{\rm c}|^{0.64^{+0.15}_{-0.14}}\) for \(\tau_{\rm s} / \tau_{\rm c} \gg 10^{-1}\) in this work.
Marginalizing over \(\tau_{\rm s} / \tau_{\rm c}\), the \(\sigma_{\rm TN}\) scaling is marginally consistent with the theoretical prediction \(\sigma_{\rm TN} \propto \Omega_{\rm c}^{-0.5} |\dot{\Omega}_{\rm c}|^{0.5} T_{\rm obs}^{3/2}\) for noises originating only from electromagnetic torques \citep{AntonelliEtAl2023}.
In future work, we aim to introduce mixture models as population-level constraints (e.g.\ a hybrid of Gaussian-Gaussian and Gaussian-uniform mixture models for \(Q_{\rm s}\) and \(Q_{\rm c}\), and/or a Gaussian mixture model for \(\tau_{\rm s} / \tau_{\rm c}\)) to check the bimodality in \(Q_{\rm s}\) and estimate the scaling of \(\sigma_{\rm TN}\) for different \(\tau_{\rm s} / \tau_{\rm c}\) consistently.

The Kalman filter in this paper does not model chromatic TOA fluctuations caused by interstellar propagation, e.g.\ DM variations.
It only models achromatic timing noise caused by the stochastic torques acting on the crust and superfluid in the two-component model.
A preliminary and approximate estimate of the impact of DM variations is presented in Appendix~\ref{app:DM-variations}, where it is shown that DM variations measured by the MeerTime program amount to \(\sim 10\%\) of the achromatic timing noise attributed to stochastic torques in the two-component model for 103 of the pulsars with \(\ln \mathfrak{B}_{\rm BF} > 5\) analyzed in Sections~\ref{sec:model_selection_psr_level}--\ref{sec:hierarchical_regression}.
To robustly model chromatic noise such as DM variations in future work, TOAs at multiple wavelengths across a wide band (\(\sim\text{GHz}\)) must be analyzed simultaneously, e.g.\ by combining the UTMOST data with MeerTime and pulsar timing array datasets \citep{BailesEtAl2020,AgazieEtAl2023a,AntoniadisEtAl2023,MilesEtAl2023,ReardonEtAl2023,XuEtAl2023,MilesEtAl2025}.
Generalizing the single-band Kalman filter in this paper to handle multi-band data [e.g.\ through expanding the measurement matrix \(\mathbfss{H}\) in \eqref{eq:measurement_equation} and \eqref{eq:measurement_matrix}] is an important but substantial task, which we defer to future work.

Glitching pulsars are excluded from the analysis in this paper, except for a brief discussion in Section~\ref{sec:glitch_relaxation_comparison}, because \eqref{eq:EOM_c} and \eqref{eq:EOM_s} do not model impulsive changes in \(\Omega_{\rm c}\).
In the future, we plan to extend \eqref{eq:EOM_c} and \eqref{eq:EOM_s} and the associated Kalman filter to handle glitching pulsars \citep{MelatosEtAl2020,DunnEtAl2021,LowerEtAl2021,DunnEtAl2022,DunnEtAl2023,DunnEtAl2025}.
The Kalman filter results will be compared with other, complementary ways of studying the mutual friction force and its relation to vortex pinning, e.g.\ via the glitch rate in individual pulsars \citep{MelatosMillhouse2023}.
We also hope to apply the Kalman filter and associated hierarchical Bayesian scheme to larger data sets, such as MeerKAT timing data \citep{MilesEtAl2023}.
Long baseline (\(T_{\rm obs}\gtrsim 5\,\)yr) data with daily to weekly cadence and \(N_{\rm TOA} \sim 10^3\) are desirable, as the accuracy of the estimates increases with increasing \(N_{\rm TOA}\).
For example, the accuracies with which \(Q_{\rm c}\) and \(Q_{\rm s}\) are inferred improve by \(\approx 20\%\), when one quadruples \(N_{\rm TOA}\) \citep{MeyersEtAl2021}.

\section*{Acknowledgements}
The authors thank Robin Evans and Farhad Farokhi for helpful discussions about the Kalman filter and Marcus Lower for valuable discussions about pulsar classification.
This research is supported by the Australian Research Council (ARC) through the Centre of Excellence for Gravitational Wave Discovery (OzGrav) (grant number CE230100016).
W.\ Dong and N.\ J.\ O'Neill acknowledge the support of the Melbourne Research Scholarship (MRS).
W.\ Dong is also a recipient of the John Tyndall Scholarship -- 2024, and the Dieul-Kurzweil Scholarship -- 2024.
The computations were performed on the OzSTAR supercomputer facility at Swinburne University of Technology. The OzSTAR program receives funding in part from the Astronomy National Collaborative Research Infrastructure Strategy (NCRIS) allocation provided by the Australian Government, and from the Victorian Higher Education State Investment Fund (VHESIF) provided by the Victorian Government.
The authors also thank the anonymous referee for their valuable and constructive feedback, which has improved the completeness of the paper.

\section*{Data Availability}
The UTMOST data underlying this work are publicly available at \url{
    https://github.com/Molonglo/TimingDataRelease1/
}.
The data produced in this work, as well as the numerical code, will be made available in the repository at \url{
    https://github.com/wdong-eric
} and shared upon reasonable request to the corresponding author.



\bibliographystyle{mnras}
\bibliography{psrpop_utmost-manu}



\appendix

\section{Population-informed posteriors from reweighted samples}
\label{app:derivation_popinform_posterior_reweighting}
In this appendix, we derive \eqref{eq:indiv_posterior_reweighting_weight}, which is an approximation to the population-informed posterior for the per-pulsar parameters \(\bm{\theta}^{(i)}\) of the \(i\)-th pulsar given \(n_{\rm s, \Lambda}\) hyperparameter samples \(\{\bm{\Lambda}_{k}\}_{k=1}^{n_{\rm s, \Lambda}}\).

From the definition of \(p[\{\bm{\theta}^{(i)}\}, \bm{\Lambda} \,|\, \{\bm{d}^{(i)}\}_{i=1}^{N_{\rm psr}}]\) [c.f.\ \eqref{eq:full_hyperposterior-decomposition-2}], we write \eqref{eq:marginalised_hyperposterior_def} and \eqref{eq:marginalised_individual_posterior_def} as
\begin{align}
    p[\bm{\Lambda} \,|\, \{\bm{d}^{(i)}\}_{i=1}^{N_{\rm psr}}]
    =& \mathcal{Z}_{\Lambda}[\{\bm{d}^{(i)}\}_{i=1}^{N_{\rm psr}}]^{-1} \pi(\bm{\Lambda}) 
    \nonumber \\
    &\times \prod_{i = 1}^{N_{\rm psr}} \int \mathrm{d} \bm{\theta}^{(i)} \, 
    \mathcal{L}[\bm{d}^{(i)} \,|\, \bm{\theta}^{(i)}] p[\bm{\theta}^{(i)} \,|\, \bm{\Lambda}] ,
    \label{eq:marginalised_hyperposterior_fullform}
\end{align}
and
\begin{align}
    p[\bm{\theta}^{(i)} \,|\, \{\bm{d}^{(i)}\}_{i=1}^{N_{\rm psr}}]
    =&\, \mathcal{Z}_{\Lambda}[\{\bm{d}^{(i)}\}_{i=1}^{N_{\rm psr}}]^{-1}
    \int \mathrm{d}\bm{\Lambda} \, \int \mathrm{d}\{\bm{\theta}^{(j)}\}_{j\neq i}
    \nonumber \\
    & \times \pi(\bm{\Lambda}) \prod_{j = 1}^{N_{\rm psr}} \mathcal{L}[\bm{d}^{(j)} \,|\, \bm{\theta}^{(j)}] \, p[\bm{\theta}^{(j)} \,|\, \bm{\Lambda}],
    \label{eq:marginalised_individual_posterior_fullform}
\end{align}
where \(\mathcal{Z}_{\Lambda}[\{\bm{d}^{(i)}\}_{i=1}^{N_{\rm psr}}]\) is the evidence of the hierarchical model with data \(\{\bm{d}^{(i)}\}_{i=1}^{N_{\rm psr}}\).
With \eqref{eq:marginalised_hyperposterior_fullform}, we rewrite \eqref{eq:marginalised_individual_posterior_fullform} as
\begin{align}
    &p[\bm{\theta}^{(i)} \,|\, \{\bm{d}^{(i)}\}_{i=1}^{N_{\rm psr}}]
    \nonumber \\
    &= \int \mathrm{d}\bm{\Lambda} \, 
    \frac{
        p[\bm{\Lambda} \,|\, \{\bm{d}^{(i)}\}_{i=1}^{N_{\rm psr}}] \mathcal{L}[\bm{d}^{(i)} \,|\, \bm{\theta}^{(i)}] \, p[\bm{\theta}^{(i)} \,|\, \bm{\Lambda}]
    }{\int \mathrm{d} \bm{\theta}^{(i)} \, \mathcal{L}[\bm{d}^{(i)} \,|\, \bm{\theta}^{(i)}] p[\bm{\theta}^{(i)} \,|\, \bm{\Lambda}]} ,
\end{align}
which, according to Bayes's theorem, can be expressed in terms of the per-pulsar posterior \(p[\bm{\theta}^{(i)} \,|\, \bm{d}^{(i)}]\) as
\begin{align}
    &p[\bm{\theta}^{(i)} \,|\, \{\bm{d}^{(i)}\}_{i=1}^{N_{\rm psr}}]
    \nonumber \\
    &= \frac{p[\bm{\theta}^{(i)} | \bm{d}^{(i)}]}{\pi[\bm{\theta}^{(i)}]}
    \int \mathrm{d}\bm{\Lambda} \, 
    \frac{
        p[\bm{\Lambda} \,|\, \{\bm{d}^{(i)}\}_{i=1}^{N_{\rm psr}}] \,
        p[\bm{\theta}^{(i)} \,|\, \bm{\Lambda}]
    }{
        \int \mathrm{d} \bm{\theta}^{(i)} \,p[\bm{\theta}^{(i)} \,|\, \bm{d}^{(i)}] \, 
        p[\bm{\theta}^{(i)} \,|\, \bm{\Lambda}]  \pi[\bm{\theta}^{(i)}]^{-1}
    } .
    \label{eq:popinform_individual_posterior_reweighted_form}
\end{align}
For an arbitrary univariate probability density function \(p(x)\), integrals of the form \(\int dx \, p(x) f(x)\)  can be approximated as \(\int dx \, p(x) f(x) \approx \sum_{k=1}^{n_{\rm s}} f(x_{k}) / n_{\rm s}\), in terms of \(n_{\rm s}\) equally weighted random samples \(x_{k}\) drawn from \(p(x)\), where the subscript \(k\) denotes the \(k\)-th random sample.
With this approximation, \eqref{eq:popinform_individual_posterior_reweighted_form} reads
\begin{align}
    &p[\bm{\theta}^{(i)} \,|\, \{\bm{d}^{(i)}\}_{i=1}^{N_{\rm psr}}]
    \nonumber \\
    &= \frac{
        n_{\mathrm{s}, i} \, p[\bm{\theta}^{(i)} | \bm{d}^{(i)}]
    }{n_{\rm s, \Lambda} \, \pi[\bm{\theta}^{(i)}]} 
    \sum_{k=1}^{n_{\rm s, \Lambda}} 
    \frac{
        p[\bm{\theta}^{(i)} \,|\, \bm{\Lambda}_{k}]
    }{
        \sum_{j=1}^{n_{\mathrm{s}, i}} p[\bm{\theta}^{(i)}_{j} \,|\, \bm{\Lambda}_{k}] \,/\, \pi[\bm{\theta}^{(i)}_{j}]
    },
\end{align}
where \(\bm{\Lambda}_{k}\) is the \(k\)-th random sample drawn from \(p[\bm{\Lambda} \,|\, \{\bm{d}^{(i)}\}_{i=1}^{N_{\rm psr}}]\), and \(\bm{\theta}^{(i)}_{j}\) is the \(j\)-th sample drawn from the pulsar-level posterior \(p[\bm{\theta}^{(i)} \,|\, \bm{d}^{(i)}]\) before overlaying the population constraints.
Equation~\eqref{eq:hyperparam_posterior_reweighting} is derived in a similar manner \citep{ThraneTalbot2019}.
Readers are referred to \citet{MooreGerosa2021} for a detailed discussion about population-informed posteriors including selection effects in the context of population studies of gravitational-wave sources.

\onecolumn

\section{Preferred models and associated parameters at the pulsar level}
\label{app:psr_level-model_comparison-parameter}
In this appendix, we present the preferred model, log Bayes factor \(\ln \mathfrak{B}_{\rm BF}\), and estimates of \(\tau, Q_{\rm c}, Q_{\rm s}\) and \(\sigma_{\rm TN}^2\) at the pulsar level (i.e.\ before overlaying population-level constraints) for all 286 nonglitching pulsars in the UTMOST dataset in Table~\ref{tab:model_comparison_canonical}--\ref{tab:model_comparison_magnetar}.
Preferred models and \(\ln \mathfrak{B}_{\rm BF}\) values from a previous UTMOST analysis \citep{LowerEtAl2020} are also included for comparison.
The estimates of the two-component model parameters \(\tau, Q_{\rm c}, Q_{\rm s}\) and \(\sigma_{\rm TN}^2\) are quoted for all 286 objects for completeness, but naturally they are more meaningful for the 105 objects that favor the two-component model.
The objects are classified as canonical, i.e.\ located in the populous core of the \(\Omega_{\rm c}\)-\(\dot{\Omega}_{\rm c}\) plane (Table~\ref{tab:model_comparison_canonical}); recycled, i.e.\ satisfying the condition \(\dot{P}/10^{-17} \leq 3.23 (P/100 \, \textrm{ms})^{-2.34}\) proposed by \citet{LeeEtAl2012} (Table~\ref{tab:model_comparison_msp}); and magnetar-like, i.e.\ with dipole magnetic field strength exceeding \(4.4\times10^{13} \, \textrm{G}\) (Table~\ref{tab:model_comparison_magnetar}).
Every object is characterized by an indicator of whether the posterior distribution is peaky for \(\tau\).
The posterior is termed peaky arbitrarily, if the half-maximum points of the posterior are at least 0.5 dex away from the boundaries of the prior support.
The reader is advised to treat the indicator as a first pass and check themselves whether the corner plot of the posterior distribution is peaky directly from the UTMOST data.

\begin{longtable}{lcccccccrr}
    \caption{
        Canonical pulsars in the populous core of the \(\Omega_{\rm c}\)-\(\dot{\Omega}_{\rm c}\) plane. 
        Preferred model (column~2, with 2C short for the two-component model), \(\ln \mathfrak{B}_{\rm BF}\) (column~3), pulsar-level estimates of \(\tau, Q_{\rm c}, Q_{\rm s}\) and \(\sigma_{\rm TN}^2\) (columns~4, 6--8), and preferred model (PLRN: power-law red noise; PLRN+F2: PLRN with \(\ddot{\Omega}_{\rm c}\) fitted) and \(\ln \mathfrak{B}_{\rm BF}\) from the previous UTMOST analysis (columns~9 and 10) \citep{LowerEtAl2020}.
        Estimates are quantified by the median and 68\% credible interval.
        Estimates of the two-component model parameters are quoted for all 286 nonglitching UTMOST pulsars for the sake of completeness, but naturally they are more meaningful for the 105 objects that favor the two-component model.
        Column 5 indicates whether the one-dimensional marginalized posterior of \(\tau\) in column 4 is peaked, with Y for yes, N for no, and RR as a special case of N, when the posterior rails against the right edge of the prior support.
        Objects with \(\ln \mathfrak{B}_{\rm BF} \geq 5\) in this paper and \(\ln \mathfrak{B}_{\rm BF} < 5\) in \citet{LowerEtAl2020} are marked with asterisks (\(^*\)).
        Objects with \(\ln \mathfrak{B}_{\rm BF} < 5\) in this paper and \(\ln \mathfrak{B}_{\rm BF} \geq 5\) in \citet{LowerEtAl2020} are marked with daggers (\(^\dagger\)). 
        The superscript \(^{\rm g}\) marks objects that have been observed to glitch before or after the UTMOST observations.
        Pulsars with \(\ln \mathfrak{B} \geq 5\) in this paper are considered as two-component candidates and fed into the population-level hierarchical analysis in Section~\ref{sec:hierarchical_regression}.}
    \label{tab:model_comparison_canonical} \\
    
 \\
\hline
 & \multicolumn{7}{c}{} & \multicolumn{2}{c}{\citet{LowerEtAl2020}} \\
 & Model & $\ln\mathfrak{B}_{\rm BF}$ & log$_{10}$($\tau$) & Peaky? & log$_{10}$($Q_{\rm c}$) & log$_{10}$($Q_{\rm s}$) & log$_{10} \sigma_{\rm TN}^2$ & Model & $\ln \mathfrak{B}_{\rm BF}$ \\
PSR J & & & [s] & & [rad$^2$s$^{-3}]$ & [rad$^2$s$^{-3}]$ & [s$^2]$ & &  \\
\hline
\endfirsthead
\caption[]{continued from previous page.} \\
\hline
 & \multicolumn{7}{c}{} & \multicolumn{2}{c}{\citet{LowerEtAl2020}} \\
 & Model & $\ln\mathfrak{B}_{\rm BF}$ & log$_{10}$($\tau$) & Peaky? & log$_{10}$($Q_{\rm c}$) & log$_{10}$($Q_{\rm s}$) & log$_{10} \sigma_{\rm TN}^2$ & Model & $\ln \mathfrak{B}_{\rm BF}$ \\
PSR J & & & [s] & & [rad$^2$s$^{-3}]$ & [rad$^2$s$^{-3}]$ & [s$^2]$ & &  \\
\hline
\endhead
\hline
\endfoot
\hline
\endlastfoot
0134$-$2937 & WTN & -2.6 & $4.9^{+3.1}_{-1.9}$ & N & $-28.3^{+1.8}_{-1.1}$ & $-27.6^{+3.5}_{-1.6}$ & $-8.5^{+1.1}_{-1.0}$ & WTN & - \\
0151$-$0635 & WTN & -2.4 & $5.2^{+2.5}_{-2.2}$ & N & $-28.1^{+1.7}_{-1.3}$ & $-27.5^{+3.3}_{-1.7}$ & $-6.2^{+1.1}_{-1.1}$ & WTN & - \\
0152$-$1637 & WTN & -2.0 & $5.2^{+2.8}_{-2.2}$ & N & $-27.9^{+1.6}_{-1.4}$ & $-27.2^{+3.6}_{-1.9}$ & $-6.5^{+1.0}_{-1.2}$ & WTN & - \\
0206$-$4028 & WTN & -1.0 & $5.0^{+2.9}_{-2.0}$ & N & $-27.1^{+1.4}_{-1.7}$ & $-26.7^{+3.6}_{-2.0}$ & $-5.9^{+0.6}_{-1.1}$ & WTN & - \\
0255$-$5304 & WTN & -1.1 & $5.0^{+2.5}_{-2.0}$ & N & $-27.1^{+1.9}_{-1.7}$ & $-27.1^{+3.1}_{-1.9}$ & $-6.3^{+0.7}_{-1.2}$ & WTN & - \\
0401$-$7608 & 2C & 7.2 & $5.0^{+2.7}_{-2.0}$ & N & $-25.0^{+0.8}_{-2.3}$ & $-24.9^{+3.5}_{-2.9}$ & $-4.0^{+0.4}_{-0.4}$ & PLRN & 10.6 \\
0418$-$4154 & WTN & 0.3 & $5.7^{+2.3}_{-2.4}$ & N & $-26.3^{+1.3}_{-2.2}$ & $-26.1^{+3.6}_{-2.4}$ & $-4.9^{+0.7}_{-1.0}$ & WTN & - \\
0450$-$1248 & WTN & 3.5 & $5.1^{+2.4}_{-2.0}$ & N & $-24.5^{+2.0}_{-2.2}$ & $-24.9^{+3.4}_{-3.3}$ & $-4.0^{+0.6}_{-0.8}$ & WTN & - \\
0452$-$1759 & 2C & 26.0 & $5.2^{+2.4}_{-2.2}$ & N & $-25.2^{+0.6}_{-2.4}$ & $-25.0^{+3.5}_{-2.6}$ & $-4.0^{+0.3}_{-0.3}$ & PLRN & 22.0 \\
0525+1115$^{\rm g}$ & WTN & -0.9 & $5.5^{+2.4}_{-2.3}$ & N & $-26.7^{+1.7}_{-2.1}$ & $-26.3^{+3.4}_{-2.5}$ & $-6.0^{+1.0}_{-1.7}$ & WTN & - \\
0529$-$6652 & WTN & -1.3 & $5.5^{+2.4}_{-2.3}$ & N & $-26.7^{+2.6}_{-2.2}$ & $-26.2^{+3.6}_{-2.5}$ & $-5.5^{+1.7}_{-2.0}$ & WTN & - \\
0533+0402 & WTN & -1.8 & $5.1^{+2.9}_{-2.1}$ & N & $-27.7^{+2.0}_{-1.6}$ & $-26.9^{+3.6}_{-2.1}$ & $-6.2^{+1.3}_{-1.4}$ & WTN & - \\
0536$-$7543 & WTN & -2.2 & $5.0^{+2.9}_{-2.0}$ & N & $-28.0^{+2.0}_{-1.3}$ & $-27.4^{+3.5}_{-1.8}$ & $-6.2^{+1.2}_{-1.2}$ & WTN & - \\
0601$-$0527$^{\rm g}$ & WTN & -2.5 & $4.9^{+3.0}_{-1.9}$ & N & $-28.2^{+1.7}_{-1.2}$ & $-27.4^{+3.3}_{-1.8}$ & $-7.4^{+1.1}_{-1.1}$ & WTN & - \\
0624$-$0424 & WTN & -2.4 & $5.2^{+2.9}_{-2.1}$ & N & $-28.4^{+1.5}_{-1.1}$ & $-27.4^{+3.4}_{-1.8}$ & $-6.9^{+1.1}_{-1.1}$ & WTN & - \\
0627+0706 & 2C & 73.0 & $5.0^{+2.8}_{-2.0}$ & N & $-24.6^{+0.7}_{-2.2}$ & $-24.5^{+3.4}_{-3.1}$ & $-3.8^{+0.3}_{-0.3}$ & PLRN & 60.2 \\
0630$-$2834 & WTN & -1.9 & $5.1^{+2.8}_{-2.1}$ & N & $-27.4^{+2.0}_{-1.7}$ & $-26.8^{+3.5}_{-2.2}$ & $-5.4^{+1.3}_{-1.6}$ & WTN & - \\
0646+0905 & WTN & 4.0 & $4.9^{+2.4}_{-1.9}$ & N & $-24.6^{+2.2}_{-2.5}$ & $-24.8^{+3.4}_{-3.2}$ & $-3.4^{+0.5}_{-0.7}$ & WTN & - \\
0659+1414$^{\rm g}$ & 2C & 25.2 & $5.9^{+2.1}_{-2.7}$ & N & $-23.9^{+0.6}_{-3.1}$ & $-23.6^{+3.6}_{-3.8}$ & $-3.1^{+0.5}_{-0.4}$ & PLRN & 26.8 \\
0729$-$1836$^{\rm g}$ & 2C & 196.4 & $6.9^{+1.2}_{-0.3}$ & Y & $-25.5^{+1.7}_{-3.1}$ & $-23.3^{+3.7}_{-3.1}$ & $-2.6^{+1.8}_{-1.9}$ & PLRN & 191.6 \\
0738$-$4042 & 2C & 832.1 & $8.2^{+0.5}_{-0.6}$ & RR & $-27.2^{+1.9}_{-1.9}$ & $-19.9^{+3.4}_{-1.7}$ & $-0.9^{+0.3}_{-0.5}$ & PLRN+F2 & 5.4 \\
0742$-$2822$^{\rm g}$ & 2C & 519.3 & $6.5^{+1.3}_{-2.3}$ & Y & $-23.1^{+1.5}_{-4.7}$ & $-20.9^{+3.4}_{-1.7}$ & $-1.5^{+0.8}_{-0.2}$ & PLRN & 512.3 \\
0758$-$1528$^{\rm g}$$^*$ & 2C & 5.4 & $5.0^{+3.0}_{-2.0}$ & N & $-25.9^{+1.0}_{-1.7}$ & $-25.6^{+3.6}_{-2.4}$ & $-4.8^{+0.5}_{-0.4}$ & PLRN & 3.2 \\
0809$-$4753$^*$ & 2C & 11.2 & $5.2^{+2.6}_{-2.1}$ & N & $-25.4^{+1.0}_{-1.9}$ & $-25.3^{+3.3}_{-2.7}$ & $-4.5^{+0.4}_{-0.5}$ & PLRN & 3.5 \\
0820$-$4114 & WTN & -1.5 & $4.9^{+2.9}_{-2.0}$ & N & $-27.0^{+2.4}_{-2.0}$ & $-26.2^{+3.7}_{-2.5}$ & $-5.7^{+1.5}_{-1.9}$ & WTN & - \\
0837+0610 & WTN & -0.9 & $5.0^{+3.0}_{-2.0}$ & N & $-27.8^{+1.2}_{-1.2}$ & $-27.2^{+3.8}_{-1.6}$ & $-5.9^{+0.6}_{-0.7}$ & WTN & - \\
0837$-$4135 & 2C & 149.4 & $5.3^{+2.6}_{-2.2}$ & N & $-26.3^{+0.4}_{-1.4}$ & $-26.0^{+3.4}_{-2.0}$ & $-4.9^{+0.2}_{-0.2}$ & PLRN & 138.6 \\
0840$-$5332 & WTN & -1.4 & $5.4^{+2.6}_{-2.3}$ & N & $-27.1^{+1.6}_{-1.9}$ & $-26.3^{+3.6}_{-2.4}$ & $-5.5^{+1.0}_{-1.7}$ & WTN & - \\
0842$-$4851 & WTN & 3.5 & $5.3^{+2.1}_{-2.1}$ & N & $-23.4^{+1.3}_{-2.8}$ & $-23.7^{+3.2}_{-4.2}$ & $-2.5^{+0.3}_{-0.7}$ & WTN & - \\
0846$-$3533$^{\rm g}$ & WTN & -1.8 & $5.1^{+2.8}_{-2.1}$ & N & $-27.6^{+1.8}_{-1.6}$ & $-26.9^{+3.3}_{-2.1}$ & $-5.9^{+1.2}_{-1.5}$ & WTN & - \\
0855$-$3331 & WTN & -2.3 & $5.1^{+2.8}_{-2.1}$ & N & $-28.1^{+1.7}_{-1.3}$ & $-27.3^{+3.4}_{-1.8}$ & $-6.4^{+1.2}_{-1.2}$ & WTN & - \\
0856$-$6137 & WTN & -1.7 & $5.2^{+2.7}_{-2.2}$ & N & $-27.4^{+1.8}_{-1.8}$ & $-26.8^{+3.4}_{-2.2}$ & $-5.7^{+1.2}_{-1.6}$ & WTN & - \\
0904$-$4246 & WTN & -1.9 & $5.1^{+2.8}_{-2.1}$ & N & $-27.6^{+2.0}_{-1.6}$ & $-26.9^{+3.4}_{-2.1}$ & $-6.0^{+1.3}_{-1.5}$ & WTN & - \\
0904$-$7459 & WTN & 0.2 & $5.3^{+2.6}_{-2.2}$ & N & $-25.1^{+1.6}_{-3.1}$ & $-24.7^{+3.6}_{-3.3}$ & $-4.0^{+0.7}_{-1.2}$ & WTN & - \\
0907$-$5157 & 2C & 93.5 & $5.0^{+2.7}_{-2.1}$ & N & $-24.3^{+0.5}_{-2.8}$ & $-24.1^{+3.5}_{-3.5}$ & $-3.8^{+0.3}_{-0.3}$ & PLRN & 92.7 \\
0908$-$1739 & WTN & -1.3 & $5.1^{+3.1}_{-2.1}$ & N & $-26.7^{+2.0}_{-2.2}$ & $-26.0^{+3.6}_{-2.7}$ & $-5.7^{+1.1}_{-1.9}$ & WTN & - \\
0908$-$4913$^{\rm g}$ & 2C & 564.7 & $8.1^{+0.6}_{-0.8}$ & RR & $-22.9^{+0.2}_{-0.6}$ & $-18.8^{+2.4}_{-1.5}$ & $-0.8^{+0.4}_{-0.7}$ & PLRN & 523.7 \\
0909$-$7212 & WTN & 1.9 & $5.2^{+2.2}_{-2.1}$ & N & $-24.7^{+2.1}_{-2.4}$ & $-25.2^{+3.2}_{-3.1}$ & $-3.2^{+0.6}_{-1.0}$ & WTN & - \\
0922+0638$^{\rm g}$ & 2C & 109.1 & $6.9^{+0.8}_{-0.8}$ & Y & $-26.1^{+2.9}_{-2.6}$ & $-21.2^{+3.9}_{-1.5}$ & $-1.7^{+1.1}_{-0.5}$ & PLRN & 101.1 \\
0924$-$5814 & WTN & -0.3 & $5.0^{+2.8}_{-2.0}$ & N & $-25.6^{+1.5}_{-2.5}$ & $-25.3^{+3.4}_{-3.0}$ & $-4.2^{+0.7}_{-1.2}$ & WTN & - \\
0934$-$5249 & WTN & 5.0 & $4.9^{+2.8}_{-1.9}$ & N & $-25.9^{+1.4}_{-1.8}$ & $-25.8^{+3.3}_{-2.4}$ & $-4.2^{+0.5}_{-0.5}$ & WTN & - \\
0942$-$5657 & 2C & 37.5 & $5.7^{+2.6}_{-2.6}$ & N & $-25.7^{+0.7}_{-2.6}$ & $-24.6^{+4.1}_{-2.6}$ & $-4.1^{+0.8}_{-0.4}$ & PLRN & 26.5 \\
0944$-$1354 & WTN & -2.4 & $5.1^{+2.8}_{-2.1}$ & N & $-28.1^{+1.7}_{-1.3}$ & $-27.3^{+3.5}_{-1.8}$ & $-7.2^{+1.1}_{-1.1}$ & WTN & - \\
0953+0755 & WTN & -2.2 & $5.0^{+2.9}_{-2.0}$ & N & $-27.8^{+1.8}_{-1.5}$ & $-27.1^{+3.5}_{-2.0}$ & $-7.3^{+1.2}_{-1.3}$ & WTN & - \\
0955$-$5304 & WTN & -2.2 & $5.1^{+2.8}_{-2.1}$ & N & $-28.1^{+1.5}_{-1.3}$ & $-27.2^{+3.4}_{-1.9}$ & $-6.6^{+1.1}_{-1.2}$ & WTN & - \\
0959$-$4809 & WTN & -1.9 & $5.1^{+2.8}_{-2.1}$ & N & $-27.6^{+1.9}_{-1.6}$ & $-27.0^{+3.5}_{-2.0}$ & $-6.3^{+1.3}_{-1.4}$ & WTN & - \\
1001$-$5507 & 2C & 534.7 & $7.1^{+0.7}_{-0.3}$ & Y & $-27.4^{+1.7}_{-1.7}$ & $-21.1^{+4.0}_{-1.6}$ & $-0.6^{+1.1}_{-0.5}$ & PLRN & 492.9 \\
1003$-$4747 & WTN & -2.1 & $5.2^{+2.9}_{-2.2}$ & N & $-27.7^{+1.9}_{-1.6}$ & $-27.0^{+3.3}_{-2.0}$ & $-7.2^{+1.2}_{-1.4}$ & WTN & - \\
1012$-$5857 & WTN & -0.5 & $5.5^{+2.7}_{-2.4}$ & N & $-26.7^{+1.6}_{-2.1}$ & $-26.2^{+3.4}_{-2.3}$ & $-5.0^{+0.8}_{-1.2}$ & WTN & - \\
1013$-$5934 & WTN & -2.1 & $5.1^{+2.9}_{-2.1}$ & N & $-27.7^{+1.7}_{-1.5}$ & $-27.0^{+3.4}_{-2.0}$ & $-6.8^{+1.1}_{-1.4}$ & WTN & - \\
1016$-$5345 & WTN & -2.3 & $4.9^{+2.9}_{-2.0}$ & N & $-27.9^{+1.9}_{-1.4}$ & $-27.2^{+3.6}_{-1.9}$ & $-6.7^{+1.3}_{-1.3}$ & WTN & - \\
1017$-$5621 & WTN & 4.4 & $5.1^{+2.8}_{-2.2}$ & N & $-26.5^{+0.9}_{-1.5}$ & $-26.2^{+3.7}_{-2.1}$ & $-5.6^{+0.4}_{-0.4}$ & WTN & - \\
1032$-$5911 & WTN & -1.4 & $5.1^{+2.9}_{-2.1}$ & N & $-26.8^{+2.4}_{-2.1}$ & $-26.2^{+3.9}_{-2.5}$ & $-6.0^{+1.6}_{-2.0}$ & WTN & - \\
1034$-$3224 & WTN & -1.9 & $5.0^{+2.9}_{-2.1}$ & N & $-27.5^{+2.0}_{-1.7}$ & $-26.8^{+3.4}_{-2.2}$ & $-5.6^{+1.3}_{-1.5}$ & WTN & - \\
1036$-$4926 & WTN & -0.5 & $5.4^{+2.7}_{-2.3}$ & N & $-27.3^{+2.1}_{-1.8}$ & $-26.5^{+3.6}_{-2.4}$ & $-6.5^{+1.5}_{-1.7}$ & WTN & - \\
1041$-$1942 & WTN & -0.8 & $5.3^{+2.7}_{-2.2}$ & N & $-27.6^{+2.0}_{-1.6}$ & $-27.0^{+3.4}_{-2.0}$ & $-5.7^{+1.3}_{-1.4}$ & WTN & - \\
1042$-$5521 & WTN & -2.1 & $5.0^{+2.9}_{-2.0}$ & N & $-27.9^{+1.8}_{-1.4}$ & $-27.3^{+3.4}_{-1.9}$ & $-6.2^{+1.2}_{-1.3}$ & WTN & - \\
1043$-$6116$^*$ & 2C & 5.4 & $5.1^{+2.7}_{-2.2}$ & N & $-23.5^{+0.8}_{-3.5}$ & $-23.2^{+3.8}_{-3.8}$ & $-4.0^{+0.5}_{-0.4}$ & WTN & - \\
1046$-$5813$^\dagger$ & WTN & 3.4 & $4.8^{+2.7}_{-1.8}$ & N & $-25.2^{+1.2}_{-2.1}$ & $-25.1^{+3.6}_{-2.9}$ & $-4.6^{+0.4}_{-0.5}$ & PLRN & 7.0 \\
1047$-$6709 & WTN & -1.1 & $5.3^{+2.7}_{-2.2}$ & N & $-27.4^{+2.0}_{-1.8}$ & $-26.6^{+3.6}_{-2.3}$ & $-7.5^{+1.4}_{-1.6}$ & WTN & - \\
1048$-$5832$^{\rm g}$ & 2C & 1295.7 & $8.0^{+0.6}_{-0.5}$ & RR & $-25.9^{+2.8}_{-2.7}$ & $-17.5^{+1.4}_{-1.2}$ & $1.2^{+0.2}_{-0.6}$ & PLRN & 1258.2 \\
1056$-$6258 & 2C & 321.3 & $5.3^{+2.4}_{-2.2}$ & N & $-23.7^{+0.3}_{-1.5}$ & $-23.8^{+3.0}_{-4.0}$ & $-2.8^{+0.2}_{-0.2}$ & PLRN & 297.3 \\
1057$-$5226 & 2C & 280.5 & $6.7^{+1.2}_{-3.2}$ & Y & $-24.2^{+1.1}_{-3.8}$ & $-22.1^{+4.0}_{-3.5}$ & $-2.8^{+1.0}_{-0.2}$ & PLRN & 267.5 \\
1057$-$7914 & WTN & 0.9 & $5.4^{+2.5}_{-2.3}$ & N & $-25.0^{+1.2}_{-3.0}$ & $-24.6^{+3.7}_{-3.2}$ & $-3.2^{+0.6}_{-0.9}$ & WTN & - \\
1059$-$5742 & WTN & 0.9 & $5.1^{+2.7}_{-2.0}$ & N & $-25.7^{+2.2}_{-1.9}$ & $-25.9^{+3.6}_{-2.5}$ & $-4.3^{+0.7}_{-1.0}$ & WTN & - \\
1105$-$6107$^{\rm g}$ & 2C & 417.7 & $5.9^{+2.0}_{-2.7}$ & N & $-20.9^{+0.2}_{-4.3}$ & $-20.7^{+2.7}_{-5.6}$ & $-1.8^{+0.2}_{-0.2}$ & PLRN & 347.5 \\
1110$-$5637 & 2C & 57.4 & $5.6^{+2.6}_{-2.5}$ & N & $-25.0^{+0.5}_{-3.0}$ & $-23.8^{+4.2}_{-3.1}$ & $-3.8^{+0.8}_{-0.3}$ & PLRN & 49.0 \\
1112$-$6613 & 2C & 30.4 & $5.8^{+2.0}_{-2.6}$ & N & $-23.8^{+0.6}_{-3.3}$ & $-23.6^{+3.2}_{-3.6}$ & $-3.3^{+0.3}_{-0.3}$ & PLRN & 35.2 \\
1112$-$6926 & WTN & 2.6 & $6.5^{+1.5}_{-3.1}$ & RR & $-26.1^{+1.1}_{-2.6}$ & $-25.5^{+3.6}_{-2.8}$ & $-4.3^{+0.6}_{-0.9}$ & WTN & - \\
1114$-$6100$^*$ & 2C & 8.5 & $5.2^{+2.6}_{-2.1}$ & N & $-23.7^{+1.8}_{-3.1}$ & $-23.7^{+3.6}_{-3.9}$ & $-2.7^{+0.6}_{-0.6}$ & WTN & - \\
1116$-$4122$^*$ & 2C & 10.9 & $6.5^{+1.5}_{-3.1}$ & RR & $-26.4^{+0.8}_{-2.2}$ & $-25.9^{+3.5}_{-2.4}$ & $-4.5^{+0.6}_{-0.6}$ & PLRN & 4.8 \\
1121$-$5444 & 2C & 184.3 & $6.8^{+1.0}_{-2.7}$ & Y & $-25.6^{+2.2}_{-3.0}$ & $-22.2^{+3.9}_{-1.7}$ & $-2.4^{+1.0}_{-0.3}$ & PLRN & 111.1 \\
1123$-$6259$^{\rm g}$ & WTN & -0.6 & $5.5^{+2.3}_{-2.4}$ & N & $-25.5^{+2.0}_{-3.0}$ & $-24.9^{+3.5}_{-3.3}$ & $-4.7^{+1.1}_{-2.3}$ & WTN & - \\
1126$-$6942 & WTN & 0.6 & $5.6^{+2.4}_{-2.4}$ & N & $-25.4^{+1.8}_{-3.1}$ & $-24.8^{+3.6}_{-3.4}$ & $-4.3^{+1.0}_{-2.2}$ & WTN & - \\
1133$-$6250 & WTN & -0.9 & $5.1^{+2.9}_{-2.1}$ & N & $-25.5^{+3.0}_{-3.0}$ & $-24.8^{+4.1}_{-3.4}$ & $-4.6^{+1.9}_{-2.7}$ & WTN & - \\
1136+1551 & 2C & 17.7 & $5.2^{+2.5}_{-2.1}$ & N & $-25.3^{+1.3}_{-1.5}$ & $-25.4^{+3.2}_{-2.8}$ & $-3.6^{+0.3}_{-0.3}$ & PLRN & 6.5 \\
1136$-$5525 & 2C & 223.0 & $7.0^{+1.2}_{-2.1}$ & Y & $-25.5^{+2.3}_{-3.0}$ & $-22.2^{+3.6}_{-1.1}$ & $-2.2^{+1.5}_{-0.4}$ & PLRN & 174.3 \\
1141$-$3322$^{\rm g}$ & WTN & -1.5 & $5.3^{+2.7}_{-2.2}$ & N & $-27.6^{+2.0}_{-1.6}$ & $-26.9^{+3.3}_{-2.1}$ & $-7.1^{+1.3}_{-1.4}$ & WTN & - \\
1141$-$6545$^{\rm g}$ & 2C & 196.3 & $6.2^{+1.1}_{-0.6}$ & Y & $-25.2^{+0.7}_{-2.9}$ & $-24.8^{+0.6}_{-2.6}$ & $-4.1^{+0.3}_{-0.8}$ & PLRN & 186.8 \\
1146$-$6030 & WTN & 1.5 & $5.3^{+2.7}_{-2.2}$ & N & $-26.3^{+1.0}_{-2.1}$ & $-25.8^{+3.5}_{-2.3}$ & $-5.7^{+0.6}_{-0.6}$ & WTN & - \\
1157$-$6224 & 2C & 218.8 & $5.0^{+2.8}_{-2.0}$ & N & $-24.3^{+0.4}_{-2.0}$ & $-24.3^{+3.3}_{-3.5}$ & $-3.5^{+0.2}_{-0.2}$ & PLRN & 97.5 \\
1202$-$5820 & 2C & 134.9 & $5.2^{+2.5}_{-2.2}$ & N & $-24.6^{+0.4}_{-2.9}$ & $-24.3^{+3.8}_{-3.1}$ & $-3.7^{+0.3}_{-0.2}$ & PLRN & 69.2 \\
1210$-$5559$^*$ & 2C & 26.0 & $5.3^{+2.3}_{-2.2}$ & N & $-26.3^{+0.9}_{-1.4}$ & $-26.3^{+3.3}_{-2.3}$ & $-5.9^{+0.3}_{-0.4}$ & PLRN & 4.5 \\
1224$-$6407 & 2C & 424.4 & $6.3^{+1.6}_{-3.0}$ & RR & $-25.2^{+0.4}_{-3.1}$ & $-23.7^{+4.1}_{-2.7}$ & $-4.5^{+0.8}_{-0.2}$ & PLRN & 372.6 \\
1231$-$6303 & WTN & -0.9 & $5.6^{+2.4}_{-2.5}$ & N & $-25.6^{+2.2}_{-2.9}$ & $-25.1^{+3.5}_{-3.2}$ & $-3.1^{+1.3}_{-2.6}$ & WTN & - \\
1239$-$6832$^*$ & 2C & 5.7 & $5.2^{+2.4}_{-2.1}$ & N & $-24.2^{+1.0}_{-2.4}$ & $-24.4^{+3.2}_{-3.6}$ & $-2.6^{+0.3}_{-0.5}$ & WTN & - \\
1243$-$6423 & 2C & 997.2 & $6.1^{+1.6}_{-2.6}$ & Y & $-24.8^{+0.3}_{-3.2}$ & $-24.0^{+4.1}_{-2.6}$ & $-3.8^{+0.3}_{-0.1}$ & PLRN & 950.4 \\
1253$-$5820 & 2C & 133.8 & $5.5^{+2.2}_{-2.4}$ & N & $-24.7^{+0.4}_{-2.7}$ & $-24.5^{+3.4}_{-3.1}$ & $-4.3^{+0.3}_{-0.3}$ & PLRN & 67.0 \\
1257$-$1027$^{\rm g}$ & WTN & -1.7 & $5.3^{+2.6}_{-2.2}$ & N & $-27.7^{+1.8}_{-1.5}$ & $-27.0^{+3.5}_{-2.0}$ & $-6.6^{+1.3}_{-1.4}$ & WTN & - \\
1259$-$6741 & WTN & -1.9 & $5.3^{+2.7}_{-2.3}$ & N & $-27.5^{+2.0}_{-1.7}$ & $-26.7^{+3.5}_{-2.2}$ & $-6.5^{+1.4}_{-1.5}$ & WTN & - \\
1305$-$6455 & 2C & 215.7 & $6.0^{+1.8}_{-2.7}$ & N & $-23.9^{+0.5}_{-3.9}$ & $-23.4^{+3.7}_{-2.9}$ & $-2.6^{+0.4}_{-0.2}$ & PLRN & 197.0 \\
1306$-$6617 & 2C & 11.9 & $5.7^{+2.3}_{-2.5}$ & N & $-24.6^{+1.0}_{-2.8}$ & $-24.3^{+3.6}_{-3.3}$ & $-3.5^{+0.6}_{-0.5}$ & PLRN & 7.8 \\
1312$-$5402 & WTN & 0.3 & $5.5^{+2.4}_{-2.5}$ & N & $-25.8^{+1.5}_{-2.7}$ & $-25.3^{+3.8}_{-3.0}$ & $-4.2^{+0.8}_{-1.4}$ & WTN & - \\
1312$-$5516 & WTN & -1.5 & $5.3^{+2.6}_{-2.3}$ & N & $-26.9^{+2.0}_{-2.1}$ & $-26.3^{+3.6}_{-2.5}$ & $-5.3^{+1.3}_{-1.9}$ & WTN & - \\
1319$-$6056 & 2C & 21.5 & $4.9^{+2.7}_{-1.9}$ & N & $-24.4^{+1.5}_{-1.8}$ & $-24.5^{+3.1}_{-3.4}$ & $-4.1^{+0.3}_{-0.4}$ & PLRN & 24.8 \\
1320$-$5359$^{\rm g}$ & 2C & 38.9 & $6.3^{+1.7}_{-3.0}$ & RR & $-25.1^{+0.7}_{-2.9}$ & $-24.9^{+3.6}_{-3.3}$ & $-4.4^{+0.4}_{-0.7}$ & PLRN & 43.1 \\
1326$-$5859 & 2C & 869.3 & $7.1^{+1.1}_{-0.4}$ & Y & $-27.5^{+1.7}_{-1.7}$ & $-23.0^{+3.6}_{-0.9}$ & $-2.5^{+1.4}_{-0.6}$ & PLRN & 718.2 \\
1326$-$6408$^*$ & 2C & 29.4 & $5.0^{+2.8}_{-2.0}$ & N & $-25.6^{+0.8}_{-1.5}$ & $-25.4^{+3.5}_{-2.6}$ & $-4.5^{+0.3}_{-0.3}$ & WTN & - \\
1326$-$6700 & 2C & 120.3 & $5.3^{+2.1}_{-2.3}$ & N & $-23.2^{+0.4}_{-2.4}$ & $-23.4^{+3.0}_{-4.4}$ & $-2.2^{+0.2}_{-0.4}$ & PLRN & 107.9 \\
1327$-$6222 & 2C & 1121.6 & $5.6^{+2.4}_{-2.4}$ & N & $-23.0^{+0.1}_{-3.5}$ & $-22.7^{+3.7}_{-4.0}$ & $-1.9^{+0.2}_{-0.1}$ & PLRN & 946.9 \\
1327$-$6301 & WTN & -1.0 & $5.1^{+3.0}_{-2.1}$ & N & $-26.4^{+1.6}_{-2.4}$ & $-25.7^{+3.6}_{-2.8}$ & $-5.8^{+0.8}_{-1.8}$ & WTN & - \\
1328$-$4357$^{\rm g}$ & 2C & 9.1 & $5.7^{+2.3}_{-2.5}$ & N & $-25.4^{+0.9}_{-2.7}$ & $-25.0^{+3.7}_{-2.9}$ & $-4.3^{+0.6}_{-0.5}$ & PLRN & 11.0 \\
1338$-$6204 & WTN & -1.5 & $4.9^{+3.1}_{-2.0}$ & N & $-26.7^{+2.3}_{-2.1}$ & $-25.9^{+3.8}_{-2.8}$ & $-5.0^{+1.6}_{-2.0}$ & WTN & - \\
1350$-$5115 & WTN & -2.1 & $5.1^{+3.0}_{-2.1}$ & N & $-27.6^{+2.0}_{-1.6}$ & $-26.8^{+3.6}_{-2.2}$ & $-7.4^{+1.3}_{-1.5}$ & WTN & - \\
1355$-$5153 & WTN & 4.9 & $5.0^{+2.6}_{-2.0}$ & N & $-25.8^{+1.2}_{-1.9}$ & $-25.7^{+3.5}_{-2.6}$ & $-4.8^{+0.4}_{-0.6}$ & PLRN & 4.1 \\
1356$-$5521$^*$ & 2C & 6.2 & $5.2^{+2.4}_{-2.1}$ & N & $-23.2^{+1.4}_{-3.3}$ & $-23.4^{+3.3}_{-4.2}$ & $-2.7^{+0.5}_{-0.7}$ & WTN & - \\
1359$-$6038 & 2C & 1637.3 & $7.6^{+0.8}_{-0.7}$ & RR & $-23.7^{+0.3}_{-1.8}$ & $-19.9^{+3.3}_{-2.0}$ & $-2.0^{+0.7}_{-0.7}$ & PLRN & 1556.6 \\
1401$-$6357 & 2C & 789.2 & $7.4^{+0.9}_{-0.4}$ & Y & $-27.3^{+1.9}_{-1.9}$ & $-20.0^{+3.4}_{-1.9}$ & $-0.1^{+1.0}_{-0.7}$ & PLRN & 693.7 \\
1413$-$6307 & 2C & 143.5 & $6.1^{+1.6}_{-2.5}$ & Y & $-24.3^{+1.4}_{-3.8}$ & $-22.8^{+3.6}_{-2.2}$ & $-2.7^{+0.5}_{-0.2}$ & PLRN & 143.4 \\
1418$-$3921 & WTN & -1.6 & $5.2^{+2.7}_{-2.1}$ & N & $-27.5^{+2.0}_{-1.7}$ & $-26.9^{+3.1}_{-2.1}$ & $-5.7^{+1.3}_{-1.5}$ & WTN & - \\
1420$-$5416 & WTN & -1.9 & $5.2^{+2.8}_{-2.1}$ & N & $-27.7^{+2.0}_{-1.6}$ & $-26.9^{+3.4}_{-2.1}$ & $-6.2^{+1.3}_{-1.4}$ & WTN & - \\
1424$-$5822 & WTN & -1.3 & $4.8^{+3.2}_{-1.9}$ & N & $-26.7^{+2.5}_{-2.2}$ & $-25.8^{+3.9}_{-2.8}$ & $-7.0^{+1.7}_{-2.1}$ & WTN & - \\
1428$-$5530 & WTN & -0.5 & $5.3^{+2.7}_{-2.3}$ & N & $-27.0^{+1.2}_{-1.9}$ & $-26.5^{+3.4}_{-2.1}$ & $-5.6^{+0.7}_{-1.1}$ & WTN & - \\
1430$-$6623 & 2C & 49.5 & $5.0^{+2.7}_{-2.0}$ & N & $-26.9^{+0.9}_{-1.0}$ & $-26.7^{+3.3}_{-1.7}$ & $-5.5^{+0.3}_{-0.3}$ & PLRN & 26.4 \\
1435$-$5954 & WTN & -0.6 & $5.3^{+2.9}_{-2.2}$ & N & $-26.1^{+1.6}_{-2.5}$ & $-25.4^{+3.7}_{-2.8}$ & $-4.7^{+0.8}_{-1.5}$ & WTN & - \\
1452$-$6036$^{\rm g}$$^*$ & 2C & 40.4 & $5.7^{+2.2}_{-2.5}$ & N & $-22.2^{+0.7}_{-4.8}$ & $-21.7^{+3.8}_{-4.7}$ & $-3.2^{+0.6}_{-0.3}$ & WTN & - \\
1453$-$6413$^{\rm g}$ & 2C & 175.8 & $5.2^{+2.5}_{-2.2}$ & N & $-25.2^{+0.7}_{-1.3}$ & $-25.3^{+3.0}_{-2.9}$ & $-5.2^{+0.2}_{-0.2}$ & PLRN & 156.9 \\
1456$-$6843 & WTN & -1.9 & $5.2^{+2.8}_{-2.2}$ & N & $-27.7^{+1.8}_{-1.5}$ & $-27.1^{+3.2}_{-1.9}$ & $-7.0^{+1.1}_{-1.3}$ & WTN & - \\
1457$-$5122 & WTN & -1.9 & $5.2^{+2.8}_{-2.1}$ & N & $-28.0^{+1.8}_{-1.4}$ & $-27.3^{+3.4}_{-1.8}$ & $-6.0^{+1.2}_{-1.3}$ & WTN & - \\
1507$-$4352 & 2C & 30.1 & $5.4^{+2.3}_{-2.3}$ & N & $-24.2^{+0.5}_{-1.9}$ & $-24.3^{+3.0}_{-3.7}$ & $-3.9^{+0.3}_{-0.3}$ & PLRN & 16.2 \\
1507$-$6640 & WTN & -1.8 & $5.7^{+2.2}_{-2.5}$ & N & $-27.9^{+1.6}_{-1.4}$ & $-27.3^{+3.3}_{-1.8}$ & $-7.4^{+1.2}_{-1.3}$ & WTN & - \\
1511$-$5414 & WTN & -1.7 & $5.1^{+3.0}_{-2.1}$ & N & $-27.2^{+2.1}_{-1.9}$ & $-26.2^{+3.8}_{-2.6}$ & $-8.0^{+1.5}_{-1.8}$ & WTN & - \\
1512$-$5759 & 2C & 276.5 & $6.4^{+1.3}_{-3.0}$ & Y & $-23.3^{+1.4}_{-4.5}$ & $-21.6^{+3.8}_{-3.7}$ & $-2.4^{+0.7}_{-0.2}$ & PLRN & 254.7 \\
1514$-$4834 & WTN & -2.0 & $5.4^{+2.6}_{-2.3}$ & N & $-27.6^{+1.9}_{-1.6}$ & $-26.9^{+3.4}_{-2.1}$ & $-7.0^{+1.4}_{-1.5}$ & WTN & - \\
1522$-$5829 & 2C & 21.4 & $5.0^{+2.6}_{-1.9}$ & N & $-24.4^{+1.8}_{-2.0}$ & $-24.5^{+3.2}_{-3.3}$ & $-3.8^{+0.3}_{-0.4}$ & PLRN & 28.9 \\
1527$-$3931 & WTN & -1.3 & $5.3^{+2.7}_{-2.3}$ & N & $-27.9^{+1.9}_{-1.4}$ & $-27.3^{+3.3}_{-1.8}$ & $-5.7^{+1.2}_{-1.3}$ & WTN & - \\
1527$-$5552 & 2C & 17.5 & $5.6^{+2.6}_{-2.4}$ & N & $-25.9^{+0.9}_{-2.1}$ & $-25.4^{+3.7}_{-2.4}$ & $-4.3^{+0.6}_{-0.4}$ & PLRN & 11.8 \\
1534$-$5334 & WTN & -3.4 & $4.7^{+3.1}_{-1.9}$ & N & $-28.7^{+1.5}_{-0.9}$ & $-27.9^{+3.6}_{-1.5}$ & $-6.9^{+0.9}_{-0.8}$ & WTN & - \\
1534$-$5405 & 2C & 37.9 & $5.7^{+2.2}_{-2.6}$ & N & $-23.1^{+0.5}_{-4.0}$ & $-22.7^{+3.6}_{-4.1}$ & $-2.9^{+0.4}_{-0.3}$ & PLRN & 37.8 \\
1539$-$5626$^{\rm g}$ & 2C & 16.9 & $6.2^{+1.2}_{-2.9}$ & Y & $-23.7^{+0.4}_{-1.9}$ & $-24.1^{+3.0}_{-3.9}$ & $-3.5^{+0.3}_{-0.9}$ & PLRN & 17.2 \\
1542$-$5034$^\dagger$ & WTN & 4.6 & $5.4^{+2.3}_{-2.3}$ & N & $-23.4^{+0.7}_{-3.0}$ & $-23.4^{+3.4}_{-4.1}$ & $-2.8^{+0.4}_{-0.4}$ & PLRN & 10.0 \\
1543+0929 & WTN & -0.2 & $5.1^{+2.8}_{-2.1}$ & N & $-27.0^{+2.3}_{-2.1}$ & $-26.4^{+3.6}_{-2.4}$ & $-5.7^{+1.5}_{-1.8}$ & WTN & - \\
1544$-$5308 & WTN & -2.2 & $4.9^{+3.0}_{-2.0}$ & N & $-27.9^{+2.0}_{-1.4}$ & $-27.2^{+3.4}_{-1.9}$ & $-7.8^{+1.2}_{-1.2}$ & WTN & - \\
1549$-$4848 & WTN & -0.9 & $5.1^{+2.9}_{-2.1}$ & N & $-25.5^{+2.4}_{-3.0}$ & $-25.1^{+3.8}_{-3.3}$ & $-5.7^{+1.6}_{-2.6}$ & WTN & - \\
1553$-$5456 & WTN & -1.6 & $5.2^{+2.9}_{-2.2}$ & N & $-27.3^{+2.0}_{-1.8}$ & $-26.4^{+3.6}_{-2.4}$ & $-6.8^{+1.5}_{-1.7}$ & WTN & - \\
1555$-$3134 & WTN & -1.7 & $5.8^{+2.1}_{-2.6}$ & N & $-27.6^{+1.8}_{-1.6}$ & $-27.0^{+3.2}_{-2.0}$ & $-6.6^{+1.2}_{-1.5}$ & WTN & - \\
1557$-$4258 & 2C & 33.4 & $5.0^{+2.8}_{-2.0}$ & N & $-25.8^{+1.1}_{-1.6}$ & $-25.7^{+3.2}_{-2.5}$ & $-5.2^{+0.3}_{-0.3}$ & PLRN & 10.7 \\
1559$-$4438$^\dagger$ & WTN & 3.6 & $5.2^{+2.4}_{-2.1}$ & N & $-25.2^{+1.4}_{-1.7}$ & $-25.4^{+3.4}_{-2.9}$ & $-5.0^{+0.5}_{-0.8}$ & PLRN & 5.7 \\
1600$-$5044 & 2C & 287.2 & $7.4^{+1.0}_{-2.5}$ & RR & $-26.2^{+2.3}_{-2.6}$ & $-21.1^{+3.8}_{-2.6}$ & $-2.8^{+1.0}_{-0.8}$ & PLRN & 270.2 \\
1603$-$2531 & WTN & 3.3 & $6.4^{+1.5}_{-3.0}$ & Y & $-25.9^{+1.1}_{-2.6}$ & $-25.3^{+3.5}_{-2.7}$ & $-5.0^{+0.7}_{-0.8}$ & WTN & - \\
1603$-$2712 & WTN & -1.0 & $5.3^{+2.6}_{-2.3}$ & N & $-27.5^{+1.9}_{-1.7}$ & $-26.9^{+3.3}_{-2.1}$ & $-6.3^{+1.3}_{-1.5}$ & WTN & - \\
1604$-$4909 & 2C & 134.9 & $6.6^{+1.2}_{-3.0}$ & Y & $-24.9^{+1.0}_{-3.3}$ & $-23.2^{+3.7}_{-2.9}$ & $-3.4^{+0.9}_{-0.2}$ & PLRN & 133.1 \\
1605$-$5257 & WTN & -2.2 & $4.9^{+3.0}_{-2.0}$ & N & $-27.9^{+1.8}_{-1.4}$ & $-27.0^{+3.6}_{-2.0}$ & $-6.5^{+1.3}_{-1.3}$ & WTN & - \\
1613$-$4714 & WTN & -2.0 & $5.2^{+2.8}_{-2.1}$ & N & $-27.7^{+1.9}_{-1.6}$ & $-27.0^{+3.4}_{-2.0}$ & $-6.9^{+1.3}_{-1.4}$ & WTN & - \\
1623$-$0908 & WTN & -2.1 & $5.2^{+2.8}_{-2.2}$ & N & $-27.8^{+1.8}_{-1.5}$ & $-27.0^{+3.5}_{-2.0}$ & $-6.0^{+1.2}_{-1.3}$ & WTN & - \\
1623$-$4256 & 2C & 17.7 & $6.1^{+2.0}_{-2.8}$ & N & $-24.7^{+0.8}_{-3.1}$ & $-24.2^{+3.7}_{-3.2}$ & $-3.7^{+0.6}_{-0.5}$ & PLRN & 21.2 \\
1626$-$4537 & WTN & -0.9 & $5.3^{+2.8}_{-2.2}$ & N & $-26.5^{+2.5}_{-2.4}$ & $-25.7^{+3.9}_{-2.9}$ & $-6.7^{+1.7}_{-2.2}$ & WTN & - \\
1633$-$4453 & WTN & -0.3 & $5.3^{+2.7}_{-2.2}$ & N & $-26.3^{+2.6}_{-2.4}$ & $-25.6^{+3.8}_{-3.0}$ & $-6.3^{+1.8}_{-2.3}$ & WTN & - \\
1633$-$5015 & WTN & -2.0 & $5.0^{+3.0}_{-2.1}$ & N & $-27.8^{+1.6}_{-1.5}$ & $-27.0^{+3.6}_{-2.0}$ & $-6.9^{+1.0}_{-1.4}$ & WTN & - \\
1639$-$4604 & WTN & 4.5 & $5.3^{+2.6}_{-2.2}$ & N & $-25.2^{+0.9}_{-2.2}$ & $-25.0^{+3.5}_{-2.9}$ & $-4.6^{+0.4}_{-0.4}$ & WTN & - \\
1644$-$4559$^{\rm g}$ & 2C & 2536.9 & $7.7^{+0.7}_{-0.4}$ & Y & $-27.7^{+1.5}_{-1.5}$ & $-20.4^{+3.5}_{-1.9}$ & $-1.0^{+0.6}_{-0.7}$ & PLRN & 2519.7 \\
1646$-$6831 & WTN & -0.2 & $5.4^{+2.5}_{-2.3}$ & N & $-27.7^{+1.7}_{-1.5}$ & $-27.0^{+3.4}_{-2.0}$ & $-5.5^{+1.1}_{-1.4}$ & WTN & - \\
1651$-$4246 & 2C & 118.2 & $5.3^{+2.6}_{-2.3}$ & N & $-24.4^{+0.6}_{-3.1}$ & $-24.1^{+3.5}_{-3.3}$ & $-2.8^{+0.3}_{-0.3}$ & PLRN & 125.1 \\
1651$-$5222 & WTN & 1.3 & $6.8^{+1.1}_{-2.3}$ & Y & $-27.5^{+1.4}_{-1.7}$ & $-27.1^{+2.8}_{-2.0}$ & $-5.8^{+0.9}_{-1.6}$ & WTN & - \\
1651$-$5255 & 2C & 17.6 & $5.6^{+2.1}_{-2.5}$ & N & $-24.4^{+0.9}_{-2.9}$ & $-24.3^{+3.2}_{-3.6}$ & $-2.9^{+0.5}_{-0.6}$ & PLRN & 23.6 \\
1652$-$2404 & WTN & -1.6 & $5.2^{+2.7}_{-2.2}$ & N & $-27.7^{+1.9}_{-1.6}$ & $-26.9^{+3.5}_{-2.1}$ & $-5.7^{+1.3}_{-1.5}$ & WTN & - \\
1700$-$3312 & WTN & -1.5 & $5.2^{+2.8}_{-2.2}$ & N & $-27.3^{+2.0}_{-1.8}$ & $-26.6^{+3.5}_{-2.2}$ & $-5.2^{+1.3}_{-1.7}$ & WTN & - \\
1701$-$3726 & WTN & 1.1 & $5.2^{+2.7}_{-2.2}$ & N & $-25.9^{+1.4}_{-2.1}$ & $-25.6^{+3.4}_{-2.4}$ & $-3.6^{+0.6}_{-0.7}$ & WTN & - \\
1703$-$1846 & WTN & -1.1 & $5.4^{+2.5}_{-2.3}$ & N & $-27.5^{+1.9}_{-1.7}$ & $-26.9^{+3.3}_{-2.1}$ & $-6.2^{+1.4}_{-1.5}$ & WTN & - \\
1703$-$3241 & WTN & -2.5 & $5.0^{+2.9}_{-2.0}$ & N & $-28.3^{+1.7}_{-1.1}$ & $-27.6^{+3.3}_{-1.7}$ & $-6.7^{+1.1}_{-1.0}$ & WTN & - \\
1703$-$4851$^{\rm g}$ & WTN & -1.9 & $5.3^{+2.6}_{-2.2}$ & N & $-27.9^{+1.7}_{-1.4}$ & $-27.0^{+3.3}_{-2.0}$ & $-5.9^{+1.2}_{-1.3}$ & WTN & - \\
1705$-$3423$^{\rm g}$ & 2C & 25.4 & $5.3^{+2.7}_{-2.3}$ & N & $-24.6^{+0.8}_{-2.9}$ & $-24.1^{+4.0}_{-3.3}$ & $-4.0^{+0.4}_{-0.4}$ & PLRN & 27.3 \\
1707$-$4053 & WTN & -0.5 & $5.1^{+2.5}_{-2.1}$ & N & $-25.3^{+2.4}_{-3.1}$ & $-25.2^{+3.4}_{-3.1}$ & $-4.0^{+1.4}_{-2.3}$ & WTN & - \\
1708$-$3426 & WTN & -0.8 & $5.4^{+2.5}_{-2.3}$ & N & $-27.2^{+1.9}_{-1.9}$ & $-26.5^{+3.4}_{-2.3}$ & $-5.7^{+1.3}_{-1.7}$ & WTN & - \\
1709$-$1640 & 2C & 59.6 & $5.2^{+2.7}_{-2.2}$ & N & $-24.4^{+0.4}_{-2.4}$ & $-24.3^{+3.2}_{-3.2}$ & $-3.2^{+0.3}_{-0.2}$ & PLRN & 48.5 \\
1711$-$5350 & 2C & 7.6 & $5.3^{+2.4}_{-2.2}$ & N & $-24.5^{+1.6}_{-2.7}$ & $-24.6^{+3.4}_{-3.2}$ & $-3.2^{+0.5}_{-0.7}$ & PLRN & 8.4 \\
1715$-$4034 & WTN & -1.5 & $5.1^{+2.8}_{-2.1}$ & N & $-27.0^{+2.1}_{-2.0}$ & $-26.4^{+3.2}_{-2.4}$ & $-4.6^{+1.4}_{-1.8}$ & WTN & - \\
1717$-$3425 & 2C & 8.3 & $5.2^{+2.5}_{-2.2}$ & N & $-23.6^{+0.7}_{-3.2}$ & $-23.4^{+3.5}_{-3.9}$ & $-2.8^{+0.4}_{-0.4}$ & PLRN & 10.9 \\
1717$-$4054 & 2C & 43.1 & $5.1^{+2.7}_{-2.1}$ & N & $-23.4^{+0.5}_{-3.1}$ & $-23.3^{+3.4}_{-4.0}$ & $-1.7^{+0.3}_{-0.2}$ & PLRN & 40.7 \\
1720$-$1633$^{\rm g}$ & WTN & -1.2 & $5.5^{+2.4}_{-2.4}$ & N & $-27.4^{+1.8}_{-1.8}$ & $-26.8^{+3.2}_{-2.2}$ & $-5.4^{+1.3}_{-1.6}$ & WTN & - \\
1720$-$2933 & WTN & -1.0 & $5.2^{+2.7}_{-2.1}$ & N & $-27.8^{+1.8}_{-1.5}$ & $-27.1^{+3.5}_{-2.0}$ & $-6.7^{+1.3}_{-1.4}$ & WTN & - \\
1722$-$3207$^\dagger$ & WTN & 2.3 & $5.3^{+2.8}_{-2.3}$ & N & $-26.4^{+1.2}_{-1.8}$ & $-25.9^{+3.5}_{-2.2}$ & $-5.4^{+0.6}_{-0.6}$ & PLRN & 6.6 \\
1722$-$3712 & 2C & 397.7 & $5.3^{+2.4}_{-2.3}$ & N & $-22.0^{+0.3}_{-4.5}$ & $-21.8^{+3.4}_{-4.3}$ & $-1.8^{+0.2}_{-0.1}$ & PLRN & 372.6 \\
1727$-$2739 & WTN & -0.8 & $5.3^{+2.6}_{-2.3}$ & N & $-27.0^{+2.3}_{-2.1}$ & $-26.3^{+3.7}_{-2.5}$ & $-5.4^{+1.6}_{-1.9}$ & WTN & - \\
1733$-$2228 & WTN & -1.7 & $5.2^{+2.8}_{-2.2}$ & N & $-26.9^{+1.9}_{-2.1}$ & $-26.2^{+3.7}_{-2.6}$ & $-5.3^{+1.3}_{-1.9}$ & WTN & - \\
1736$-$2457 & WTN & 0.8 & $5.4^{+2.5}_{-2.3}$ & N & $-25.3^{+2.4}_{-3.2}$ & $-24.5^{+3.9}_{-3.7}$ & $-3.4^{+1.3}_{-2.9}$ & WTN & - \\
1739$-$2903$^{\rm g}$ & 2C & 12.4 & $5.1^{+2.5}_{-2.1}$ & N & $-24.3^{+0.8}_{-2.3}$ & $-24.4^{+3.3}_{-3.4}$ & $-4.0^{+0.4}_{-0.5}$ & PLRN & 19.7 \\
1741$-$3927 & 2C & 182.1 & $8.6^{+0.3}_{-0.4}$ & RR & $-27.7^{+1.6}_{-1.6}$ & $-20.3^{+3.8}_{-1.1}$ & $-1.8^{+0.3}_{-0.2}$ & PLRN & 159.5 \\
1743$-$3150$^{\rm g}$ & WTN & -2.2 & $5.1^{+2.9}_{-2.0}$ & N & $-28.1^{+1.7}_{-1.3}$ & $-27.2^{+3.5}_{-1.9}$ & $-5.8^{+1.1}_{-1.2}$ & WTN & - \\
1745$-$3040 & 2C & 75.1 & $7.3^{+1.1}_{-3.3}$ & RR & $-26.5^{+0.8}_{-2.2}$ & $-25.3^{+4.0}_{-2.3}$ & $-5.1^{+1.0}_{-0.5}$ & PLRN & 68.7 \\
1751$-$4657 & WTN & -1.4 & $5.2^{+2.7}_{-2.1}$ & N & $-27.3^{+1.7}_{-1.8}$ & $-26.8^{+3.1}_{-2.2}$ & $-5.9^{+1.1}_{-1.5}$ & WTN & - \\
1752$-$2806 & 2C & 349.7 & $6.3^{+1.2}_{-2.8}$ & Y & $-23.4^{+0.2}_{-0.2}$ & $-24.7^{+2.5}_{-3.5}$ & $-2.3^{+0.2}_{-0.9}$ & PLRN & 292.9 \\
1757$-$2421$^{\rm g}$ & WTN & -0.8 & $5.3^{+2.7}_{-2.2}$ & N & $-25.7^{+3.1}_{-2.9}$ & $-25.1^{+3.9}_{-3.3}$ & $-6.1^{+2.0}_{-2.6}$ & WTN & - \\
1759$-$2205 & 2C & 51.0 & $6.5^{+1.4}_{-3.0}$ & Y & $-25.1^{+1.2}_{-3.3}$ & $-23.1^{+3.9}_{-2.6}$ & $-3.1^{+1.0}_{-0.3}$ & PLRN & 54.9 \\
1759$-$3107 & WTN & -1.4 & $5.4^{+2.6}_{-2.3}$ & N & $-27.7^{+1.9}_{-1.6}$ & $-26.9^{+3.5}_{-2.1}$ & $-6.4^{+1.4}_{-1.4}$ & WTN & - \\
1801$-$0357$^{\rm g}$ & WTN & -1.0 & $5.4^{+2.6}_{-2.3}$ & N & $-27.7^{+1.9}_{-1.5}$ & $-26.9^{+3.3}_{-2.1}$ & $-6.6^{+1.4}_{-1.5}$ & WTN & - \\
1801$-$2920 & WTN & 2.8 & $6.8^{+0.9}_{-2.2}$ & Y & $-27.4^{+1.4}_{-1.7}$ & $-27.1^{+2.7}_{-2.0}$ & $-5.3^{+0.9}_{-1.6}$ & WTN & - \\
1803$-$2137$^{\rm g}$ & 2C & 73.4 & $8.1^{+0.6}_{-1.1}$ & RR & $-24.9^{+3.2}_{-3.5}$ & $-17.6^{+1.6}_{-2.1}$ & $-1.1^{+0.4}_{-0.8}$ & PLRN & 41.3 \\
1805$-$1504 & WTN & 0.9 & $5.5^{+2.5}_{-2.5}$ & N & $-25.2^{+2.9}_{-3.2}$ & $-24.3^{+4.2}_{-3.8}$ & $-3.8^{+1.8}_{-3.0}$ & WTN & - \\
1807$-$0847 & WTN & -1.6 & $5.1^{+2.8}_{-2.1}$ & N & $-27.6^{+1.4}_{-1.4}$ & $-27.0^{+3.5}_{-1.9}$ & $-7.5^{+0.8}_{-1.1}$ & WTN & - \\
1807$-$2715$^\dagger$ & WTN & 4.5 & $6.8^{+0.7}_{-1.8}$ & Y & $-27.2^{+1.6}_{-1.9}$ & $-26.9^{+3.0}_{-2.1}$ & $-5.3^{+1.1}_{-1.9}$ & PLRN & 5.2 \\
1808$-$0813 & WTN & 2.0 & $5.9^{+2.1}_{-2.6}$ & N & $-25.8^{+1.3}_{-2.6}$ & $-25.3^{+3.6}_{-2.9}$ & $-4.1^{+0.7}_{-0.9}$ & WTN & - \\
1809$-$2109 & WTN & -0.3 & $5.5^{+2.5}_{-2.4}$ & N & $-26.5^{+2.4}_{-2.4}$ & $-25.8^{+3.7}_{-2.8}$ & $-5.3^{+1.8}_{-2.1}$ & WTN & - \\
1810$-$5338 & WTN & -1.7 & $5.2^{+2.8}_{-2.2}$ & N & $-27.6^{+1.9}_{-1.6}$ & $-26.8^{+3.5}_{-2.2}$ & $-7.2^{+1.3}_{-1.5}$ & WTN & - \\
1816$-$2650 & WTN & -1.9 & $5.3^{+2.7}_{-2.2}$ & N & $-27.3^{+2.0}_{-1.8}$ & $-26.6^{+3.5}_{-2.3}$ & $-6.1^{+1.4}_{-1.7}$ & WTN & - \\
1818$-$1422$^{\rm g}$ & WTN & -1.2 & $5.3^{+2.7}_{-2.2}$ & N & $-26.3^{+2.4}_{-2.5}$ & $-25.5^{+3.6}_{-3.0}$ & $-5.4^{+1.6}_{-2.3}$ & WTN & - \\
1820$-$0427 & 2C & 98.8 & $5.6^{+2.2}_{-2.4}$ & N & $-24.1^{+0.3}_{-2.7}$ & $-23.9^{+3.6}_{-3.6}$ & $-2.8^{+0.2}_{-0.2}$ & PLRN & 99.8 \\
1822$-$2256 & WTN & -2.1 & $5.1^{+2.8}_{-2.1}$ & N & $-27.8^{+1.9}_{-1.5}$ & $-27.2^{+3.3}_{-1.9}$ & $-5.8^{+1.2}_{-1.3}$ & WTN & - \\
1823$-$0154 & WTN & -1.4 & $5.3^{+2.7}_{-2.2}$ & N & $-28.0^{+1.8}_{-1.4}$ & $-27.2^{+3.4}_{-1.9}$ & $-6.7^{+1.2}_{-1.3}$ & WTN & - \\
1823$-$1115 & WTN & -1.4 & $5.1^{+2.8}_{-2.1}$ & N & $-26.5^{+2.4}_{-2.3}$ & $-25.7^{+3.7}_{-2.8}$ & $-5.7^{+1.7}_{-2.1}$ & WTN & - \\
1823$-$3106 & 2C & 33.8 & $5.9^{+2.2}_{-2.7}$ & N & $-24.2^{+0.5}_{-3.5}$ & $-23.7^{+4.1}_{-3.0}$ & $-3.7^{+0.6}_{-0.3}$ & PLRN & 33.1 \\
1824$-$0127 & WTN & -0.6 & $5.4^{+2.6}_{-2.3}$ & N & $-27.1^{+2.0}_{-2.0}$ & $-26.5^{+3.4}_{-2.4}$ & $-5.0^{+1.4}_{-1.7}$ & WTN & - \\
1824$-$1945 & 2C & 327.4 & $7.3^{+0.8}_{-0.4}$ & Y & $-26.9^{+2.0}_{-2.0}$ & $-20.2^{+3.7}_{-1.8}$ & $-1.8^{+1.0}_{-0.6}$ & PLRN & 327.7 \\
1827$-$0750 & 2C & 46.5 & $5.7^{+2.3}_{-2.5}$ & N & $-22.7^{+0.5}_{-4.2}$ & $-22.4^{+3.8}_{-4.4}$ & $-2.6^{+0.4}_{-0.3}$ & PLRN & 30.5 \\
1829$-$1751 & 2C & 199.6 & $7.6^{+0.8}_{-0.8}$ & RR & $-27.0^{+2.0}_{-2.0}$ & $-19.6^{+3.2}_{-2.3}$ & $-0.6^{+0.6}_{-1.3}$ & PLRN & 187.9 \\
1830$-$1135$^{\rm g}$ & WTN & -1.7 & $5.2^{+2.8}_{-2.2}$ & N & $-27.4^{+2.0}_{-1.7}$ & $-26.6^{+3.5}_{-2.3}$ & $-4.2^{+1.4}_{-1.6}$ & WTN & - \\
1832$-$0827 & 2C & 66.9 & $5.5^{+2.4}_{-2.4}$ & N & $-24.8^{+0.5}_{-2.9}$ & $-24.5^{+3.7}_{-2.7}$ & $-3.6^{+0.4}_{-0.3}$ & PLRN & 67.8 \\
1833$-$0338 & 2C & 242.2 & $7.2^{+0.8}_{-0.4}$ & Y & $-27.1^{+1.9}_{-1.9}$ & $-20.3^{+3.6}_{-1.9}$ & $-0.8^{+1.1}_{-0.6}$ & PLRN & 254.7 \\
1833$-$0827$^{\rm g}$ & 2C & 25.9 & $5.2^{+2.5}_{-2.2}$ & N & $-22.7^{+0.6}_{-3.7}$ & $-22.7^{+3.2}_{-4.5}$ & $-3.5^{+0.3}_{-0.3}$ & PLRN & 28.4 \\
1834$-$0426 & WTN & 1.2 & $5.8^{+2.1}_{-2.6}$ & N & $-25.7^{+1.3}_{-2.6}$ & $-25.3^{+3.3}_{-2.9}$ & $-5.0^{+0.8}_{-1.0}$ & WTN & - \\
1836$-$0436 & WTN & -0.3 & $5.4^{+2.6}_{-2.3}$ & N & $-26.5^{+2.6}_{-2.4}$ & $-25.7^{+3.9}_{-2.9}$ & $-5.8^{+1.8}_{-2.2}$ & WTN & - \\
1836$-$1008$^{\rm g}$ & 2C & 91.8 & $5.6^{+2.4}_{-2.5}$ & N & $-24.2^{+0.4}_{-3.2}$ & $-23.8^{+3.9}_{-3.4}$ & $-3.0^{+0.3}_{-0.2}$ & PLRN & 85.4 \\
1837$-$0653 & WTN & -1.4 & $5.2^{+2.8}_{-2.1}$ & N & $-26.7^{+2.4}_{-2.2}$ & $-26.0^{+3.6}_{-2.7}$ & $-4.8^{+1.7}_{-2.0}$ & WTN & - \\
1840$-$0809 & WTN & -1.9 & $5.2^{+2.8}_{-2.1}$ & N & $-27.7^{+1.9}_{-1.5}$ & $-27.0^{+3.4}_{-2.0}$ & $-6.1^{+1.3}_{-1.4}$ & WTN & - \\
1840$-$0815 & WTN & -1.7 & $5.2^{+2.8}_{-2.1}$ & N & $-27.6^{+1.9}_{-1.6}$ & $-26.8^{+3.5}_{-2.2}$ & $-6.2^{+1.4}_{-1.5}$ & WTN & - \\
1841+0912 & 2C & 21.0 & $5.3^{+2.5}_{-2.2}$ & N & $-23.6^{+0.7}_{-3.1}$ & $-23.6^{+3.2}_{-4.1}$ & $-2.9^{+0.3}_{-0.3}$ & PLRN & 19.3 \\
1841$-$0425$^{\rm g}$ & WTN & 1.4 & $5.6^{+2.4}_{-2.4}$ & N & $-24.5^{+1.3}_{-3.4}$ & $-24.1^{+3.8}_{-3.8}$ & $-4.5^{+0.7}_{-1.2}$ & WTN & - \\
1842$-$0359 & WTN & -1.6 & $5.2^{+2.8}_{-2.2}$ & N & $-27.2^{+2.1}_{-1.9}$ & $-26.6^{+3.5}_{-2.3}$ & $-5.0^{+1.4}_{-1.7}$ & WTN & - \\
1843$-$0000 & 2C & 19.2 & $5.4^{+2.5}_{-2.3}$ & N & $-24.5^{+0.8}_{-3.0}$ & $-24.1^{+3.7}_{-3.0}$ & $-3.0^{+0.5}_{-0.3}$ & PLRN & 6.3 \\
1844$-$0433$^{\rm g}$ & WTN & -2.0 & $5.2^{+2.8}_{-2.2}$ & N & $-27.9^{+1.7}_{-1.4}$ & $-27.2^{+3.3}_{-1.9}$ & $-6.6^{+1.2}_{-1.3}$ & WTN & - \\
1848$-$0123 & 2C & 45.9 & $5.1^{+2.6}_{-2.0}$ & N & $-25.2^{+1.6}_{-2.0}$ & $-25.2^{+3.3}_{-2.9}$ & $-4.0^{+0.4}_{-0.4}$ & PLRN & 46.7 \\
1849$-$0636 & 2C & 12.3 & $7.0^{+0.7}_{-2.3}$ & Y & $-27.4^{+1.5}_{-1.7}$ & $-27.1^{+2.8}_{-2.0}$ & $-4.9^{+0.9}_{-1.9}$ & PLRN & 13.7 \\
1852$-$0635$^{\rm g}$ & WTN & -0.1 & $6.0^{+2.1}_{-2.7}$ & N & $-26.1^{+1.3}_{-2.4}$ & $-25.6^{+3.7}_{-2.8}$ & $-4.8^{+0.8}_{-1.5}$ & WTN & - \\
1852$-$2610 & WTN & -0.3 & $5.3^{+2.6}_{-2.2}$ & N & $-27.2^{+2.1}_{-1.9}$ & $-26.5^{+3.4}_{-2.4}$ & $-6.8^{+1.5}_{-1.8}$ & WTN & - \\
1857+0212 & WTN & -1.4 & $5.3^{+2.8}_{-2.2}$ & N & $-27.0^{+2.1}_{-2.0}$ & $-26.1^{+3.6}_{-2.7}$ & $-6.1^{+1.5}_{-1.9}$ & WTN & - \\
1900$-$2600 & WTN & 2.2 & $5.6^{+2.4}_{-2.5}$ & N & $-26.6^{+1.0}_{-1.9}$ & $-26.0^{+3.9}_{-2.1}$ & $-5.3^{+0.6}_{-0.5}$ & WTN & - \\
1901+0331 & 2C & 283.4 & $5.8^{+2.0}_{-2.6}$ & N & $-23.8^{+0.3}_{-3.6}$ & $-23.5^{+3.6}_{-3.5}$ & $-2.6^{+0.3}_{-0.2}$ & PLRN & 277.3 \\
1901+0716$^{\rm g}$ & WTN & 1.2 & $5.4^{+2.5}_{-2.2}$ & N & $-24.7^{+1.8}_{-3.6}$ & $-24.5^{+3.5}_{-3.7}$ & $-3.4^{+1.0}_{-2.2}$ & PLRN & 4.1 \\
1901$-$0906 & WTN & -2.2 & $5.2^{+2.9}_{-2.2}$ & N & $-28.2^{+1.6}_{-1.2}$ & $-27.5^{+3.3}_{-1.7}$ & $-6.2^{+1.0}_{-1.1}$ & WTN & - \\
1902+0556 & WTN & -0.7 & $5.4^{+2.6}_{-2.3}$ & N & $-27.3^{+2.0}_{-1.8}$ & $-26.6^{+3.6}_{-2.3}$ & $-6.2^{+1.4}_{-1.6}$ & WTN & - \\
1902+0615$^{\rm g}$ & WTN & -1.5 & $5.5^{+2.4}_{-2.3}$ & N & $-27.3^{+2.1}_{-1.9}$ & $-26.6^{+3.4}_{-2.3}$ & $-6.2^{+1.5}_{-1.7}$ & WTN & - \\
1903+0135 & 2C & 65.9 & $5.7^{+2.2}_{-2.6}$ & N & $-25.0^{+0.4}_{-2.9}$ & $-24.6^{+3.9}_{-2.8}$ & $-3.7^{+0.4}_{-0.2}$ & PLRN & 74.4 \\
1903$-$0632 & 2C & 9.7 & $5.3^{+1.9}_{-1.9}$ & Y & $-24.2^{+2.6}_{-1.5}$ & $-25.4^{+2.9}_{-3.0}$ & $-4.1^{+0.5}_{-1.0}$ & PLRN & 10.2 \\
1905$-$0056 & WTN & -0.7 & $5.3^{+2.7}_{-2.3}$ & N & $-27.6^{+2.1}_{-1.6}$ & $-26.8^{+3.6}_{-2.2}$ & $-6.6^{+1.4}_{-1.6}$ & WTN & - \\
1909+0007$^{\rm g}$ & 2C & 65.3 & $5.3^{+2.6}_{-2.2}$ & N & $-24.9^{+0.5}_{-2.3}$ & $-24.8^{+3.3}_{-2.8}$ & $-3.5^{+0.3}_{-0.2}$ & PLRN & 64.3 \\
1909+0254 & WTN & -0.3 & $5.9^{+2.1}_{-2.6}$ & N & $-27.0^{+1.3}_{-1.9}$ & $-26.6^{+3.4}_{-2.2}$ & $-5.3^{+0.8}_{-1.3}$ & WTN & - \\
1909+1102$^{\rm g}$ & 2C & 192.3 & $7.3^{+1.0}_{-2.0}$ & RR & $-26.4^{+2.4}_{-2.4}$ & $-21.5^{+3.7}_{-2.2}$ & $-2.7^{+1.2}_{-0.9}$ & PLRN & 183.9 \\
1910+0358$^{\rm g}$ & WTN & -1.2 & $5.3^{+2.7}_{-2.3}$ & N & $-27.1^{+2.2}_{-2.0}$ & $-26.2^{+3.7}_{-2.5}$ & $-4.9^{+1.5}_{-1.9}$ & WTN & - \\
1910$-$0309$^{\rm g}$ & WTN & 2.1 & $6.6^{+1.4}_{-3.2}$ & RR & $-26.2^{+1.2}_{-2.5}$ & $-25.5^{+3.6}_{-2.7}$ & $-4.9^{+0.7}_{-1.0}$ & WTN & - \\
1913+1400 & WTN & -2.3 & $5.2^{+2.8}_{-2.2}$ & N & $-28.0^{+1.7}_{-1.4}$ & $-27.2^{+3.5}_{-1.9}$ & $-7.3^{+1.2}_{-1.3}$ & WTN & - \\
1913$-$0440 & 2C & 188.7 & $6.8^{+0.9}_{-1.8}$ & Y & $-26.9^{+2.2}_{-2.1}$ & $-22.9^{+4.0}_{-1.7}$ & $-2.9^{+1.1}_{-0.4}$ & PLRN & 175.6 \\
1915+1009$^{\rm g}$ & WTN & -1.7 & $5.0^{+2.9}_{-2.0}$ & N & $-27.2^{+2.0}_{-1.9}$ & $-26.7^{+3.3}_{-2.2}$ & $-6.5^{+1.3}_{-1.6}$ & WTN & - \\
1916+0951 & 2C & 13.3 & $5.3^{+2.5}_{-2.3}$ & N & $-24.8^{+0.9}_{-2.7}$ & $-24.6^{+3.4}_{-3.1}$ & $-4.4^{+0.5}_{-0.5}$ & PLRN & 19.7 \\
1916+1312 & 2C & 109.1 & $7.5^{+0.9}_{-0.6}$ & Y & $-25.9^{+2.4}_{-2.7}$ & $-19.7^{+3.1}_{-1.9}$ & $-1.3^{+0.6}_{-0.8}$ & PLRN & 111.7 \\
1917+1353 & 2C & 80.4 & $5.3^{+2.4}_{-2.2}$ & N & $-23.6^{+0.4}_{-3.6}$ & $-23.4^{+3.4}_{-4.0}$ & $-3.8^{+0.2}_{-0.2}$ & PLRN & 74.0 \\
1919+0021$^{\rm g}$ & WTN & -2.2 & $5.1^{+2.8}_{-2.2}$ & N & $-27.9^{+1.6}_{-1.4}$ & $-27.2^{+3.3}_{-1.9}$ & $-6.2^{+1.1}_{-1.3}$ & WTN & - \\
1926+0431$^{\rm g}$ & WTN & -2.3 & $5.0^{+3.0}_{-2.0}$ & N & $-28.0^{+1.8}_{-1.3}$ & $-27.3^{+3.5}_{-1.9}$ & $-6.5^{+1.1}_{-1.2}$ & WTN & - \\
1932+1059 & 2C & 259.1 & $6.7^{+1.0}_{-2.4}$ & Y & $-25.8^{+2.2}_{-2.9}$ & $-22.4^{+3.9}_{-1.5}$ & $-3.1^{+1.0}_{-0.3}$ & PLRN & 206.5 \\
1932$-$3655 & WTN & -0.7 & $5.3^{+2.7}_{-2.2}$ & N & $-27.2^{+2.2}_{-1.9}$ & $-26.4^{+3.5}_{-2.4}$ & $-6.4^{+1.5}_{-1.8}$ & WTN & - \\
1935+1616 & 2C & 30.6 & $5.3^{+2.6}_{-2.2}$ & N & $-25.9^{+0.5}_{-1.8}$ & $-25.7^{+3.5}_{-2.3}$ & $-5.8^{+0.3}_{-0.3}$ & PLRN & 31.3 \\
1941$-$2602 & WTN & -2.3 & $4.9^{+3.0}_{-2.0}$ & N & $-28.2^{+1.7}_{-1.2}$ & $-27.5^{+3.5}_{-1.8}$ & $-7.6^{+1.1}_{-1.1}$ & WTN & - \\
1943$-$1237 & WTN & -1.9 & $5.2^{+2.8}_{-2.2}$ & N & $-28.0^{+1.8}_{-1.4}$ & $-27.2^{+3.3}_{-1.9}$ & $-6.5^{+1.2}_{-1.2}$ & WTN & - \\
1945$-$0040 & WTN & -1.5 & $5.1^{+3.0}_{-2.1}$ & N & $-26.7^{+2.5}_{-2.2}$ & $-25.6^{+3.9}_{-3.0}$ & $-5.8^{+1.8}_{-2.1}$ & WTN & - \\
1946$-$2913 & WTN & -1.9 & $5.1^{+2.7}_{-2.1}$ & N & $-27.4^{+2.4}_{-1.7}$ & $-27.0^{+3.3}_{-2.0}$ & $-6.0^{+1.3}_{-1.5}$ & WTN & - \\
2006$-$0807 & WTN & -1.5 & $4.7^{+2.9}_{-1.8}$ & N & $-26.9^{+2.0}_{-2.1}$ & $-26.3^{+3.5}_{-2.4}$ & $-5.7^{+1.3}_{-1.8}$ & WTN & - \\
2033+0042 & WTN & -2.1 & $5.1^{+2.9}_{-2.1}$ & N & $-27.7^{+1.8}_{-1.6}$ & $-26.9^{+3.5}_{-2.1}$ & $-4.7^{+1.3}_{-1.5}$ & WTN & - \\
2038$-$3816 & WTN & -1.6 & $5.3^{+2.7}_{-2.2}$ & N & $-27.2^{+2.1}_{-1.9}$ & $-26.4^{+3.6}_{-2.4}$ & $-5.4^{+1.5}_{-1.8}$ & WTN & - \\
2046+1540 & WTN & -2.2 & $5.1^{+2.9}_{-2.1}$ & N & $-27.8^{+1.8}_{-1.5}$ & $-27.2^{+3.4}_{-1.9}$ & $-6.1^{+1.1}_{-1.3}$ & WTN & - \\
2046$-$0421 & WTN & -2.7 & $4.9^{+3.0}_{-1.9}$ & N & $-28.4^{+1.7}_{-1.1}$ & $-27.7^{+3.4}_{-1.6}$ & $-6.5^{+1.0}_{-1.0}$ & WTN & - \\
2048$-$1616 & WTN & -2.2 & $5.0^{+2.7}_{-2.0}$ & N & $-27.8^{+2.1}_{-1.5}$ & $-27.3^{+3.2}_{-1.9}$ & $-5.6^{+1.4}_{-1.3}$ & WTN & - \\
2053$-$7200 & WTN & -2.1 & $5.3^{+2.8}_{-2.2}$ & N & $-27.8^{+1.8}_{-1.4}$ & $-27.1^{+3.5}_{-2.0}$ & $-7.2^{+1.3}_{-1.3}$ & WTN & - \\
2116+1414$^{\rm g}$$^*$ & 2C & 6.7 & $5.3^{+2.6}_{-2.3}$ & N & $-25.6^{+1.0}_{-2.2}$ & $-25.3^{+3.6}_{-2.7}$ & $-4.7^{+0.5}_{-0.5}$ & WTN & - \\
2144$-$3933 & WTN & -2.7 & $5.2^{+2.8}_{-2.1}$ & N & $-28.5^{+1.6}_{-1.0}$ & $-27.6^{+3.4}_{-1.6}$ & $-5.2^{+1.0}_{-0.9}$ & WTN & - \\
2155$-$3118 & WTN & -2.1 & $5.0^{+3.0}_{-2.1}$ & N & $-27.8^{+1.9}_{-1.5}$ & $-27.0^{+3.5}_{-2.0}$ & $-6.3^{+1.3}_{-1.3}$ & WTN & - \\
2248$-$0101 & WTN & -2.1 & $5.1^{+2.7}_{-2.1}$ & N & $-27.7^{+2.0}_{-1.6}$ & $-27.0^{+3.5}_{-2.1}$ & $-7.1^{+1.4}_{-1.4}$ & WTN & - \\
2324$-$6054 & WTN & -1.3 & $5.3^{+2.8}_{-2.2}$ & N & $-27.2^{+1.5}_{-1.8}$ & $-26.7^{+3.5}_{-2.1}$ & $-5.0^{+0.9}_{-1.4}$ & WTN & - \\
2330$-$2005 & WTN & -1.6 & $5.0^{+2.8}_{-2.1}$ & N & $-27.6^{+1.5}_{-1.5}$ & $-27.0^{+3.4}_{-1.9}$ & $-5.6^{+0.9}_{-1.2}$ & WTN & - \\

\end{longtable}
\begin{table*}
    \caption{Same as Table~\ref{tab:model_comparison_canonical}, but for recycled pulsars satisfying \(\dot{P}/10^{-17} \leq 3.23 (P/100 \, \textrm{ms})^{-2.34}\) \citep{LeeEtAl2012}.}
    \begin{tabular}{lcccccccrr}
\hline
 & \multicolumn{7}{c}{} & \multicolumn{2}{c}{\citet{LowerEtAl2020}} \\
 & Model & $\ln\mathfrak{B}_{\rm BF}$ & log$_{10}$($\tau$) & Peaky? & log$_{10}$($Q_{\rm c}$) & log$_{10}$($Q_{\rm s}$) & log$_{10} \sigma_{\rm TN}^2$ & Model & $\ln \mathfrak{B}_{\rm BF}$ \\
PSR J & & & [s] & & [rad$^2$s$^{-3}]$ & [rad$^2$s$^{-3}]$ & [s$^2]$ & &  \\
\hline
0030+0451 & WTN & -0.4 & $5.4^{+2.7}_{-2.3}$ & N & $-26.2^{+2.6}_{-2.6}$ & $-25.3^{+4.1}_{-3.2}$ & $-10.2^{+1.8}_{-2.4}$ & WTN & - \\
0348+0432 & WTN & -1.1 & $5.4^{+2.6}_{-2.3}$ & N & $-27.3^{+2.1}_{-1.8}$ & $-26.4^{+3.5}_{-2.4}$ & $-8.4^{+1.5}_{-1.8}$ & WTN & - \\
0437$-$4715$^*$ & 2C & 65.7 & $4.9^{+0.5}_{-1.0}$ & Y & $-21.9^{+2.0}_{-0.9}$ & $-27.7^{+1.6}_{-1.5}$ & $-10.5^{+0.8}_{-0.4}$ & PLRN & 4.2 \\
0711$-$6830 & WTN & 1.2 & $6.0^{+1.9}_{-2.7}$ & N & $-25.4^{+1.5}_{-3.0}$ & $-24.8^{+3.6}_{-3.4}$ & $-8.0^{+0.8}_{-1.8}$ & WTN & - \\
0737$-$3039A & WTN & -1.5 & $5.0^{+2.8}_{-2.0}$ & N & $-26.5^{+2.0}_{-2.4}$ & $-25.9^{+3.6}_{-2.8}$ & $-7.8^{+1.2}_{-2.1}$ & WTN & - \\
1017$-$7156 & WTN & 0.3 & $5.2^{+2.4}_{-2.1}$ & N & $-23.2^{+1.6}_{-4.2}$ & $-24.1^{+3.4}_{-3.8}$ & $-7.0^{+0.7}_{-2.3}$ & WTN & - \\
1022+1001 & WTN & -0.7 & $5.2^{+2.7}_{-2.1}$ & N & $-26.4^{+2.3}_{-2.5}$ & $-25.8^{+3.5}_{-2.8}$ & $-8.0^{+1.4}_{-2.2}$ & WTN & - \\
1045$-$4509 & WTN & -0.9 & $5.3^{+2.7}_{-2.2}$ & N & $-26.6^{+2.4}_{-2.3}$ & $-25.9^{+3.7}_{-2.8}$ & $-9.0^{+1.6}_{-2.1}$ & WTN & - \\
1528$-$3146 & WTN & -1.0 & $5.3^{+2.6}_{-2.3}$ & N & $-27.1^{+2.3}_{-2.0}$ & $-26.3^{+3.5}_{-2.5}$ & $-7.8^{+1.5}_{-1.9}$ & WTN & - \\
1600$-$3053 & WTN & -1.1 & $5.1^{+2.7}_{-2.2}$ & N & $-26.5^{+2.5}_{-2.3}$ & $-26.0^{+3.6}_{-2.7}$ & $-9.6^{+1.7}_{-2.1}$ & WTN & - \\
1603$-$7202 & WTN & -0.5 & $5.2^{+2.8}_{-2.2}$ & N & $-27.2^{+2.2}_{-1.8}$ & $-26.5^{+3.3}_{-2.4}$ & $-9.2^{+1.4}_{-1.7}$ & WTN & - \\
1730$-$2304 & WTN & -1.0 & $5.3^{+2.6}_{-2.3}$ & N & $-26.5^{+2.2}_{-2.4}$ & $-25.8^{+3.4}_{-2.9}$ & $-8.7^{+1.4}_{-2.1}$ & WTN & - \\
1909$-$3744 & WTN & 3.9 & $4.8^{+1.0}_{-0.9}$ & Y & $-19.9^{+1.8}_{-3.1}$ & $-26.8^{+2.4}_{-2.1}$ & $-8.3^{+1.1}_{-0.7}$ & WTN & - \\
2051$-$0827 & WTN & 4.0 & $4.9^{+2.9}_{-2.0}$ & N & $-23.4^{+1.3}_{-3.4}$ & $-23.3^{+3.7}_{-4.1}$ & $-6.5^{+0.5}_{-0.5}$ & WTN & - \\
2129$-$5721 & WTN & -1.8 & $4.9^{+3.1}_{-2.0}$ & N & $-27.3^{+2.1}_{-1.8}$ & $-26.3^{+3.8}_{-2.4}$ & $-11.0^{+1.4}_{-1.6}$ & WTN & - \\
2145$-$0750 & 2C & 25.4 & $5.4^{+2.3}_{-2.4}$ & N & $-24.1^{+0.4}_{-3.0}$ & $-24.0^{+3.4}_{-3.7}$ & $-6.2^{+0.3}_{-0.3}$ & PLRN & 33.1 \\
2222$-$0137 & WTN & -2.0 & $4.9^{+3.0}_{-1.9}$ & N & $-27.7^{+2.2}_{-1.5}$ & $-27.0^{+3.5}_{-2.1}$ & $-9.3^{+1.5}_{-1.3}$ & WTN & - \\
2241$-$5236 & 2C & 15.4 & $4.6^{+0.5}_{-0.6}$ & Y & $-19.6^{+1.3}_{-1.0}$ & $-27.4^{+1.8}_{-1.8}$ & $-9.2^{+0.7}_{-0.4}$ & PLRN & 8.7 \\
\hline
\end{tabular}

    \label{tab:model_comparison_msp}
\end{table*}
\begin{table*}
    \caption{Same as Table~\ref{tab:model_comparison_canonical}, but for magnetars with \(B_{\rm surf} \geq 4.4 \times 10^{13} \rm{G}\).}
    \begin{tabular}{lcccccccrr}
\hline
 & \multicolumn{7}{c}{} & \multicolumn{2}{c}{\citet{LowerEtAl2020}} \\
 & Model & $\ln\mathfrak{B}_{\rm BF}$ & log$_{10}$($\tau$) & Peaky? & log$_{10}$($Q_{\rm c}$) & log$_{10}$($Q_{\rm s}$) & log$_{10} \sigma_{\rm TN}^2$ & Model & $\ln \mathfrak{B}_{\rm BF}$ \\
PSR J & & & [s] & & [rad$^2$s$^{-3}]$ & [rad$^2$s$^{-3}]$ & [s$^2]$ & &  \\
\hline
1622$-$4950 & 2C & 424.7 & $7.6^{+0.4}_{-0.4}$ & Y & $-18.7^{+0.3}_{-5.7}$ & $-15.1^{+0.7}_{-0.8}$ & $5.6^{+0.2}_{-0.4}$ & PLRN & 211.8 \\
\hline
\end{tabular}

    \label{tab:model_comparison_magnetar}
\end{table*}

\twocolumn

\section{Representative selection of corner plots}
\label{app:cornerplot_selections}

\begin{figure*}
    \centering
    \begin{subfigure}{0.68\columnwidth}
        \centering
        \includegraphics[width=\textwidth]{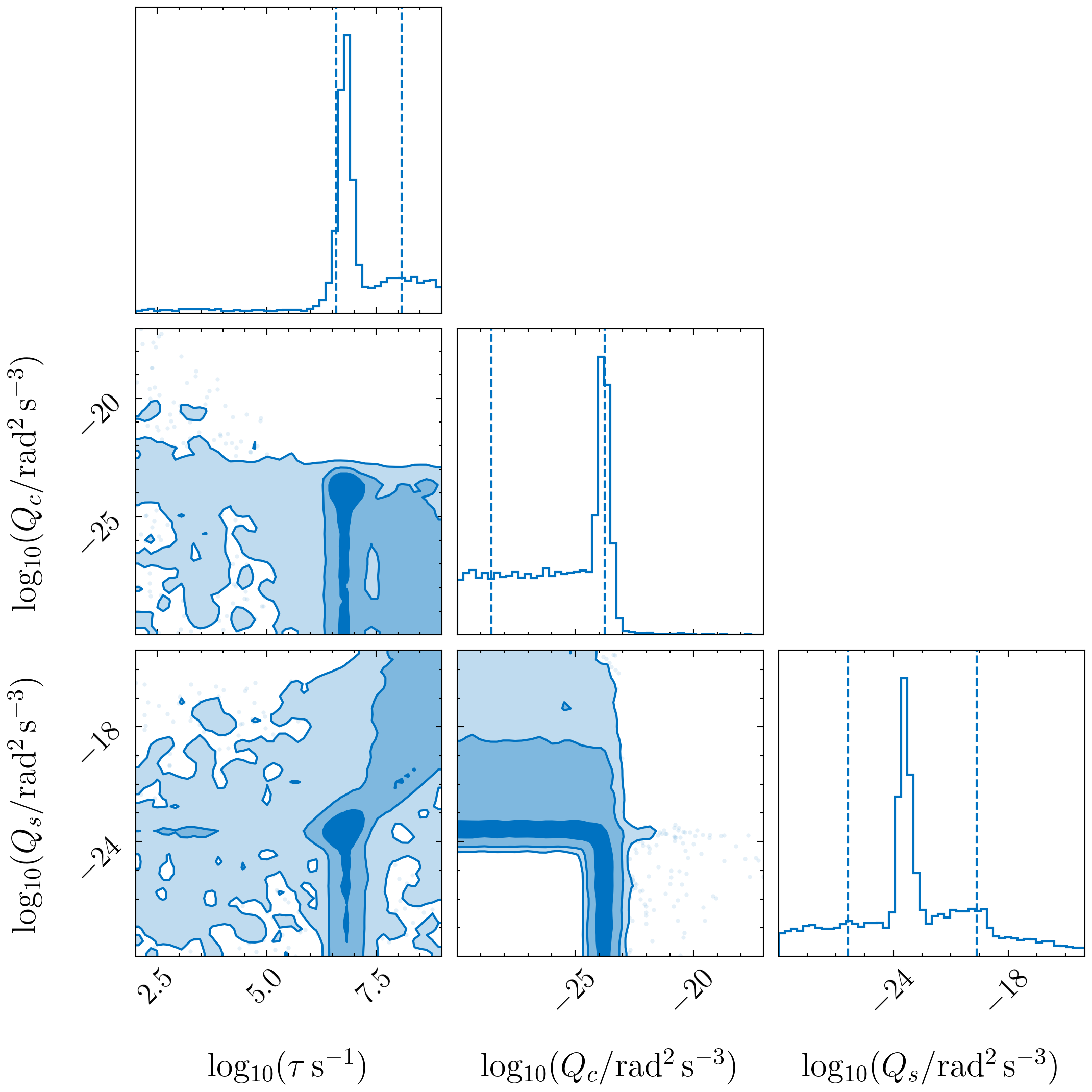}
        \caption{PSR J0729$-$1836}
        \label{subfig:corner_J0729-1836}
    \end{subfigure}
    \begin{subfigure}{0.68\columnwidth}
        \centering
        \includegraphics[width=\textwidth]{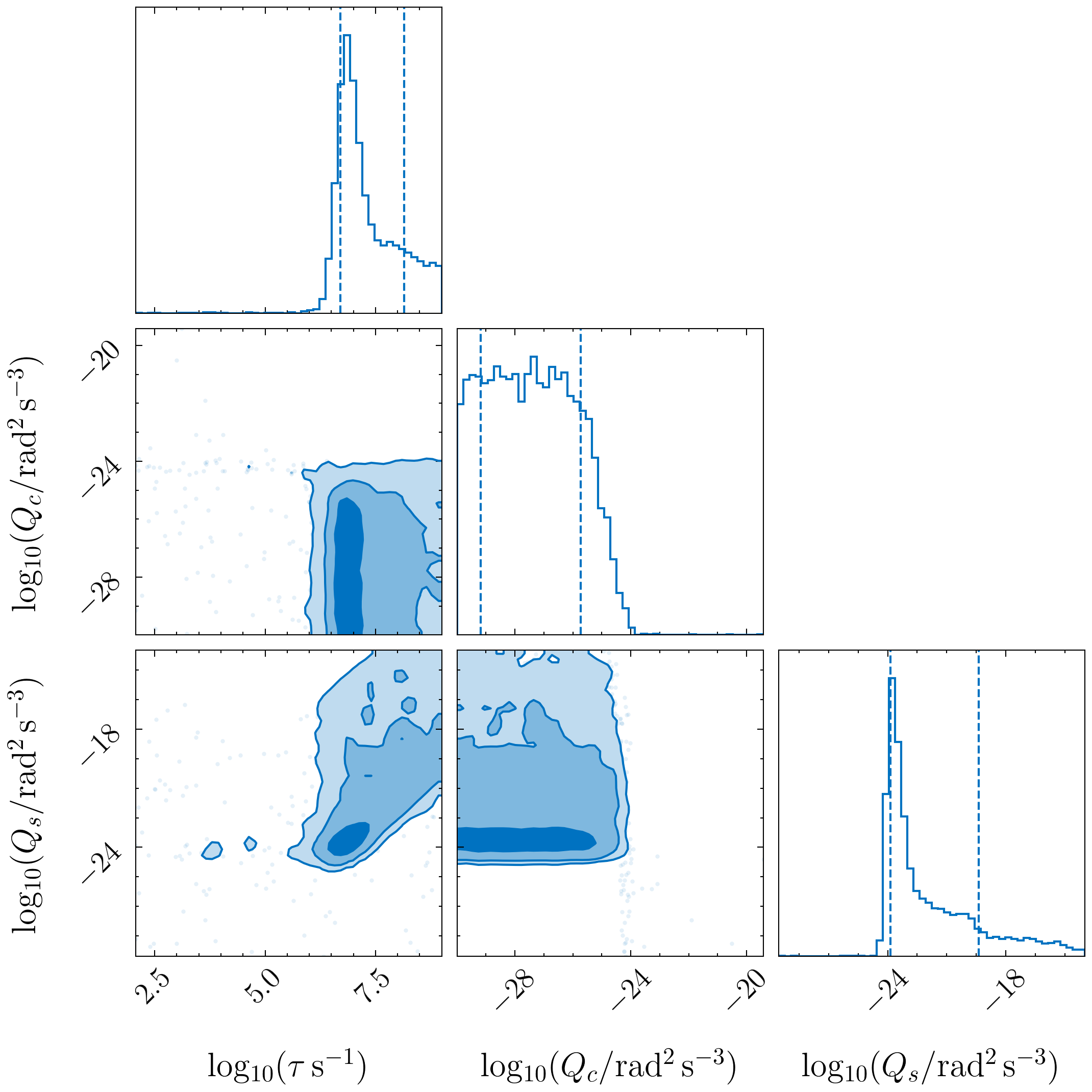}
        \caption{PSR J1326$-$5859}
        \label{subfig:corner_J1326-5859}
    \end{subfigure}
    \begin{subfigure}{0.68\columnwidth}
        \centering
        \includegraphics[width=\textwidth]{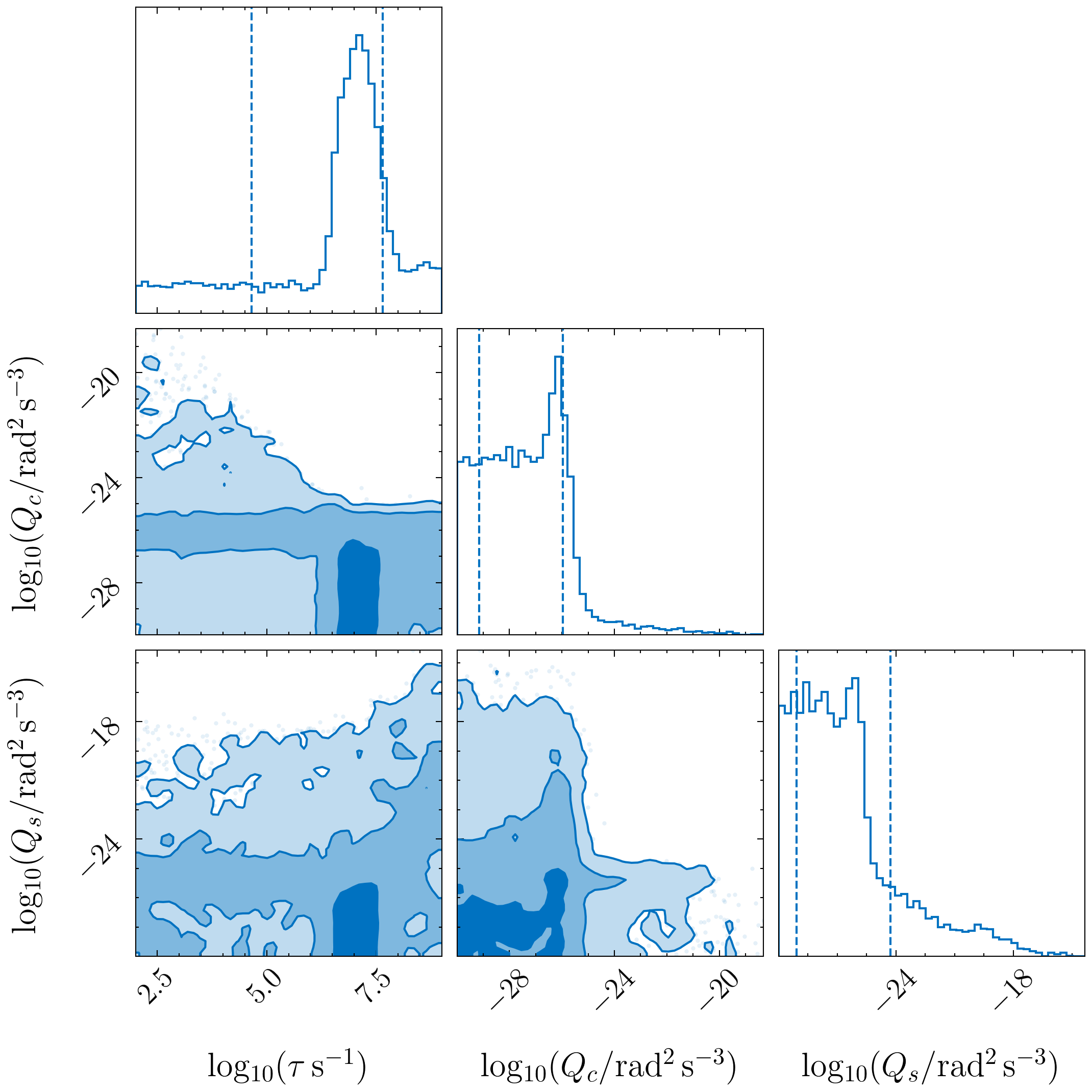}
        \caption{PSR J1849$-$0636}
        \label{subfig:corner_J1849-0636}
    \end{subfigure}
    \vfill
    \begin{subfigure}{0.68\columnwidth}
        \centering
        \includegraphics[width=\textwidth]{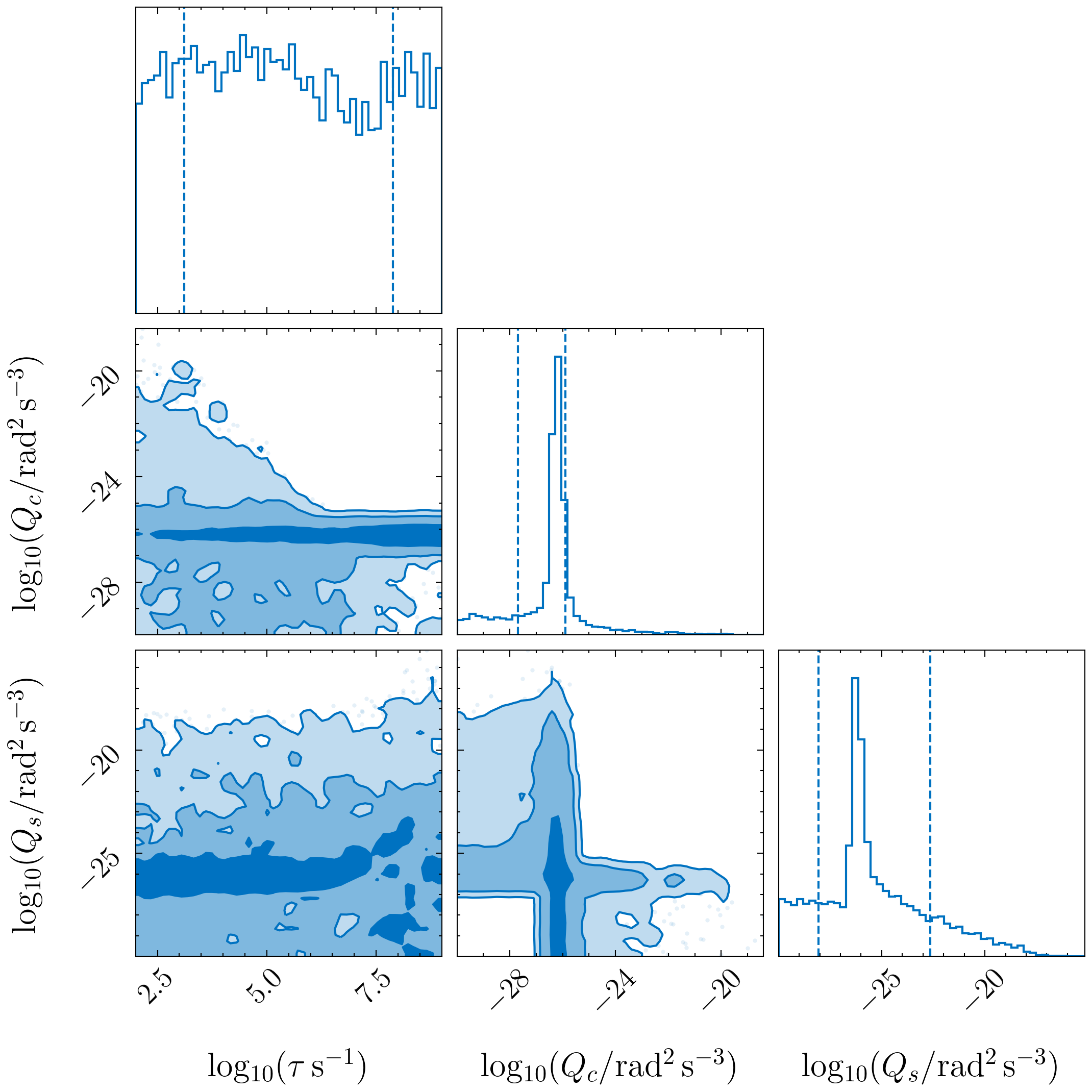}
        \caption{PSR J0837$-$4135}
        \label{subfig:corner_J0837-4135}
    \end{subfigure}
    \begin{subfigure}{0.68\columnwidth}
        \centering
        \includegraphics[width=\textwidth]{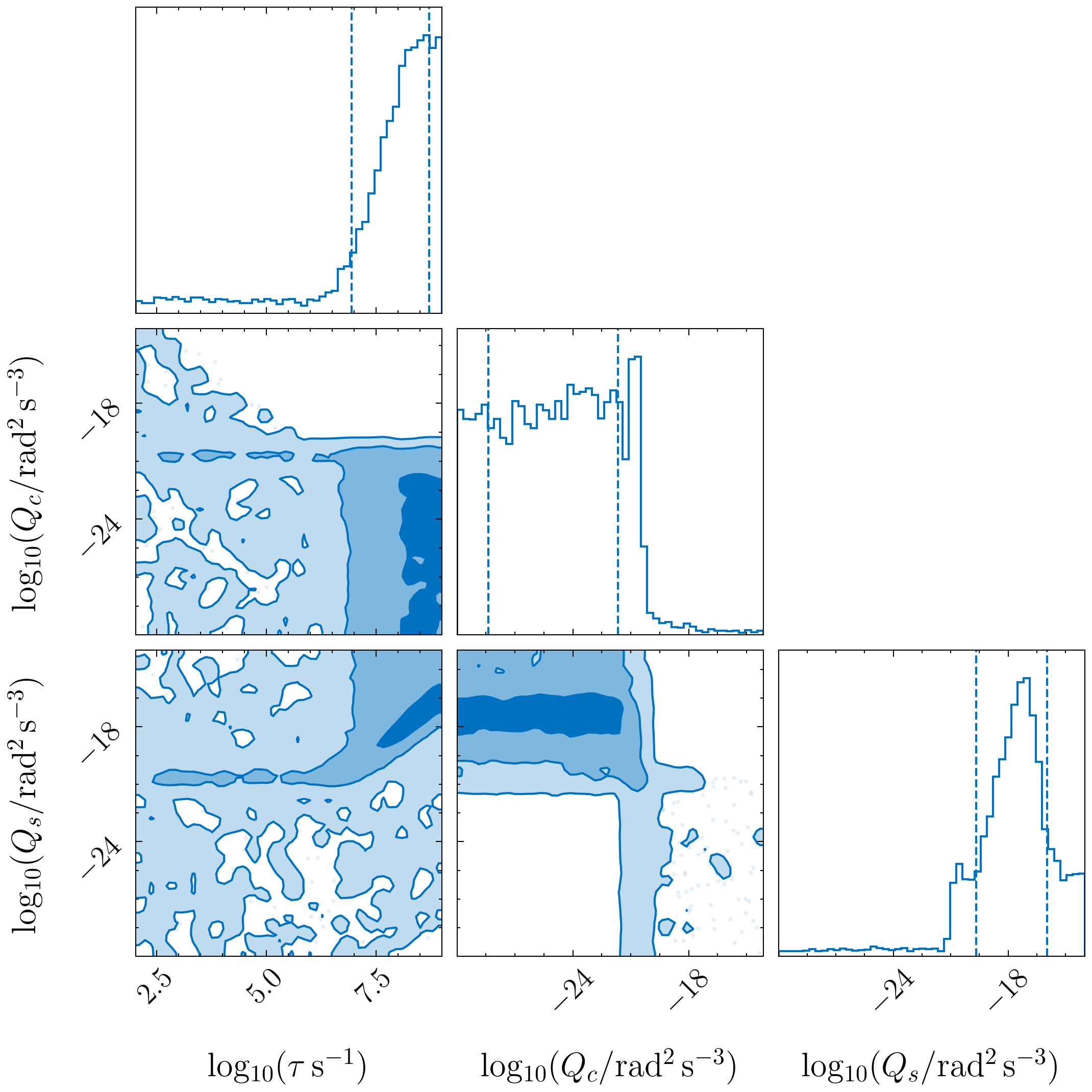}
        \caption{PSR J1803$-$2137}
        \label{subfig:corner_J1803-2137}
    \end{subfigure}

    \caption{Corner plots of the per-pulsar posterior of \(\tau\), \(Q_{\rm c}\) and \(Q_{\rm s}\) for five representative canonical pulsars (i.e.\ not recycled pulsars or magnetars) from five different subsets of the Venn diagram in Fig.~\ref{fig:venn_peaky_tau_Qc_Qs} before overlaying population-level information.
    (a) PSR J0729$-$1836: peaky one-dimensional posteriors in \(\tau\), \(Q_{\rm c}\) and \(Q_{\rm s}\).
    (b) PSR J1326$-$5859: peaky posteriors in \(\tau\) and \(Q_{\rm s}\) and left-railing posterior in \(Q_{\rm c}\).
    (c) PSR J1849$-$0636: peaky posterior in \(\tau\) and non-peaky posteriors in both \(Q_{\rm c}\) and \(Q_{\rm s}\).
    (d) PSR J0837$-$4135: peaky posteriors in both \(Q_{\rm c}\) and \(Q_{\rm s}\) and non-peaky posterior in \(\tau\).
    (e) PSR J1803$-$2137: peaky posterior in \(Q_{\rm s}\) and non-peaky posteriors in \(\tau\) and \(Q_{\rm c}\).
    Representative recycled pulsars and magnetars are displayed separately in Fig.~\ref{fig:cornerplot_msps_pop_uninformed} and \ref{fig:cornerplot_magnetar_J1622-4950-pop_uninformed}.
    The contours indicate the 1-, 2-, and 3-sigma credible regions for a two-dimensional slice of the posterior after marginalizing over the other seven two-component parameters.
    The one-dimensional histograms are obtained by marginalizing over all but one parameter.
    The vertical dashed lines in the one-dimensional histograms bracket the 68\% confidence intervals.}
    \label{fig:cornerplot_exemplary_canonical_uninformed}
\end{figure*}

\begin{figure}
    \centering
    \includegraphics[width=\columnwidth]{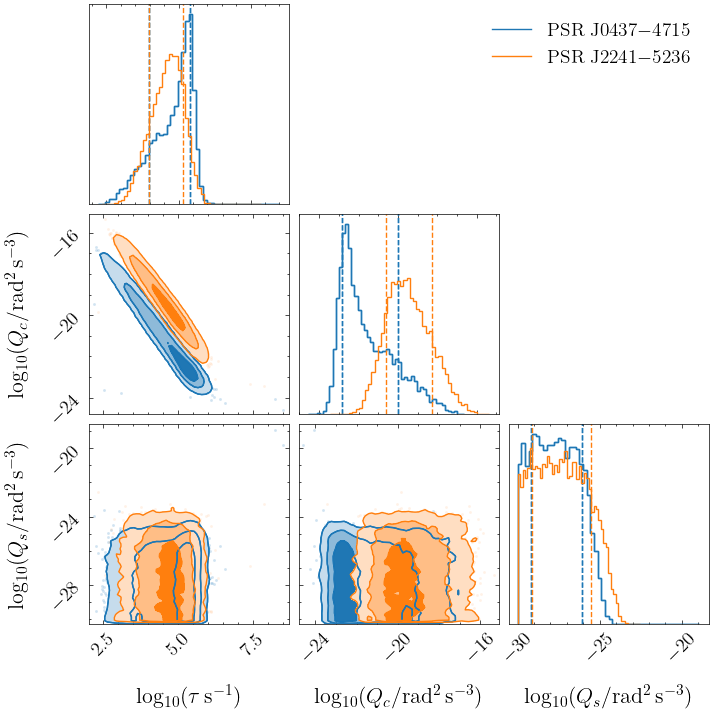}
    \caption{Corner plot of the per-pulsar posterior of \(\tau\), \(Q_{\rm c}\) and \(Q_{\rm s}\) for the recycled pulsars PSR J0437$-$4715 (blue curves) and PSR J2241$-$5236 (orange curves), which have \(\ln \mathfrak{B}_{\rm BF} \geq 5\) and peaky \(\tau\) posteriors, before overlaying population-level information.
    The two-dimensional contours and one-dimensional histograms copy the format in Fig.~\ref{fig:cornerplot_exemplary_canonical_uninformed}.}
    \label{fig:cornerplot_msps_pop_uninformed}
\end{figure}

\begin{figure}
    \centering
    \includegraphics[width=\columnwidth]{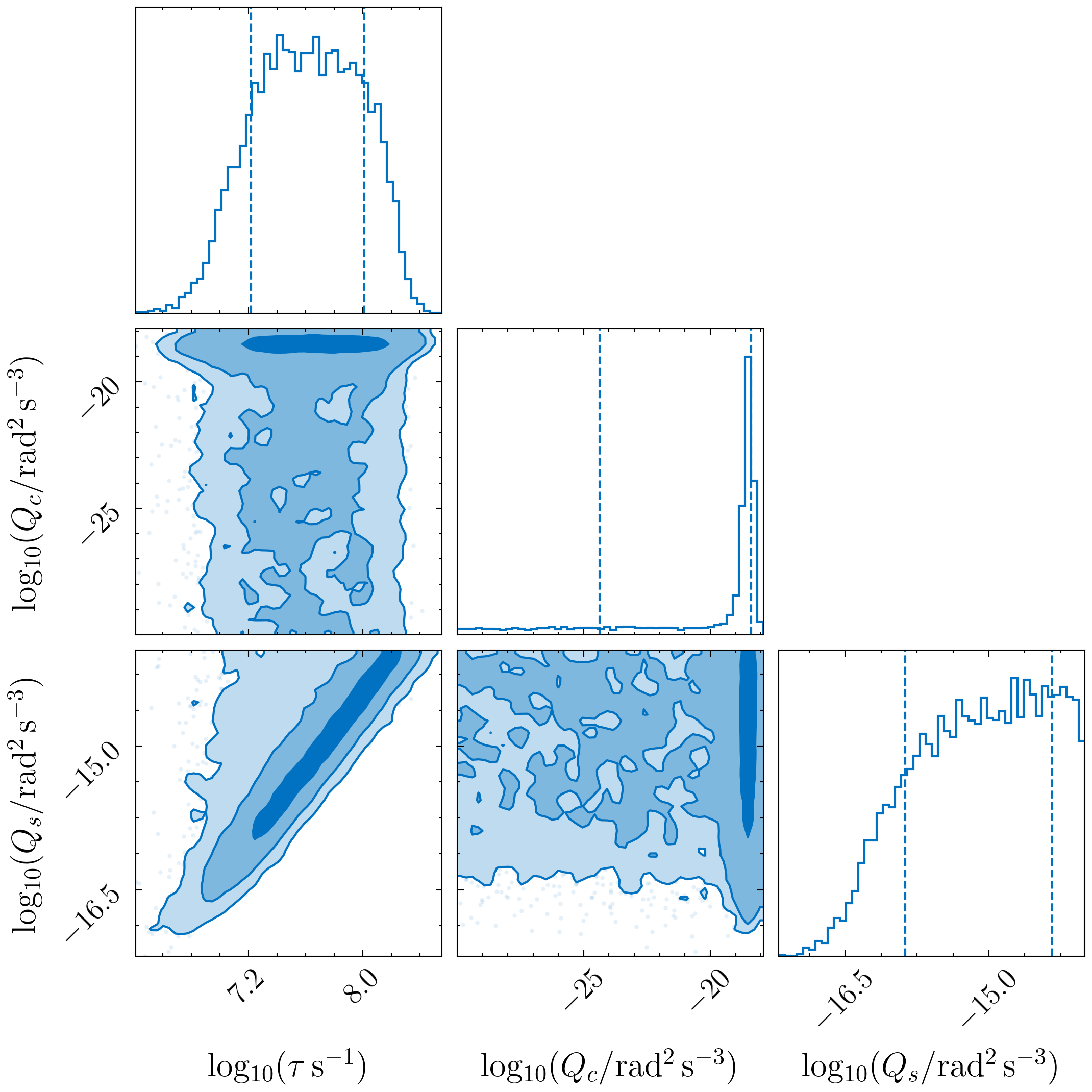}
    \caption{Same as Fig.~\ref{fig:cornerplot_msps_pop_uninformed}, but for the magnetar PSR J1622$-$4950.
    We see that \(Q_{\rm s}\) rails against the upper boundary of the prior range, \(\log_{10} (\max_{Q_{\rm s}} / \rm{rad}^2 \rm{s}^{-3}) = -14\) (see Table~\ref{tab:psrlevel_priors}). The railing accompanies the cut-off in the \(\tau\) posterior for \(\log_{10} (\tau \, \rm{s}^{-1}) \gtrsim 8.5\).}
    \label{fig:cornerplot_magnetar_J1622-4950-pop_uninformed}
\end{figure}

\begin{figure}
    \centering
    \includegraphics[width=\columnwidth]{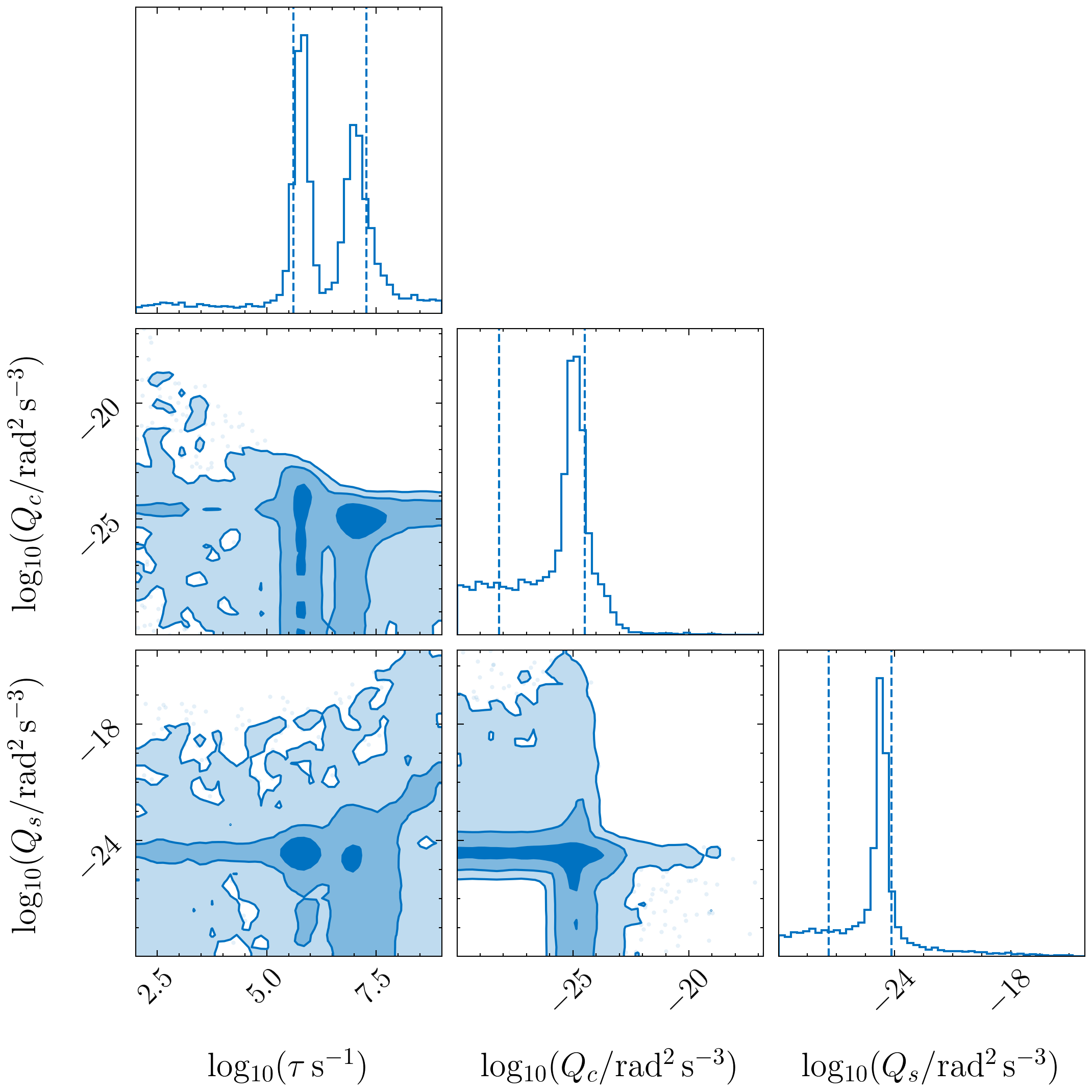}
    \includegraphics[width=\columnwidth]{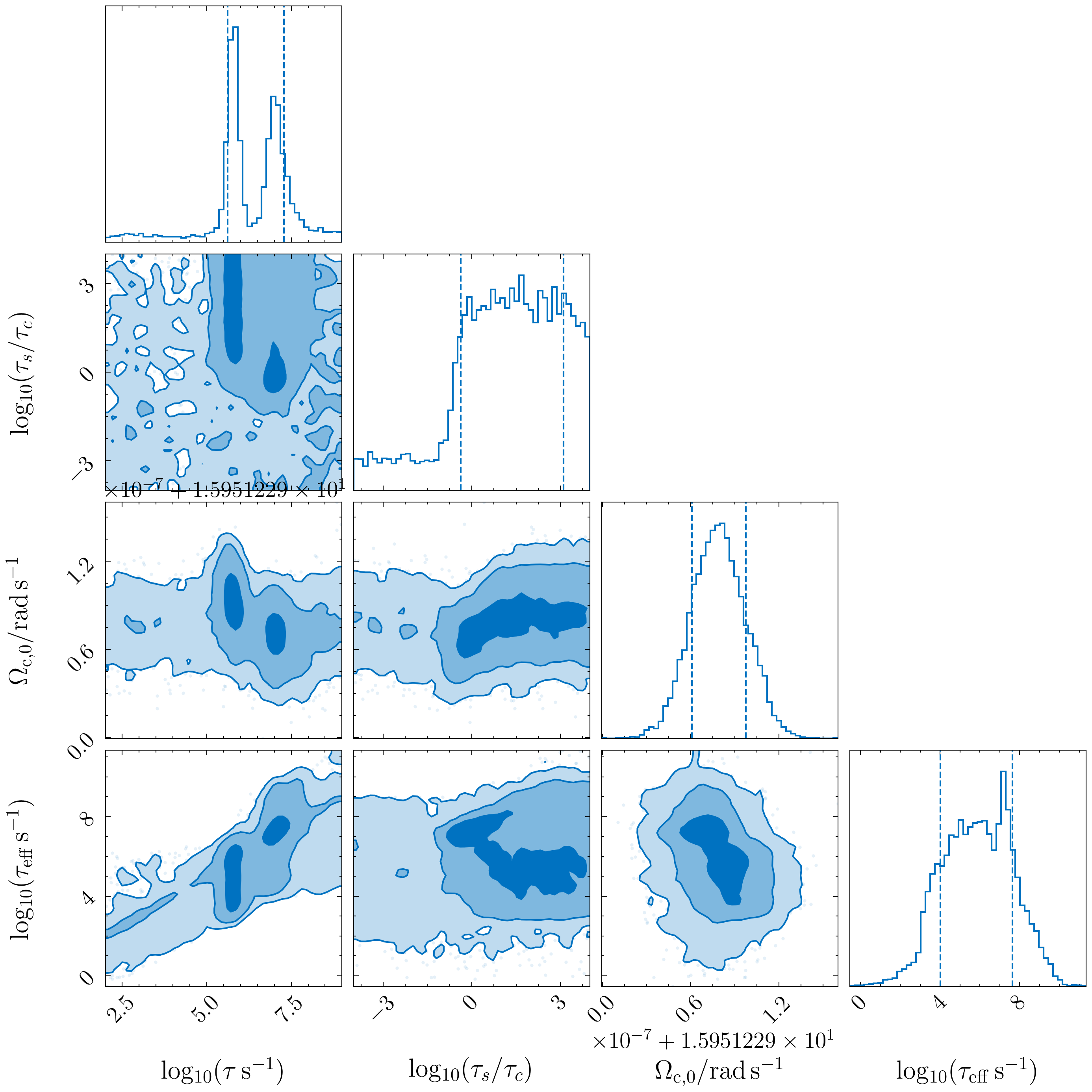}
    \caption{Corner plot of the per-pulsar posterior of \(\tau\), \(Q_{\rm c}\), and \(Q_{\rm s}\) (top panel) and \(\tau\), \(\tau_{\rm s}/\tau_{\rm c}\), \(\Omega_{\rm c,0}\), and \(\tau_{\rm eff}\) (bottom panel) before overlaying population-level information for PSR J1141$-$6545. 
    The two-dimensional contours and one-dimensional histograms copy the format in Fig.~\ref{fig:cornerplot_exemplary_canonical_uninformed}.
    We see that the \(\tau\) histogram is bimodal.
    The bottom panel showcases correlations between \(\tau\), \(\tau_{\rm s}/\tau_{\rm c}\), \(\Omega_{\rm c,0}\), and \(\tau_{\rm eff}\) for the two modes.}
    \label{fig:cornerplot_bimodal-tau_J1141-6545-pop_uninformed}
\end{figure}

\begin{figure}
    \centering
    \includegraphics[width=\columnwidth]{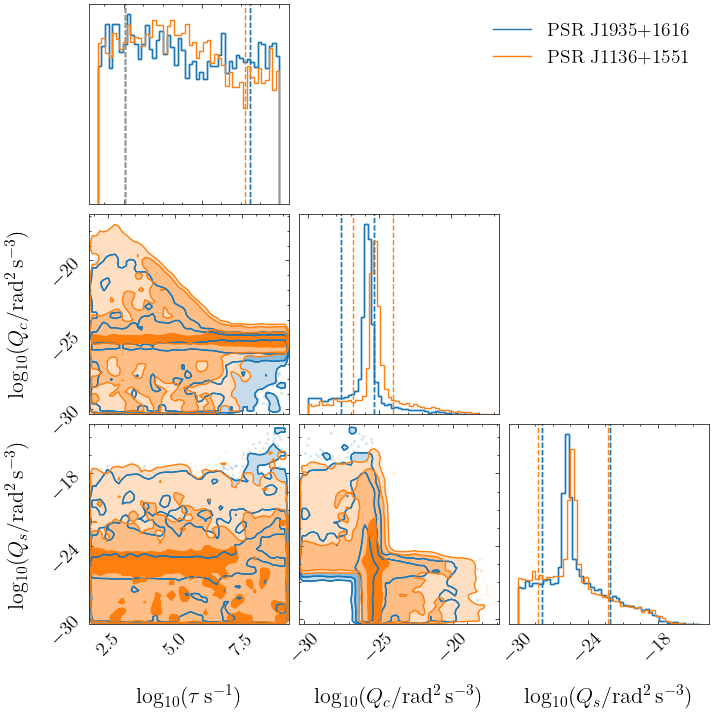}
    \caption{Same as Fig.~\ref{fig:cornerplot_msps_pop_uninformed}, but for PSR J1136+1551 (orange curves) and PSR J1935+1616 (blue curves), whose autocorrelation timescales were measured independently by \citet{PriceEtAl2012} to be 10 and 20 days respectively.}
    \label{fig:cornerplot_autocorr-based_J1136+1551-J1935+1616_pop_uninformed}
\end{figure}

In this appendix, we present corner plots of the per-pulsar posterior distributions of \(\tau\), \(Q_{\rm c}\) and \(Q_{\rm s}\) for five representative canonical pulsars (Fig.~\ref{fig:cornerplot_exemplary_canonical_uninformed}), selected from the populous core of the \(\Omega_{\rm c}\)-\(\dot{\Omega}_{\rm c}\) plane. 
We also present corner plots for the noteworthy objects discussed in Section~\ref{subsec:noteworthy_objects} (Fig.~\ref{fig:cornerplot_msps_pop_uninformed}--\ref{fig:cornerplot_autocorr-based_J1136+1551-J1935+1616_pop_uninformed}).

Each of the five panels in Fig.~\ref{fig:cornerplot_exemplary_canonical_uninformed} displays the corner plot for a canonical pulsar, (i.e.\ not a recycled pulsar or a magnetar) which is drawn from a different subset in the Venn diagram in Fig.~\ref{fig:venn_peaky_tau_Qc_Qs}.
Specifically, panel (a) displays PSR J0729$-$1836, a representative example from the subset of objects, whose \(\tau\), \(Q_{\rm c}\) and \(Q_{\rm s}\) posteriors are peaky;
panel (b) displays PSR J1326$-$5859, a representative example from the subset of objects, whose \(\tau\) and \(Q_{\rm s}\) (but not \(Q_{\rm c}\)) posteriors are peaky;
panel (c) displays PSR J1849$-$0636, a representative example from the subset of objects, whose \(\tau\) (but not \(Q_{\rm c}\) and \(Q_{\rm s}\)) posterior is peaky;
panel (d) displays PSR J0837$-$4135, a representative example from the subset of objects, that have a nearly flat \(\tau\) posterior and peaky \(Q_{\rm c}\) and \(Q_{\rm s}\) posteriors;
and panel (e) displays PSR J1803$-$2137, a representative example from the subset of objects, whose \(Q_{\rm s}\) (but not \(\tau\) and \(Q_{\rm c}\)) posterior is peaky.
Peaks in \(\tau\) posteriors in panels~(a)--(c) are at \(\log_{10} (\tau \, \rm{s}^{-1}) \approx 7.0\).
Fat high-\(\tau\) tails are common in panels~(a)--(c), as are low-\(\tau\) tails in panels~(a)--(c) and (e).
The peak in the \(Q_{\rm c}\) posterior occurs at values similar to those of the \(Q_{\rm s}\) posterior in panels~(a) and (d).
The \(Q_{\rm c}\) posterior in panel~(c) has a peak but is not classified as peaky, according to the definition in Section~\ref{subsec:tau-psr-level}, because its left half-maximum point lies within 0.5 dex of the lower prior boundary of \(Q_{\rm c}\).
Tails in the \(Q_{\rm c}\) and \(Q_{\rm s}\) posteriors are also common and appear in all five panels, except for the flat, left-railing \(Q_{\rm c}\) posterior in panel~(b).
All objects in panel~(e) have right-railing \(\tau\) posteriors, which correlate with the drop-off in the right tail of the \(Q_{\rm s}\) posterior, as illustrated by PSR J1803$-$2137.
The drop-off persists, when the upper boundaries of the prior for \(\tau\) and \(Q_{\rm s}\) are increased by two and four dex respectively.

Fig.~\ref{fig:cornerplot_msps_pop_uninformed}--\ref{fig:cornerplot_autocorr-based_J1136+1551-J1935+1616_pop_uninformed} display corner plots for PSR J0437$-$4715 and PSR J2241$-$5236 (recycled pulsars), PSR J1622$-$4950 (magnetar), PSR J1141$-$6545 (whose \(\tau\) posterior is bimodal), and PSR J1136+1551 and PSR J1935+1616 (which have independent autocorrelation-based \(\tau\) measurements), as discussed in Section~\ref{subsec:noteworthy_objects}.
In Fig.~\ref{fig:cornerplot_msps_pop_uninformed}, PSR J0437$-$4715 and PSR J2241$-$5236 have peaky \(\tau\) and \(Q_{\rm c}\) (but not \(Q_{\rm s}\)) posteriors, with peaks at \(\log_{10} (\tau \, \rm{s}^{-1}) \approx 5.5\) and \(4.8\), and \(\log_{10} (Q_{\rm c}) \approx -23.5\) and \(-20\) respectively.
In Fig.~\ref{fig:cornerplot_magnetar_J1622-4950-pop_uninformed}, the \(\tau\) posterior is truncated, while the \(Q_{\rm s}\) posterior rails right.
The truncation persists, when the upper boundaries of the prior for \(\tau\) and \(Q_{\rm s}\) are increased by two and four dex respectively.
Fig.~\ref{fig:cornerplot_bimodal-tau_J1141-6545-pop_uninformed} displays two modes in the posterior of \(\tau\) for PSR J1141$-$6545, with peaks at \(\log_{10} (\tau \, \rm{s}^{-1}) \approx 5.8\) and \(7.0\).
The primary mode is favored, when we have \(\log_{10} (Q_{\rm c} / \rm{rad}^2 \rm{s}^{-3}) \leq \log_{10} (Q_{\rm s} / \rm{rad}^2 \rm{s}^{-3}) \approx -25\).
The secondary mode is favored, when we have \(\log_{10} (Q_{\rm c} / \rm{rad}^2 \rm{s}^{-3}) \geq \log_{10} (Q_{\rm s} / \rm{rad}^2 \rm{s}^{-3}) \approx -25\).
An additional panel displaying the corner plot of \(\tau\), \(\tau_{\rm s} / \tau_{\rm c}\), and \(\Omega_{\rm c, 0}\) for PSR J1141$-$6545 is included in Fig.~\ref{fig:cornerplot_bimodal-tau_J1141-6545-pop_uninformed}, to showcase how the posteriors of \(\tau_{\rm s} / \tau_{\rm c}\), \(\Omega_{\rm c, 0}\), and \(\tau_{\rm eff}\) depend on the two modes of \(\tau\).
For example, the probability density of \(\tau_{\rm s} / \tau_{\rm c}\) concentrates at \(\log_{10}(\tau_{\rm s} / \tau_{\rm c}) \sim 0\) for the primary mode and at \(\log_{10}(\tau_{\rm s} / \tau_{\rm c}) \gtrsim 0.5\) for the secondary mode.
In Fig.~\ref{fig:cornerplot_autocorr-based_J1136+1551-J1935+1616_pop_uninformed}, the posterior of \(\tau\) is nearly flat, and the posteriors of \(Q_{\rm c}\) and \(Q_{\rm s}\) are sharply peaked at \(\log_{10} (Q_{\rm c} / \rm{rad}^2 \rm{s}^{-3}) \approx \log_{10} (Q_{\rm s} / \rm{rad}^2 \rm{s}^{-3}) \approx -25.5\).

\section{Estimating the contribution of DM variations to timing noise} \label{app:DM-variations}
UTMOST operates in a narrow band centered at \(843\,\text{MHz}\) with a bandwidth of \(\sim30\,\text{MHz}\) \citep{BailesEtAl2017,JankowskiEtAl2019,LowerEtAl2020}.
It is therefore difficult to disentangle chromatic TOA fluctuations caused by interstellar propagation, specifically DM variations, from achromatic TOA fluctuations driven by the stochastic torques acting on the crust and superfluid in the two-component model \citep{GoncharovEtAl2021}.
One rigorous way to surmount this difficulty is to generalize the Kalman filter in Section~\ref{subsec:kalman_filter} to ingest TOAs from multiple wavelengths across a wide band (\(\sim \text{GHz}\)), model the dispersive time delay between the multi-wavelength TOAs, and supplement the UTMOST TOAs with data from other radio telescopes.
This is an important but substantial programming task which lies outside the scope of this paper.
Instead, in this appendix, we estimate the impact of DM variations approximately as a first step, by comparing the DM variations measured by the MeerTime timing program with the DM variations required hypothetically to match the estimated timing noise \(\sigma_{\rm TN}\) from the two-component model.

To start, we extract the DM time series \(\text{DM}(t)\) using data from the MeerTime timing program \citep[e.g.][]{BailesEtAl2020} through the \texttt{psrdb} interface\footnote{\url{https://psrdb.readthedocs.io/en/latest/}}.
Out of a total of 300 UTMOST pulsars, 276 have DM measurements from MeerTime, and 103 out of the 276 pulsars overlap with our sample of 105 pulsars with \(\ln \mathfrak{B}_{\rm BF} > 5\).
Hereafter, we refer to the 103 pulsars as the ``common pulsars''.
The two non-overlapping objects are PSR J1651$-$5255 and PSR J1717$-$4054.
Upon assuming as a first approximation that the DM varies according to \(\text{DM}_{\rm fit}(t) = \text{DM}(0) + [\text{d(DM)}/\text{d}t]\,t\) \citep{PetroffEtAl2013,JonesEtAl2017a}, we estimate the root-mean-square residual of DM variations \(|\delta \text{DM}|_{\rm rms}\) according to 
\begin{equation}
    |\delta \text{DM}|_{\rm rms}^2 
    = \frac{1}{N_{\rm DM}} \sum_{i=1}^{N_{\rm DM}} [\text{DM}(t_i) - \text{DM}_{\rm fit}(t_i)]^2,
\end{equation}
where \(N_{\rm DM}\) is the number of DM measurements, and \(\text{DM}_{\rm fit}(t)\) is obtained via a least-squares fit to \(\text{DM}(t)\).\footnote{
    For some pulsars, sudden jumps in \(\text{DM}(t)\) are observed, which may arise from solar conjunction or extreme scintillation events.
    We consider these jumps as outliers and exclude them by sigma-clipping when fitting for \(\text{DM}(t)\) and calculating \(|\delta \text{DM}|_{\rm rms}\).
    The threshold for sigma-clipping is set to \(3\sigma\).
}
We compare \(|\delta\text{DM}|_{\rm rms}\) with the DM variations required to match the population-uninformed \(\sigma_{\rm TN}\) [equation \eqref{eq:sigma_TN_squared}], denoted by \(|\Delta \text{DM}|_{\sigma_{\rm TN}}\).
The latter quantity is calculated by inverting the dispersive time delay \citep[see e.g.][]{LamEtAl2015}, viz.
\begin{equation}
    |\Delta \text{DM}|_{\sigma_{\rm TN}} = K^{-1} \nu^{2} \sigma_{\rm TN},
\end{equation}
where \(K = 4.149\,\text{ms}\,\text{GHz}^2\,\text{pc}^{-1}\,\text{cm}^3\) is the dispersion constant, and \(\nu = 0.835\,\text{GHz}\) is the UTMOST observing frequency.
The goal is to quantify whether \(|\delta \text{DM}|_{\rm rms}\) accounts for a significant fraction of \(\sigma_{\rm TN}\), by comparing \(|\delta \text{DM}|_{\rm rms}\) with \(|\Delta \text{DM}|_{\sigma_{\rm TN}}\).

\begin{figure}
    \centering
    \includegraphics[width=\columnwidth]{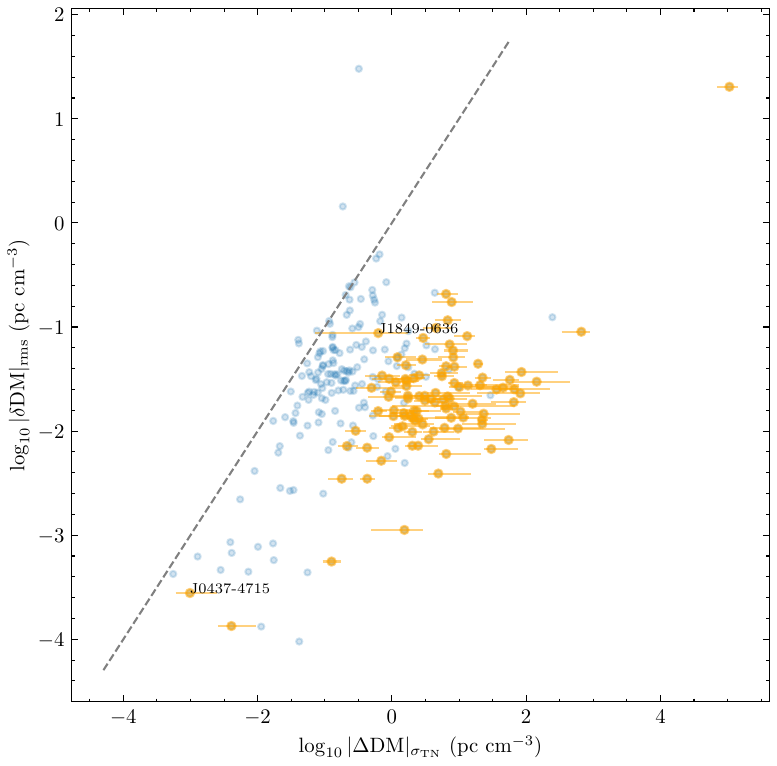}
    \caption{Scatter plot comparing \(|\Delta \text{DM}|_{\sigma_{\rm TN}}\) (in units of \(\text{pc}\,\text{cm}^{-3}\); horizontal axis), the DM variations required hypothetically to produce the population-uninformed \(\sigma_{\rm TN}\) in the absence of stochastic torques, against \(|\delta \text{DM}|_{\rm rms}\) (in units of \(\text{pc}\,\text{cm}^{-3}\); vertical axis), the estimated DM variations from the MeerTime timing program \citep{BailesEtAl2020}, for the 276 (blue symbols) MeerTime objects, of which 103 (orange symbols overlaid) are common with the 105 UTMOST objects with \(\ln \mathfrak{B}_{\rm BF} > 5\) identified in Section~4.2 and Table~B1.
    The error bars indicate the 68\% percentiles of \(|\Delta \text{DM}|_{\sigma_{\rm TN}}\) propagated from the 68\% credible intervals of \(\sigma_{\rm TN}\).
    The dashed line marks the diagonal \(|\delta \text{DM}|_{\rm rms} = |\Delta \text{DM}|_{\sigma_{\rm TN}}\).
    The two objects among the 103 common pulsars whose error bars intersect the diagonal are labelled with their PSR J names.
    In the other 101 objects, we find \(|\delta \text{DM}|_{\rm rms} \lesssim 0.1 |\Delta \text{DM}|_{\sigma_{\rm TN}}\).
    }
    \label{fig:dDM_MT_vs_dDMmax_TN}
\end{figure}

In Fig.~\ref{fig:dDM_MT_vs_dDMmax_TN}, we construct a scatter plot of \(|\delta \text{DM}|_{\rm rms}\) against \(|\Delta \text{DM}|_{\sigma_{\rm TN}}\).
The 103 common pulsars are highlighted by orange circles, with error bars propagated from the 68\% credible interval of \(\sigma_{\rm TN}\).
We find that 101 out of the 103 common pulsars have \(|\delta \text{DM}|_{\rm rms} / |\Delta \text{DM}|_{\sigma_{\rm TN}} \lesssim 0.1\), except for PSR J0437$-$4715 (recycled) and PSR J1849$-$0636 (canonical).
Specifically, for the 101 pulsars with \(|\delta \text{DM}|_{\rm rms} / |\Delta \text{DM}|_{\sigma_{\rm TN}} \lesssim 0.1\), we find \(\min(|\delta \text{DM}|_{\rm rms} / |\Delta \text{DM}|_{\sigma_{\rm TN}}) = 1.4^{+1.3}_{-0.34}\times10^{-4}\) for PSR J1048$-$5832, median \(|\delta \text{DM}|_{\rm rms} / |\Delta \text{DM}|_{\sigma_{\rm TN}} = 6.0^{+1.8}_{-2.0}\times10^{-3}\) for PSR J1836$-$1008, and \(\max(|\delta \text{DM}|_{\rm rms} / |\Delta \text{DM}|_{\sigma_{\rm TN}}) = 5.2^{+3.8}_{-3.6}\times10^{-2}\) for PSR JJ1745$-$3040.
This suggests that DM variations contribute \(\lesssim 10\%\) of the red timing noise attributed to the stochastic torques in the two-component model for most of the 103 common pulsars.
Of course, the preliminary estimates of \(|\delta \text{DM}|_{\rm rms} / |\Delta \text{DM}|_{\sigma_{\rm TN}}\) are approximate and cannot replace a rigorous multi-wavelength analysis, as noted in the first paragraph of this appendix.
With regard to the two exceptions in Fig.~\ref{fig:dDM_MT_vs_dDMmax_TN}, PSR J0437$-$4715 is a precisely timed millisecond pulsar discussed in Section~\ref{subsubsec:psr-millisec} with relatively low intrinsic, achromatic noise \citep{GoncharovEtAl2021}. 
It is excluded in the population analysis in Section~\ref{sec:hierarchical_regression}, along with other recycled pulsars. 
PSR J1849$-$0636, on the other hand, is a canonical pulsar that is included in the population analysis in Section~\ref{sec:hierarchical_regression}. 
Visual inspection of \(\text{DM}(t)\) for PSR J1849$-$0636 shows no obvious red noise structure, which is a feature of the timing noise attributed to the two-component model.
Running the hierarchical regression analysis in Section~\ref{sec:hierarchical_regression} excluding PSR J1849$-$0636, we find that the inferred values of the hyperparameters barely change, except for \(a_{Q_{\rm s}}\) and \(a_{Q_{\rm c}}\), which drop from \(a_{Q_{\rm s}} = 0.71^{+0.76}_{-0.78}\) and \(a_{Q_{\rm c}} = 1.23^{+0.80}_{-0.75}\) to \(a_{Q_{\rm s}} = 0.61^{+0.72}_{-0.73}\) and \(a_{Q_{\rm c}} = 1.04^{+0.86}_{-0.77}\).
This alters \(\sigma_{\rm TN}\) from \(\sigma_{\rm TN} \propto \Omega_{\rm c}^{-0.65^{+0.38}_{-0.39}}\) to \(\sigma_{\rm TN} \propto \Omega_{\rm c}^{-0.70^{+0.36}_{-0.39}}\) for \(\tau_{\rm s} / \tau_{\rm c} \gg 10^{-1}\), and from \(\sigma_{\rm TN} \propto \Omega_{\rm c}^{-0.39^{+0.40}_{-0.38}}\) to \(\sigma_{\rm TN} \propto \Omega_{\rm c}^{-0.48^{+0.43}_{-0.38}}\) for \(\tau_{\rm s} / \tau_{\rm c} \ll 10^{-1}\).
The changes in \(a_{Q_{\rm s}}\) and \(a_{Q_{\rm c}}\), and hence \(\sigma_{\rm TN}\), lie inside the associated \(68\%\) credible interval.

\section{Relating the inference output at the pulsar and population levels} \label{app:relating_per_pulsar_population_inference}
It is natural to ask whether the data favor the hierarchical model with population-level constraints over and above the \(N_{\rm psr}=101\) independent per-pulsar models.
In this appendix, we address this question in two ways: (1) by calculating the Bayes factor relating the hierarchical and per-pulsar models; and (2) by checking whether the per-pulsar parameter estimates shifts significantly, relative to the uncertainty, when the population-level constraints are imposed.

The Bayes factor of the hierarchical model with population-level constraints relative to the \(N_{\rm psr}=101\) independent per-pulsar models is defined by analogy with \eqref{eq:bayes_factor} as
\begin{align}
    \mathfrak{B}_{\rm BF, \Lambda}
    &= \frac{
        \int \mathrm{d}\bm{\Lambda} \int \mathrm{d}\{\bm{\theta}^{(i)}\} \, \mathcal{L}[\{\bm{d}^{(i)}\} \,|\, \{\bm{\theta}^{(i)}\}] \, p[\{\bm{\theta}^{(i)}\} | \bm{\Lambda}] \, \pi(\bm{\Lambda})
    }{
        \int \mathrm{d}\{\bm{\theta}^{(i)}\} \, \mathcal{L}[\{\bm{d}^{(i)}\} \,|\, \{\bm{\theta}^{(i)}\}] \, \pi[\{\bm{\theta}^{(i)}\}]
    }
    \label{eq:bayes_factor_hierarchical_integral}
    \\
    &= \frac{\mathcal{Z}_{\Lambda}[\{\bm{d}^{(i)}\}_{i=1}^{N_{\rm psr}}]}{
        \prod_{i=1}^{N_{\rm psr}} \mathcal{Z}[\bm{d}^{(i)}]
    } .
    \label{eq:bayes_factor_hierarchical_simplified}
\end{align}
In \eqref{eq:bayes_factor_hierarchical_integral} and \eqref{eq:bayes_factor_hierarchical_simplified}, \(\mathcal{Z}[\bm{d}^{(i)}] = \int \mathrm{d}\bm{\theta}^{(i)} \, \mathcal{L}[\bm{d}^{(i)} \,|\, \bm{\theta}^{(i)}] \pi[\bm{\theta}^{(i)}]\) and \(\mathcal{Z}_{\Lambda}[\{\bm{d}^{(i)}\}_{i=1}^{N_{\rm psr}}]\) are the evidence of the \(i\)-th per-pulsar (two-component) model and the hierarchical model with population-level constraints respectively.
The denominator of \eqref{eq:bayes_factor_hierarchical_integral} factorizes, because the \(N_{\rm psr}\) per-pulsar models are independent.

We evaluate \(\mathcal{Z}[\bm{d}^{(i)}]\) and \(\mathcal{Z}_{\Lambda}[\{\bm{d}^{(i)}\}_{i=1}^{N_{\rm psr}}]\) by nested sampling.
The nested sampling calculates the evidence with \(\mathcal{Z} = \sum_{k=1}^{n_{\rm iter}} (\min_{n_{\rm live}} \mathcal{L})_{k} \, \Delta X_{k}\), where \(\Delta X_{k}\) and \((\min_{n_{\rm live}} \mathcal{L})_{k}\) are the change in the prior volume and the minimum likelihood among \(n_{\rm live}\) likelihood samples at the \(k\)-th sampling iteration (out of a total of \(n_{\rm iter}\) iterations) respectively \citep{Skilling2006,AshtonEtAl2022}.
In this paper, we set \(n_{\rm live} = 2000\) and \(n_{\rm live} = 600\) for the per-pulsar and hierarchical population models respectively.
The result is \(\ln \mathfrak{B}_{\rm BF, \Lambda} \approx 135\).
That is, the hierarchical population level is favored strongly.

\begin{figure*}
    \centering
    \includegraphics[width=\textwidth]{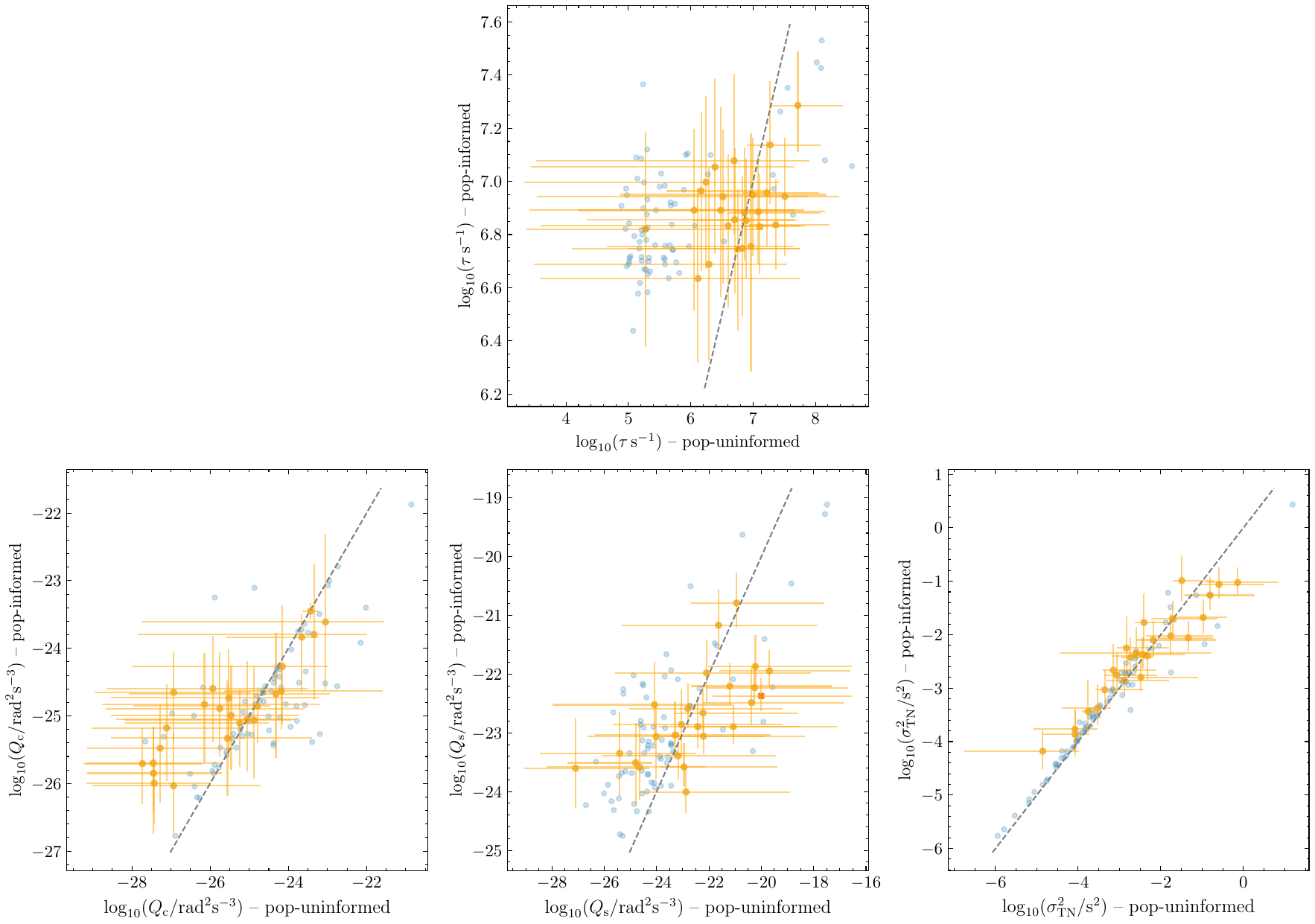}
    \caption{Scatter plots comparing population-informed and population-uninformed estimates of \(\log_{10} \tau\) (top panel; units s), \(\log_{10} Q_{\rm c}\) (bottom left panel; units \(\rm{rad}^2 \rm{s}^{-3}\)), \(\log_{10} Q_{\rm s}\) (bottom middle panel; units \(\rm{rad}^2 \rm{s}^{-3}\)), and \(\log_{10} \sigma_{\rm TN}^2\) (bottom right panel; units \(\rm{s}^{2}\)) for the 101 objects analyzed in Sections~\ref{sec:model_selection_psr_level} and \ref{sec:hierarchical_regression}.
    In all panels, the horizontal axis represents the population-uninformed estimate, while the vertical axis represents the population-informed estimate.
    Objects with peaky population-uninformed posteriors in \(\tau\) are overlaid with orange circles, with error bars indicating the 68\% credible intervals.
    The dashed line marks the diagonal.
    The object whose two-dimensional \(Q_{\rm s}\) error box does not intersect the diagonal is marked with a red cross in the bottom middle panel.}
    \label{fig:scatter_tauQcQsTN_lnBF5_canonical_informed-vs-uninformed}
\end{figure*}

To check if the parameter estimates shift significantly, when population-level constraints are imposed, we construct scatter plots for \(\tau\), \(Q_{\rm c}\), \(Q_{\rm s}\) and \(\sigma_{\rm TN}^2\) in Fig.~\ref{fig:scatter_tauQcQsTN_lnBF5_canonical_informed-vs-uninformed}, with the population-uninformed estimates on the horizontal axis and population-informed estimates on the vertical axis.
In each plot, 25 out of 28 canonical objects (i.e.\ excluding two recycled pulsars and one magnetar) with peaky, population-uninformed posteriors in \(\tau\) are plotted as orange circles, with error bars indicating the 68\% credible intervals.
Visually, it looks like the points for \(\sigma_{\rm TN}^2\) cluster near the diagonal, whereas significant scatter exists for \(\tau\), \(Q_{\rm c}\) and \(Q_{\rm s}\).
Moreover, the estimates of \(\tau\), \(Q_{\rm c}\), and \(Q_{\rm s}\) appear visually to shift towards the population medians at \(\log_{10} \tau \approx 6.8\), \(\log_{10} Q_{\rm c} \approx -25\) and \(\log_{10} Q_{\rm s} \approx -23\) (see Section~\ref{sec:hierarchical_regression}), when population-level constraints are imposed.
However, the above shifts in the central values must be interpreted within the context of the relatively large uncertainties expressed through the 68\% credible intervals.
In fact, most of the \(4\times25 = 100\) points in the four panels are consistent with the diagonal, when the uncertainties are taken into account.
To quantify this observation, we count the number of intersections between the diagonal and one-dimensional error bars and two-dimensional error boxes in Table~\ref{tab:intersections_tauQcQsTN_lnBF5_canonical_informed-vs-uninformed}.
We find that the error boxes for all 25 objects and for \(\tau\), \(Q_{\rm c}\), \(Q_{\rm s}\) and \(\sigma_{\rm TN}^2\) intersect the diagonal, except for the \(Q_{\rm s}\) error box for PSR J1401$-$6357 (marked with a red cross in Fig.~\ref{fig:scatter_tauQcQsTN_lnBF5_canonical_informed-vs-uninformed}).
In one dimension, at least 18 of the population-uninformed error bars intersect the diagonal for \(\tau\), \(Q_{\rm c}\), \(Q_{\rm s}\), and \(\sigma_{\rm TN}^2\), and the same holds for at least seven of the population-informed error bars.

\begin{table*}
    \centering
    \caption{
        The number of intersections (25 maximum) between the diagonal and the two-dimensional error boxes (column~2), one-dimensional population-uninformed error bars (column~3), and one-dimensional population-informed error bars (column~4) in the four panels in Fig.~\ref{fig:scatter_tauQcQsTN_lnBF5_canonical_informed-vs-uninformed} for \(\tau\), \(Q_{\rm c}\), \(Q_{\rm s}\) and \(\sigma_{\rm TN}^2\).}
    \label{tab:intersections_tauQcQsTN_lnBF5_canonical_informed-vs-uninformed}
    \begin{tabular}{cccc}
        \hline
        & Error boxes & Pop-uninformed error bars & Pop-informed error bars \\
        \hline
        \(\tau\) & 25 & 24 & 10 \\
        \(Q_{\rm c}\) & 25 & 22 & 13 \\
        \(Q_{\rm s}\) & 24 & 18 & 7 \\
        \(\sigma_{\rm TN}^2\) & 25 & 24 & 16 \\
        \hline
    \end{tabular}
\end{table*}

\section{Post-glitch recovery time-scale and healing parameter in the two-component model}
\label{app:healing_in_2C}
In this appendix, we review briefly the predictions of the classic two-component model \citep{BaymEtAl1969}, when it is applied to the quasi-exponential recovery following a glitch.
Specifically, we calculate the recovery time-scale \(\tau_{\rm g}\) (defined in Section~\ref{subsec:glitch_tau_comparison}) in terms of \(\tau_{\rm c}\) and \(\tau_{\rm s}\) and show that it satisfies \(\tau_{\rm g} = \tau\), where \(\tau\) is defined by \eqref{eq:tau}.
We also calculate the healing parameter \(q_{\rm heal}\) (defined in Section~\ref{subsec:healing_parameter}) in terms of \(I_{\rm c}\) and \(I_{\rm s}\) and show that it satisfies \(q_{\rm heal} < 1\), if there are additional stellar components beyond the crust and superfluid, which store angular momentum during and after the glitch.

Consider the classic two-component model in \citet{BaymEtAl1969}.
Let us assume a glitch occurs at time \(t = 0\) and the moments of inertia \(I_{\rm c}\) and \(I_{\rm s}\) do not change during the glitch.
The equations of motion for the crust and superfluid components are given by
\begin{align}
    I_{\rm c} \dot{\Omega}_{\rm c}
    &= \frac{I_{\rm s}}{\tau_{\rm s}} (\Omega_{\rm s} - \Omega_{\rm c}) - N_{\rm c} + I_{\rm c} \Delta \Omega_{\rm c, 0} \, \delta(t),
    \label{eq:EOM_c_gl}
    \\
    I_{\rm s} \dot{\Omega}_{\rm s}
    &= -\frac{I_{\rm s}}{\tau_{\rm s}} (\Omega_{\rm s} - \Omega_{\rm c}) + I_{\rm s} \Delta \Omega_{\rm s, 0} \, \delta(t),
    \label{eq:EOM_s_gl}
\end{align}
In \eqref{eq:EOM_c_gl} and \eqref{eq:EOM_s_gl}, \(\Delta \Omega_{\rm c,0}\) and \(\Delta \Omega_{\rm s,0}\) are the glitch-induced impulsive steps in \(\Omega_{\rm c}\) and \(\Omega_{\rm s}\) at \(t=0\) respectively, and \(\delta(t)\) is the Dirac delta function.
We write the solutions to \eqref{eq:EOM_c_gl} and \eqref{eq:EOM_s_gl} in the form \(\Omega_{\rm c}(t) = \Omega_{\text{c,\,no-gl}}(t) + H(t) \Delta \Omega_{\rm c}(t)\) and \(\Omega_{\rm s}(t) = \Omega_{\text{s,\,no-gl}}(t) + H(t) \Delta \Omega_{\rm s}(t)\), where \(H(t)\) is the Heaviside step function, and the subscript no-gl refers to the evolution without the glitch.

The glitch-induced changes \(\Delta \Omega_{\rm c}(t)\) and \(\Delta \Omega_{\rm s}(t)\) satisfy the initial conditions \(\Delta \Omega_{\rm c}(0) = \Delta \Omega_{\rm c,0}\) and \(\Delta \Omega_{\rm s}(0) = \Delta \Omega_{\rm s,0}\), and the equations of motion
\begin{align}
    I_{\rm c} \Delta \dot{\Omega}_{\rm c}(t) 
    &= -\frac{I_{\rm s}}{\tau_{\rm s}} [\Delta \Omega_{\rm c}(t) - \Delta \Omega_{\rm s}(t)]
    \label{eq:EOM_delta_c}
    \\
    I_{\rm s} \Delta \dot{\Omega}_{\rm s}(t) 
    &= \frac{I_{\rm s}}{\tau_{\rm s}} [\Delta \Omega_{\rm c}(t) - \Delta \Omega_{\rm s}(t)] .
    \label{eq:EOM_delta_s}
\end{align}
Upon solving \eqref{eq:EOM_delta_c} and \eqref{eq:EOM_delta_s}, we obtain \citep{AlparEtAl1993,HaskellMelatos2015}
\begin{align}
    \Delta \Omega_{\rm c}(t) 
    &= \Delta \Omega_{\rm c,0} - \frac{I_{\rm s}}{I_{\rm c} + I_{\rm s}} (\Delta \Omega_{\rm c,0} - \Delta \Omega_{\rm s,0}) \left(1 - e^{-t/\tau_{\rm g}}\right) ,
    \label{eq:Delta_Omega_c_sol}
\end{align}
where we write \(\tau_{\rm g}^{-1} = (1 + I_{\rm s}/I_{\rm c}) \tau_{\rm s}^{-1}\).
With \(I_{\rm c} / I_{\rm s} = \tau_{\rm c} / \tau_{\rm s}\), we obtain \(\tau_{\rm g} = \tau\). That is, the postglitch recovery time-scale \(\tau_{\rm g}\) is the same as the interglitch coupling time-scale \(\tau\).

From \eqref{eq:Delta_Omega_c_sol}, we extract the healing parameter \(q_{\rm heal}\) as
\begin{align}
    q_{\rm heal} = \frac{I_{\rm s}}{I_{\rm c} + I_{\rm s}} \left(1 - \frac{\Delta \Omega_{\rm s,0}}{\Delta \Omega_{\rm c,0}}\right).
\end{align}
If angular momentum is conserved between the crust and superfluid during a glitch, we must have \(-\Delta \Omega_{\rm s,0} / \Delta \Omega_{\rm c,0} = I_{\rm c} / I_{\rm s}\) and therefore \(q_{\rm heal} = 1\).
This implies that additional stellar components, beyond the crust and superfluid, are required to store angular momentum during and after the glitch, if one observes \(q_{\rm heal} < 1\) in pulsar timing data.
For example, in a region where no glitch-induced vortex moves, we have \(\Delta \Omega_{\rm s,0} = 0\) and \(q_{\rm heal} = I_{\rm s} / (I_{\rm c} + I_{\rm s}) < 1\) \citep{AlparEtAl1993,Gugercinoglu2017}.


\bsp	
\label{lastpage}
\end{document}